\documentclass{article}
\usepackage{amsmath,amssymb,pictex,cite}

\advance \topmargin by -\headheight
\advance \topmargin by -\headsep
     
\evensidemargin \oddsidemargin
\makeatletter
\@addtoreset{equation}{section}

\makeatother

\let\frac\undefined

\def\eq#1{\begin{equation}#1\end{equation}}
\long\def\subeq#1{\begin{subequations}#1\end{subequations}}
\def\Align#1{\begin{align}#1\end{align}}
\def\AlignStar#1{\begin{align*}#1\end{align*}}
\def\Aligned#1{\begin{aligned}#1\end{aligned}}
\def\Gather#1{\begin{gather}#1\end{gather}}
\def\Multline#1{\begin{multline}#1\end{multline}}

\def\lcolon{\mathopen:}
\def\rcolon{\mathclose:}

\def\Im{\mathop{\rm Im}\nolimits}

\def\Tr{\mathop{\rm Tr}\nolimits}

\def\Res{\mathop{\rm Res}\limits}
\def\const{\mathord{\rm\,const\,}}

\def\Z{{\mathbb{Z}}}
\def\cC{{\cal C}}
\def\cD{{\cal D}}

\def\cK{{\cal K}}
\def\cS{{\cal S}}
\def\cO{{\cal O}}
\def\cV{{\cal V}}
\def\rA{{\rm A}}
\def\rI{{\rm I}}
\def\rS{{\rm S}}
\def\ta{{\tilde a}}
\def\tb{{\tilde b}}

\def\bg{{\bar g}}
\def\bG{{\bar G}}
\def\hV{{\hat V}}
\def\hI{{\hat I}}
\def\bbA{{\boldsymbol A}}
\def\bbB{{\boldsymbol B}}
\def\bbxi{{\boldsymbol\xi}}
\def\bbeta{{\boldsymbol\eta}}
\def\bbj{{\boldsymbol j}}
\def\vtheta{{\skew{-1}\vec\theta}}

\def\d{\partial}
\def\e{{\rm e}}
\def\i{{\rm i}}
\def\ve{\varepsilon}
\def\vep{\varepsilon^\vphprime}
\def\phibar{{\bar\phi}}
\def\llangle{\mathopen{\langle\!\langle}}
\def\rrangle{\mathclose{\rangle\!\rangle}}
\def\ipihalf{{\textstyle{\i\pi\over2}}}
\def\ts{\textstyle}
\def\interdisplay{{\vskip -\abovedisplayskip \vskip -\belowdisplayskip
			\vskip\jot}}
\def\0{{\mskip2.35mu0\mskip2.35mu}}
\def\vphprime{{\vphantom{\prime}}}
\makeatletter
\def\section{\@startsection{section}{1}{\z@}%
                                   {-3.5ex \@plus -1ex \@minus -.2ex}%
                                   {2.3ex \@plus.2ex}%
                                   {\normalfont\normalsize\bfseries}}
\def\subsection{\@startsection{subsection}{2}{\z@}%
                                     {-3.25ex\@plus -1ex \@minus -.2ex}%
                                     {1.5ex \@plus .2ex}%
                                     {\normalfont\normalsize\bfseries}}
\def\@seccntformat#1{\csname the#1\endcsname.~~}
\long\def\@makecaption#1#2{%
  \vskip\abovecaptionskip
  \sbox\@tempboxa{\small#1. #2}%
  \ifdim \wd\@tempboxa >0.9\hsize
  {\leftskip=0.05\hsize\rightskip=0.05\hsize\relax\small
    #1. #2\par}
  \else
    \global \@minipagefalse
    \hb@xt@\hsize{\hfil\box\@tempboxa\hfil}%
  \fi
  \vskip\belowcaptionskip}
\def\Appendix{\appendix
  \def\@seccntformat##1{Appendix~\csname the##1\endcsname.~~}}

\let\over\@@over
\let\atop\@@atop
\let\above\@@above
\let\overwithdelims\@@overwithdelims
\let\atopwithdelims\@@atopwithdelims
\let\abovewithdelims\@@abovewithdelims

\makeatother

\def\Maketitle{{\def\newpage{}\maketitle}}

\allowdisplaybreaks

\def\ar#1 #2, #3 #4 {\arrow <0.3truecm> [0.1,0.3] from #1 #2 to #3 #4 }
\def\arw#1 #2, #3 #4 {\arrow <0.3truecm> [0.2,0.6] from #1 #2 to #3 #4 }
\catcode`!=11
\def\ln#1 #2, #3 #4 {\!start(#1,#2)\!ljoin(#3,#4)}
\catcode`!=12
\def\rl#1 #2, #3 #4 {\putrule from #1 #2 to #3 #4 }
\def\trl#1 #2, #3 #4 {\linethickness .8pt
	\putrule from #1 #2 to #3 #4 \linethickness .4pt}
\def\point#1 #2 {\put{\hbox{\kern -1.05pt \raise -0.6pt \hbox{.}}}
	[Bl] at #1 #2 }
\def\bfpoint#1 #2 {%
\put{\hbox{\kern -2pt \raise -2.5pt%
\hbox{$\textstyle\bullet$}}} [Bl] at #1 #2 }
\def\vdpoint#1 #2 {%
\put{\hbox{\kern -2pt \raise -2.5pt%
\hbox{$\textstyle\circ$}}} [Bl] at #1 #2 }
\begin{document}

\def\i{{\rm i}}

\title{Form factors of exponential fields\\
for two-parametric family of integrable models}

\author{V.~A.~Fateev$^{1,2}$ and M.~Lashkevich$^1$\\[\medskipamount]
$^1$~\parbox[t]{0.9\textwidth}{\normalsize\it\raggedright
Landau Institute for Theoretical Physics,
142432 Chernogolovka of Moscow Region, Russia}\\[\medskipamount]
$^2$~\parbox[t]{0.9\textwidth}{\normalsize\it\raggedright
Laboratoire de Physique Math\'ematique, Universit\'e
Montpellier~II, Pl.~E.~Bataillon, 34095 Montpellier, France}
}

\date{}

\rightline{hep-th/0402082}
\rightline{LPM--04--02}
\Maketitle

\begin{abstract}
A two-parametric family of integrable models (the $SS$ model) that
contains as particular cases several well known integrable quantum
field theories is considered. After the quantum group restriction it
describes a wide class of integrable perturbed conformal field
theories. Exponential fields in the $SS$ model are closely related to
the primary fields in these perturbed theories. We use the bosonization
approach to derive an integral representation for the form factors of
the exponential fields in the $SS$ model. The same representations for
the sausage model and the cosine-cosine model are obtained as limiting
cases. The results are tested at the special points, where the theory
contains free particles.
\end{abstract}

%%%%%%%%%%%%%%%%%%%%%%%%%%%%%%%%%%%%%%%%%%%%%%%%%%%%%%%%%%%%%%%%%%%%%%%%

\section{Introduction}
\label{sec-introduction}

A complete set of form factors of all local and quasilocal operators,
defined as matrix elements in the basis of asymptotic states,
completely determines a model of quantum field theory and provides an
important tool for studying its physical properties. In the case of an
integrable massive model the form factors can be found
exactly~\cite{KW78,Smirnov84,Smirnovbook}, as soon as the exact
scattering matrix is known, as solutions of a set of functional
(difference) equations named the form factor axioms. It is assumed
(though not proven) that all solutions of these equations provide form
factors of all local operators in the theory.

There are several techniques for solving the form factor equations. In
the case of a system of neutral particles of different masses, where
the two-particle $S$ matrices are scalar functions, any form factor
factorizes into a product of given functions (`minimal form factors')
and a trigonometric polynomial. The functional equations are reduced to
an infinite chain of linear equations for coefficients of these
polynomials~\cite{Fring:pt}. For some particular local operators these
equations can be solved for the whole set of form
factors~\cite{Yurov:1990kv,Koubek:ke}.

The situation is more complicated in the case of particles with
isotopic degrees of freedom, where the $S$ matrices are matrix
functions while the form factors are multivector functions. In this
case the form factor axioms have the form of equations in
multidimensional spaces with matrix coefficients. The historically
first approach to these systems was Smirnov's integral
representation~\cite{Smirnovbook}. It gives form factors in terms of
multiple integrals. The integrand can be represented as a particular
transcendental function times a function of a given class, which depends
on the local operator. In this way many integrable models have been
studied. The advantage of this scheme is that it provides a general
solution that corresponds to the most general local operator in the
theory. The main problem of this approach is identification of sets of
form factors with particular local operators in the model. It was only
made for several most important operators like currents and the
energy-momentum tensor.

Another approach was proposed by Lukyanov~\cite{Lukyanov93}. It is
based on the bosonization (free field representation) technique.
Bosonization was first used in conformal field theory (CFT) as a way to
construct an integral representation for correlation functions in the
Virasoro minimal models~\cite{Dotsenko:1984nm}. Lukyanov showed that
solutions to the form factor axioms can be found as traces of some
objects, namely, vertex and screening operators, very similar to those
of CFT. These operators can be expressed in terms of free bosonic
operators. This provides integral representations for form factors
different from those proposed by Smirnov. In the framework of this
approach, the form factors of a wide class of local operators, namely
the exponential fields, were found for the sine-Gordon model and the
$A$ series of the affine Toda
models~\cite{Lukyanov:1997bp,Lukyanov:1997yb,Brazhnikov:1997wn}. Note,
that this approach is intimately related to the bosonization schemes
proposed for lattice models of statistical mechanics~\cite{JMbook}.

It is worth to mention a new approach developed by Babujian and
Karowski~\cite{Babujian:2001xn}, based on what the authors call the
off-shell Bethe ansatz. The advantage of this method is that is has to
do with the same objects as the algebraic Bethe ansatz does.

In this paper we construct an integral representation for the form
factors of local fields in the two-parametric family of integrable
field theories, which is also known as the $SS$ model. This notation of
the theory is related with the form of the scattering matrix for the
fundamental particles. It is a tensor product of two sine-Gordon
soliton $S$ matrices with different coupling constants. The $SS$ model
has an explicit $U(1)\otimes U(1)$ symmetry, which can be extended up
to the symmetry generated by two quantum affine algebras
$U_{q_1}(\widehat{sl}_2)\otimes U_{q_2}(\widehat{sl}_2)$. This theory
contains as the particular cases $N=2$ supersymmetric sine-Gordon
theory, $O(4)$ and $O(3)$ non-linear sigma models and several other
interesting integrable theories. The $SS$ model possesses a dual
sigma model representation~\cite{Fateev96}, which is useful for the
short-distance renormalization group analysis of the theory. This
analysis together with the conformal perturbation theory gives us a
rather complete description of the short-distance properties of the
model. The long-distance behavior can be derived from the form factor
decomposition for correlation functions of the theory.

The study of the form factors in the $SS$ model was started by
Smirnov~\cite{Smirnov93}, who calculated the matrix elements for a
set of quantum group invariant operators as, for example, stress-energy
tensor and $U(1)$ currents. An important class of operators in the $SS$
model is formed by the exponential fields. In the restricted versions
of the theory they correspond to primary fields in the related
perturbed CFTs. Construction of an integral representation for the form
factors of the exponential fields is the main problem that we consider
in this paper. For this purpose we use a modification of the
bosonization techniques proposed by Konno~\cite{Konno97} for the
analysis of integrable lattice models.

The paper is organized as follows. In Sec.~\ref{sec-themodel} we
shortly give the necessary information about the model under
consideration. In Sec.~\ref{sec-bosonization} we describe the
bosonization procedure for the model and construct a three-parameter
family of form factors. In Sec.~\ref{sec-identification} we identify
the constructed form factors to those of the exponential operators. In
Secs.~\ref{sec-sausage} and~\ref{sec-p3zero} we consider two degenerate
limits of the model, namely, the sausage model and the cosine-cosine
model, and give some checks of our construction. The computational
details and some reference information are moved to appendices.

%%%%%%%%%%%%%%%%%%%%%%%%%%%%%%%%%%%%%%%%%%%%%%%%%%%%%%%%%%%%%%%%%%%%%%%%

\section{The model}
\label{sec-themodel}

The two-parametric family of integrable models under consideration (the
$SS$ model) possesses a Lagrangian formulation in terms of three scalar
fields
$\varphi_1$, $\varphi_2$, $\varphi_3$. The action has the
form~\cite{Fateev96}:
\eq{
\cS=\int d^2x\left(
{(\d_\mu\varphi_1)^2+(\d_\mu\varphi_2)^2+(\d_\mu\varphi_3)^2
\over8\pi}
+{\mu\over\pi}\left(
\cos(\alpha_1\varphi_1+\alpha_2\varphi_2)\e^{\beta\varphi_3}
+\cos(\alpha_1\varphi_1-\alpha_2\varphi_2)\e^{-\beta\varphi_3}
\right)
\right),
\label{sfmodel}
}
where the parameters $\alpha_1$, $\alpha_2$ and $\beta$ satisfy the
integrability condition
\eq{
\alpha_1^2+\alpha_2^2-\beta^2=1.
}
It is convenient to introduce the notation
\eq{
p_1=2\alpha_1^2,
\qquad
p_2=2\alpha_2^2,
\qquad
p_3=-2\beta^2,
\qquad
p_1+p_2+p_3=2
\label{pidef}
}
and
\eq{
\alpha_3=-\i\beta.
\label{alpha3def}
}
We have a two-parametric family of integrable models of quantum field
theory. As particular cases this family covers a number of integrable
families, like the sausage model~\cite{Fateev:1992tk} for $p_2=0$,
$p_1\ge2$, the cosine-cosine model~\cite{Bukhvostov:1980sn} for
$p_1+p_2=2$, $p_1,p_2\ge0$, the $N=2$ supersymmetric sine-Gordon
model~\cite{Kobayashi:1991st} for $p_2=2$ and $p_1\ge0$. Other models,
like the parafermion sine-Gordon and an integrable perturbation of
the $SU(2)_{p_1-2}\times SU(2)_{p_2-2}/SU(2)_{p_1+p_2-4}$ coset model,
can be obtained from this family by different restriction at
appropriate values of~$p_1$,~$p_2$~\cite{Fateev96}.

There are three essentially different regimes in the theory:
\subeq{\Align{
p_1,p_2>0,\quad p_3<0
&\qquad(\text{\bfseries Regime I});
\label{regionI}
\\
p_1,p_2,p_3>0
&\qquad(\text{\bfseries Regime II});
\label{regionII}
\\
p_1,p_2<0,
\quad
p_3>0
&\qquad(\text{\bfseries Regime III}).
\label{regionIII}
}}
The regime~I is unitary. Just this regime is the subject of this paper.
The particle content and scattering theory in the nonunitary regimes~II
and~III are rather complicated. The regime~II is of particular
interest, as it recovers the invariance under the substitutions
$\varphi_i\leftrightarrow\varphi_j$, $p_i\leftrightarrow p_j$. The
spectrum and the $S$ matrix of the model in the regime~II, which can be
extracted from the free field realization described below, is given in
the Appendix~\ref{appendix-regimeII}. A detailed description of this
regime will be given elsewhere.

In the unitary regime the model possesses the $U(1)\times U(1)$
symmetry described by the conserved topological charges%
\footnote{In this paper we use the notation different from that used
in~\cite{Fateev96}: the charges $Q_+$ and $Q_-$ here are $Q_1$ and $Q_2$
of the reference~\cite{Fateev96}.}
\eq{
Q_\pm={1\over2}(Q_1\pm Q_2),
\qquad
Q_i=\int dx^1\,j^0_i,
\qquad
j^\mu_i={\alpha_i\over\pi}\varepsilon^{\mu\nu}\d_\nu\varphi_i,
\qquad
i=1,2.
\label{Q1Q2}
}
Classically the charges satisfy the conditions
\eq{
Q_\pm\in\Z
\qquad
\text{or}
\qquad
Q_1,Q_2\in\Z,\quad Q_1+Q_2\in2\Z.
\label{Q12quantization}
}
In the quantum case these conditions are valid for the eigenvalues of
the charges. Note that the operators $Q_1$ and $Q_2$ are elements of a
wider algebra, namely the above-mentioned
$U_{q_1}(\widehat{sl}_2)\otimes U_{q_2}(\widehat{sl}_2)$ with
$q_i=-\e^{\i\pi/p_i}$.

The spectrum of the model consists of the fundamental particles
$z_{\ve\ve'}$ ($\ve,\ve'=\pm\equiv\pm1$) and their bound states. The
fundamental particles are characterized by eigenvalues of the
topological charges: $Q_1|z_{\ve\ve'}\rangle=\ve|z_{\ve\ve'}\rangle$,
$Q_2|z_{\ve\ve'}\rangle=\ve'|z_{\ve\ve'}\rangle$. The mass of the
fundamental quadruplet is proportional to the parameter~$\mu$ of the
Lagrangian. The exact relation between these quantities has a
form~\cite{Fateev96}:
\eq{
m={\mu\over\pi}\,{\Gamma\left(p_1\over2\right)\Gamma\left(p_2\over2\right)
\over\Gamma\left(p_1+p_2\over2\right)}.
\label{regimeImass}
}
The two-particle $S$ matrix of the fundamental particles as a function
of the rapidity difference $\theta$ is given by
\eq{
S_{p_1p_2}(\theta)=-S_{p_1}(\theta)\otimes S_{p_2}(\theta)
\label{sfSmatrix}
}
with $S_p(\theta)$ being the two-soliton $S$ matrix of the sine-Gordon
model with the coupling constant $\beta^2_{\rm SG}=8\pi{p\over
p+1}$~\cite{ZamZam79}:
\eq{
\Aligned{
{}\span
S_p(\theta)^{++}_{++}
=-\e^{\i\delta_p(\theta)},
\qquad
S_p(\theta)^{+-}_{+-}
=-\e^{\i\delta_p(\theta)}
{\sinh{\theta\over p}\over\sinh{\i\pi-\theta\over p}},
\qquad
S_p(\theta)^{-+}_{+-}
=-\e^{\i\delta_p(\theta)}
{\i\sin{\pi\over p}\over\sinh{\i\pi-\theta\over p}},
\\
\span
S(\theta)^{-\ve'_1,-\ve'_2}_{-\ve_1,-\ve_2}
=S(\theta)^{\ve'_1\ve'_2}_{\ve_1\ve_2},
\qquad
\delta_p(\theta)
=2\int^\infty_0{dt\over t}\,
{\sinh{\pi t\over2}\sinh{\pi(p-1)t\over2}\sin\theta t
\over\sinh\pi t\sinh{\pi pt\over2}}.
}\label{SGSmatrix}
}
The $S$ matrix (\ref{sfSmatrix}) was first considered
in~\cite{Smirnov93} and identified as the $S$ matrix of the model
(\ref{sfmodel}) in the region~I in~\cite{Fateev96}.

In the case when one of the parameters $p_1$, $p_2$ is less than one,
the scattering matrix~(\ref{sfSmatrix}) possesses poles in the interior
of the physical strip $0<\Im\theta<\pi$ at the points
\eq{
\theta=\i u_{i,n},
\qquad
u_{i,n}=\pi-\pi p_in,
\qquad
n=1,2,3,\ldots,
\quad
np_i<1,
\qquad
\text{if\quad $0<p_i<1$}.
\label{thetain}
}
These poles correspond to bound states. As in the unitary regime
$p_1+p_2\ge2$, only one of these two sets of bound states can be
present in the spectrum. The masses of the bound states are given by
\eq{
M_{i,n}=2m\sin{\pi p_in\over2},
\qquad
n=1,2,\ldots,
\quad
np_i<1.
\label{boundstatemass}
}
The bound states form quadruplets different from those of the
fundamental particles. For example, for $p_1<1$ each quadruplet
consists of one particle $b^{1,n}_+$ with $Q_2=2$, one particle
$b^{1,n}_-$ with $Q_2=-2$ and two particles $b^{1,n}_\rS$,
$b^{1,n}_\rA$ with $Q_2=0$, while $Q_1=0$ for the whole quadruplet.

%%%%%%%%%%%%%%%%%%%%%%%%%%%%%%%%%%%%%%%%%%%%%%%%%%%%%%%%%%%%%%%%%%%%%%%%

\section{Bosonization and Integral Representation for Form Factors}
\label{sec-bosonization}

To compute form factors of local operators we would like to apply the
bosonization procedure based on algebraic methods for solving the form
factor equations. It was observed by Lukyanov~\cite{Lukyanov93} that
solutions to the form factor axioms can be found in terms of
representations of the corner Hamiltonian $H$ and the vertex operators
that satisfy the commutation relations of the extended
Zamolodchikov--Faddeev (ZF) algebra. Namely, the form factors of a
given local operator are identified with traces of products of the
vertex operators with the operator $\exp(-2\pi H)$ taken in a
representation that depends on the particular form of the local
operator.%
\footnote{This picture is deeply related~\cite{Lukyanov:1993hc} to the
angular quantization procedure~\cite{Unruh:1983ms} where the operator
$\exp(-2\pi H)$ is treated as a density matrix.}
A crucial step in this approach is construction of a free field
realization of the extended algebra. This makes it possible to compute
the traces defining the form factors by application of the Wick
theorem.

Consider an integrable quantum field theory with the spectrum
consisting of a set of particles $\{z_I\}$ with the charge conjugation
matrix $C_{IJ}$.%
\footnote{The charge conjugation matrix is related to the $S$ matrix by
the crossing symmetry condition:
$\sum_{J''}C_{J'J''}S(\i\pi-\theta)^{I'J''}_{I\>\,J}
=\sum_{J''}C_{JJ''}S(\theta)^{J''I'}_{J'\>I}$.}
Assume that scattering of two particles $z_I$ with the rapidity
$\theta_1$ and $z_J$ with the rapidity $\theta_2$ into the particles
$z_{I'}$ and $z_{J'}$ is described by the $S$ matrix
$S(\theta_1-\theta_2)^{I'J'}_{I\;J}$ satisfying the unitarity relation,
crossing symmetry condition, and the Yang--Baxter equation. Then the
corresponding extended ZF algebra is generated by the {\it vertex
operators\/} $Z_I(\theta)$, the {\it corner Hamiltonian\/} $H$, and the
central elements $\Omega_I$ with the defining relations
\subeq{\label{HZalgebra}
\Align{
Z_I(\theta_1)Z_J(\theta_2)
&=\sum_{I'J'}S(\theta_1-\theta_2)^{I'J'}_{I\;J}
Z_{J'}(\theta_2)Z_{I'}(\theta_1),
\label{Zcommut}
\\*
Z_I(\theta')Z_J(\theta)
&=-{\i C_{IJ}\over\theta'-\theta-\i\pi}+O(1),
\qquad
\theta'\to\theta+\i\pi
\label{Znorm}
\\*
[H,Z_I(\theta)]
&=\i{d\over d\theta}Z_I(\theta)-\Omega_IZ_I(\theta).
\label{HZcommut}
}
If the scattering matrix for two particles $z_I$ and $z_J$ has a pole
at the point $\theta=\i u^\cK_{IJ}$ corresponding to a set $\cK$ of
bound states of the same mass, then the vertex operators should satisfy
some bootstrap conditions. Namely, the vertex operator for the
particles $z_K$ with $K\in\cK$ appear in the fusion of the operators
$Z_I$ and $Z_J$. This condition can be expressed algebraically as follows.
Let $u_\pm$ be defined by the equations
$$
u_++u_-=u^\cK_{IJ},
\qquad
{\sin u_+\over\sin u_-}={m_I\over m_J},
$$
where $m_I$, $m_J$ are masses of the particles $z_I$, $z_J$. Then we
have the relation
\eq{
Z_I(\theta'+\i u_+)
Z_J(\theta-\i u_-)
={\i\over\theta'-\theta}\sum_{K\in\cK}\Gamma^K_{IJ}Z_K(\theta)+O(1),
\qquad
\theta'\to\theta
\label{ZZboundstate}
}
}
with some constants $\Gamma^K_{IJ}$. These constants can be found from
compatibility with the conditions~(\ref{Zcommut}) and~(\ref{Znorm}) and
expressed in terms of the residues of the $S$ matrix. Therefore, an
important problem is to construct the basic vertex operators for the
fundamental particles. Then the vertex operators for bound state
particles can be derived by fusion of the basic ones.

Now we are in position to discuss application of the representation
theory of the extended algebra~(\ref{HZalgebra}) to form factors. For
any representation $\Pi$ and any element $X$ of the algebra we
introduce the following notation
\eq{
\llangle X\rrangle_\Pi
={\Tr_\Pi(\e^{-2\pi H}X)\over\Tr_\Pi(\e^{-2\pi H})}.
\label{doubleangles}
}
The main observation of the Ref.~\cite{Lukyanov93} was that the
expression of the form $\llangle Z_{I_N}(\theta_N)\ldots
Z_{I_1}(\theta_1)\rrangle_\Pi$ satisfies the form factor axioms, if we
assume that it is a meromorphic function of the variables
$\theta_1,\ldots,\theta_N$. For $\theta'_1<\ldots<\theta'_m$,
$\theta_1<\ldots<\theta_n$ the $(m+n)$-particle form factor of some
operator $\cal O$ is given by the relation
\Multline{
\langle\theta'_1I'_1,\ldots,\theta'_mI'_m|
\cO(0)|\theta_1I_1,\ldots,\theta_nI_n\rangle
\\
=N_\cO
\llangle
Z^{I'_1}(\theta'_1+\ipihalf)\ldots Z^{I'_m}(\theta'_m+\ipihalf)
Z_{I_n}(\theta_n-\ipihalf)\ldots Z_{I_1}(\theta_1-\ipihalf)
\rrangle_{\Pi(\cO)},
\label{formfactors}
}
where
$$
Z^I(\theta)=(C^{-1})^{IJ}Z_J(\theta).
$$
To obtain the form factors of a particular local operator $\cO$ we have
to choose appropriately the representation $\Pi(\cO)$ of the
algebra~(\ref{HZalgebra}). It is important that in this formulation the
form factor axioms in integrable quantum field theory are a direct
consequence of the relations~(\ref{HZalgebra}) and the cyclic property
of the trace.

The normalization factor $N_\cO$ in Eq.~(\ref{formfactors}) cannot be
established from the form factor axioms and should be calculated by
other methods. The value of the central element $\Omega_I$ on the
representation $\Pi_\cO$ gives the mutual locality index
$\e^{2\pi\i\Omega_I}$ of the operator $\cO$ and the local field
$\cV_I(x)$ that creates the particle $z_I$. It means that in the
Euclidean space
\eq{
\cO(\e^{2\pi\i}z,\e^{-2\pi\i}\bar z)\cV_I(0,0)
=\e^{2\pi\i\Omega_I}\cO(z,\bar z)\cV_I(0,0),
\qquad
z=x^1+\i x^2,
\quad
\bar z=x^1-\i x^2.
\label{mutuallocality}
}
If the operator $\cO$ has a fixed Lorentzian spin $s$ the form factors
possess an additional homogeneity property:
\eq{
\llangle Z_{I_N}(\theta_N+\vartheta)
\ldots Z_{I_1}(\theta_1+\vartheta)\rrangle
=\e^{s\vartheta}
\llangle Z_{I_N}(\theta_N)\ldots Z_{I_1}(\theta_1)\rrangle.
\label{ffhomogeneity}
}
For spinless operators (like the exponential operators, which will be
considered below) the form factors are invariant with respect to overall
shifts of rapidities and we have
\eq{
\langle\text{vac}|\cO(0)|\theta_1I_1,\ldots,\theta_NI_N\rangle
=N_\cO\llangle Z_{I_N}(\theta_N)\ldots Z_{I_1}(\theta_1)\rrangle
\qquad\text{for~~$s=0$.}
\label{vacformfactor}
}

The second important observation of the reference~\cite{Lukyanov93} is
that the extended ZF algebras can be realized in terms of free bosons
acting on a completion of a Fock space. Below we describe such a
realization for the algebra~(\ref{HZalgebra}) of the
model~(\ref{sfmodel}).

In the case of the model~(\ref{sfmodel}) we have to introduce the
vertex operators $Z_{\ve\ve'}(\theta)$ corresponding to the fundamental
particles $z_{\ve\ve'}$. The respective $S$ matrix of the fundamental
particles is $S_{p_1p_2}(\theta)$ given by Eqs.~(\ref{sfSmatrix}),
(\ref{SGSmatrix}) and the charge conjugation matrix is
\eq{
C_{\vep_1\ve'_1,\vep_2\ve'_2}
=\delta_{\vep_1,-\vep_2}\delta_{\ve'_1,-\ve'_2}.
\label{Cmatrixdef}
}
For the values of the parameters, where the bound states appear, e.~g.\
$p_1<1$, the matrix elements which contain the bound states
corresponding to the poles~(\ref{thetain}) must be taken into account.
The respective vertex operators can be expressed in terms of bilinear
combinations of the operators for the fundamental particles:
\subeq{\label{boundstateVOs}
\Align{
B^{1,n}_+(\theta)
&=-\i\Gamma_{1,n}^{-1}
\Res_{\theta'=\theta}Z_{++}(\theta'+{\ts{\i u_{1,n}\over2}})
Z_{-+}(\theta-{\ts{\i u_{1,n}\over2}}),
\label{B+def}
\\*
B^{1,n}_\rS(\theta)
&=-\i(K^{1,n}_\rS\Gamma_{1,n})^{-1}
\Res_{\theta'=\theta}(Z_{++}(\theta'+{\ts{\i u_{1,n}\over2}})
Z_{--}(\theta-{\ts{\i u_{1,n}\over2}})
+Z_{+-}(\theta'+{\ts{\i u_{1,n}\over2}})
Z_{-+}(\theta-{\ts{\i u_{1,n}\over2}})),
\label{B1def}
\\
B^{1,n}_\rA(\theta)
&=-\i(K^{1,n}_\rA\Gamma_{1,n})^{-1}
\Res_{\theta'=\theta}(Z_{++}(\theta'+{\ts{\i u_{1,n}\over2}})
Z_{--}(\theta-{\ts{\i u_{1,n}\over2}})
-Z_{+-}(\theta'+{\ts{\i u_{1,n}\over2}})
Z_{-+}(\theta-{\ts{\i u_{1,n}\over2}})),
\label{B2def}
\\*
B^{1,n}_-(\theta)
&=-\i\Gamma_{1,n}^{-1}
\Res_{\theta'=\theta}Z_{+-}(\theta'+{\ts{\i u_{1,n}\over2}})
Z_{--}(\theta-{\ts{\i u_{1,n}\over2}})
\label{B-def}
}
}
with
\eq{
\Aligned{{}
\span
\Gamma_{1,n}^2
=\i\Res_{\theta=\i u_{1,n}}(-S_{p_1}(\theta)^{+-}_{-+}
S_{p_2}(\theta)^{++}_{++})
=(-1)^{n+1}p_1\sin{\pi\over p_1}\cdot
S_{p_1}(\i u_{1,n})^{++}_{++}S_{p_2}(\i u_{1,n})^{++}_{++},
\\
\span
K^{1,n}_\rS
=\left(2{\sin{\pi\over p_2}+\sin{u_{1,n}\over p_2}
\over\sin{\pi-u_{1,n}\over p_2}}\right)^{1/2},
\qquad
K^{1,n}_\rA
=\left(2{\sin{\pi\over p_2}-\sin{u_{1,n}\over p_2}
\over\sin{\pi-u_{1,n}\over p_2}}\right)^{1/2}.
}\label{GammanKdef}
}
The nonzero elements of the charge conjugation matrix for any
quadruplet of this type read
$$
C_{+-}=C_{-+}=C_{\rS\rS}=C_{\rA\rA}=1.
$$
In the case $p_2<1$ the vertex operators $B^{2,n}_\pm(\theta)$,
$B^{2,n}_\rS(\theta)$, $B^{2,n}_\rA(\theta)$ are defined similarly.
From the physical point of view both cases $p_1<1$ and $p_2<1$ are
equivalent, but the bosonization we develop below is not symmetric with
respect to exchange of $p_1$ and $p_2$. This is why we use both cases
below while considering different degenerate limiting cases. The $S$
matrices involving bound states can be found from
Eqs.~(\ref{HZalgebra}) and~(\ref{boundstateVOs}). We do not discuss
them here in general because of their complicacy. Some important
particular cases will be considered in
Secs.~\ref{sec-sausage},~\ref{sec-p3zero}.

Note that the vertex operators $B^{i,n}_\nu$ ($\nu=+,\rS,\rA,-$) given
by Eq.~(\ref{B-def}) is not the unique choice of vertex operators for
the bound states. Linear transformations that do not affect the
normalization condition~(\ref{Znorm}) do not affect the $N$-particle
form factor contributions into the correlation functions. We shall use
this fact to get rid of cumbersome coefficients in the vertex operators
of the sausage model.

Turn our attention to the operator contents of the
theory~(\ref{sfmodel}). A generic operator local with respect to the
fields $\varphi_i(x)$ is a linear combination of the monomials
\eq{
\d_{\mu_1}^{k_1}\varphi_{i_1}\,\ldots\,
\d_{\mu_N}^{k_N}\varphi_{i_N}\,
\e^{\i a_1\varphi_1+\i a_2\varphi_2+b\varphi_3}
\label{Oa1a2a3descendants}
}
with arbitrary nonnegative integers $k_j$. Finding form factors of all
these operators is a difficult problem. Here we address a simpler
problem of finding form factors for the exponential fields:
\eq{
O^b_{a_1a_2}(x)
=\e^{\i a_1\varphi_1(x)+\i a_2\varphi_2(x)+b\varphi_3(x)}.
\label{Oa1a2b}
}
To make the notations more symmetric it is convenient to introduce an
imaginary parameter
\eq{
a_3=-\i b.
\label{a3def}
}
With this notation we have
\eq{
O_{a_1a_2a_3}(x)
=O^{\i a_3}_{a_1a_2}(x)
=\e^{\i a_1\varphi_1(x)+\i a_2\varphi_2(x)+\i a_3\varphi_3(x)}.
\label{Oa1a2a3}
}
The normalization factors for these operators were found
in Ref.~\cite{Baseilhac:1998eq}:
\Align{
N_{a_1a_2a_3}
&=\left(\mu\over2\right)^{\sum_ia_i^2}
\left(\prod_i\gamma({1\over2}+P_i)
\over\gamma({1\over2}+P)\right)^{1/2}
\notag
\\*
&\quad\times
\exp\int^\infty_0{dt\over t}\left(
{1\over\sinh t}\sum_i{\sinh^2\alpha_ia_it\over\tanh\alpha_i^2t}
+{2\sinh Pt\prod_i\sinh P_it\over\sinh^2(t/2)}-\e^{-t}\sum_ia_i^2
\right),
\label{Na1a2a3}
}
where
$$
P=\sum^3_{i=1}\alpha_ia_i,
\quad
P_i=P-2\alpha_ia_i,
\quad
i=1,2,3;
\qquad
\gamma(z)={\Gamma(z)\over\Gamma(1-z)}.
$$
The products and sums are assumed to be over $i=1,2,3$.

Before describing the whole bosonization procedure, we give the
resulting expressions for form factors. Consider a trace function
$\llangle Z_{\vep_N\ve'_N}(\theta_N)\ldots
Z_{\vep_1\ve'_1}(\theta_1)\rrangle_{\Pi(\cO_{a_1a_2a_3})}$, which,
according to Eqs.~(\ref{formfactors}),~(\ref{vacformfactor}), gives the
form factors of the operator $\cO_{a_1a_2a_3}(x)$ for $N$ fundamental
particles in the model~(\ref{sfmodel}). This function is nonzero for
even $N$ with $U(1)$ charges subject to the neutrality condition
$\sum_j\ve_j=\sum_j\ve'_j=0$. For $N=2n$ we have
\Multline{
\llangle Z_{\vep_{2n}\ve'_{2n}}(\theta_{2n})
\ldots Z_{\vep_1\ve'_1}(\theta_1)\rrangle_{\Pi(\cO_{a_1a_2a_3})}
=\tilde C^nG(k_1,k_2|\vtheta)
\\*
\times
\sum_{A_1,\ldots,A_n=\pm\atop B_1,\ldots,B_n=\pm}
\prod^n_{s=1}\int_{\cD^1_s}{d\xi_s\over2\pi}
\prod^n_{s=1}\int_{\cD^2_s}{d\eta_s\over2\pi}\,
W_\bbj(k_1,p_1|\vtheta;\bbxi)W_{\bbj'}(k_2,p_2|\vtheta;\bbeta)
\bG^{(\bbA,\bbB)}_{\bbj,\bbj'}(k_1,k_2,k_3|\bbxi;\bbeta),
\label{sfformfactors}
}
where $\vtheta=(\theta_1,\ldots,\theta_{2n})$, while the bold letters
$\bbxi=(\xi_1,\ldots,\xi_n)$ etc.\ are $n$-tuples.
The values $k_i$ (and $\kappa_i$ which appear below) are defined by
Eq.~(\ref{kdefsf}). The numbers $j_1<\ldots<j_n$ ($j'_1<\ldots<j'_n$)
are defined so that $\ve_j=-$ ($\ve'_j=-$) for $j\in\{j_s\}^n_{s=1}$
($j\in\{j'_s\}^n_{s=1}$) and $\ve_j=+$ ($\ve'_j=+$) otherwise. The
functions in the r.~h.~s.\ read
\Align{
\tilde C
&=c_1c_2\bar C_1\bar C_2C_3^2,
\label{tildeCconstdef}
\\
G(k_1,k_2|\vtheta)
&=\e^{{1\over2}(k_1+k_2)\sum^{2n}_{j=1}\theta_j}
\prod_{1\le j<j'\le 2n}G_{33}(\theta_j-\theta_{j'}),
\label{Gvthetadef}
\\
W_\bbj(k,p|\vtheta;\bbxi)
&=\prod^n_{s=1}
\left({(-1)^{j_s-1}\pi\e^{-k\xi_s}
\over\cosh{\xi_s-\theta_{j_s}-\i\pi/2\over p}}
\prod^{j_s-1}_{j=1}W(p;\theta_j-\xi_s)
\prod^{2n}_{j=j_s}W(p;\xi_s-\theta_j)\right),
\label{Wvthetadef}
\\
\bG^{(\bbA,\bbB)}_{\bbj,\bbj'}(k_1,k_2,k_3|\bbxi;\bbeta)
&=\prod^n_{s=1}\i^{{1\over2}(B_s-A_s)}
\e^{(B_s\kappa_2-A_s\kappa_3)}
\notag
\\*
&\quad\times
\prod_{1\le s<s'\le n}
\bG_{11}^{(-A_{s'},-A_s)}(\xi_s-\xi_{s'})
\bG_{22}^{(B_{s'}B_s)}(\eta_s-\eta_{s'})
\notag
\\*
&\quad\times
\prod_{1\le s,s'\le n}
\bG_{12}^{(A_sB_{s'})}(\eta_{s'}-\xi_s+{\ts{\i\pi p_1\over2}}A_s).
\label{barGvthetadef}
}
Here the signs $\pm$ are identified with $\pm1$. The constants $C_3$,
$\bar C_i$ and functions $G_{33}(\theta)$, $W(p;\theta)$,
$\bG_{ij}^{(AB)}(\theta)$ can be found in
Appendix~\ref{appendix-traces}. The constants $c_i$ are given by
Eq.~(\ref{cidef}). The integration contours
$\cD^i_s=\cD^i_s(\bbA,\bbB)$ in Eq.~(\ref{sfformfactors}) depend on
summation term and are chosen as follows. The contour $\cD^i_s$ goes
from $-\infty$ to $+\infty$ below the poles of the integrand at the
points $\theta_j+{\i\pi\over2}(p_i-1+2Mp_i)+2\pi\i N$
($M,N=0,1,2,\ldots$) and above the poles at
$\theta_j-{\i\pi\over2}(p_i-1+2Mp_i)-2\pi\i N$. Note that the poles at
$\theta_j+{\i\pi\over2}(p_i-1+2Mp_i)$ are absent for $j<j_s$ and the
poles at $\theta_j-{\i\pi\over2}(p_i-1+2Mp_i)$ are absent for $j>j_s$.
The contour $\cD^1_s$ goes above the points $\xi_{s'}-\i\pi-2\pi\i N$
and below the points $\xi_{s'}+\i\pi+2\pi\i N$, if $A_s=A_{s'}$. A
similar rule is valid for $\cD^2_s$. Besides, the contour $\cD^2_s$
goes below the poles at $\xi_{s'}-{\i\pi\over2}(p_1+p_2)+2\pi\i N$ and
above the poles at $\xi_{s'}-{\i\pi\over2}(p_1+p_2)-2\pi\i(N+1)$, if
$B_s=A_{s'}=+$, and below the poles at
$\xi_{s'}+{\i\pi\over2}(p_1+p_2)+2\pi\i N$ and above the poles
$\xi_{s'}+{\i\pi\over2}(p_1+p_2)-2\pi\i(N+1)$, if $B_s=A_{s'}=-$.

Now we describe the construction of the form
factors~(\ref{sfformfactors})--(\ref{barGvthetadef}) based on
application of the bosonization procedure. Up to now, there is no
general way to derive bosonization directly from the model and we need
to guess it. So we introduce the construction in a mathematical manner,
mostly without substantiations. The definitions will be justified by
the check of the relations~(\ref{HZalgebra}) for the final boson vertex
operators~(\ref{HZsf}). As the construction is based on Konno's
bosonization for the $U_{q,p}(\widehat{sl}_2)$ algebra~\cite{Konno97},
some motivations can be found in his paper.

Consider the boson operators $a_i(t)$ ($i\in\Z_3$) that depend on the
real parameter $t$ and satisfy the commutation relations
\eq{
[a_i(t),a_j(t')]
=t{\sinh^2{\pi t\over2}\over\sinh\pi t\sinh{\pi p_it\over2}}
\delta(t+t')\delta_{ij}.
\label{acommut}
}
It is useful to introduce the fields
\subeq{\label{allphidef}
\Align{
\phi_i(\theta;v)
&=\int^\infty_{-\infty}{dt\over\i t}\,
a_i(t)\e^{\i\theta t+\pi v|t|/4},
\label{phidef}
\\*
\phibar_i(\theta;v)
&=\int^\infty_{-\infty}{dt\over\i t}\,{\sinh\pi t\over\sinh{\pi t\over2}}
a_i(t)\e^{\i\theta t+\pi v|t|/4},
\label{phibardef}
\\
\phi^{(\pm)}_i(\theta;v)
&=2\int^\infty_0{dt\over\i t}\,
\sinh{\pi p_it\over2}
a_i(\pm t)\e^{\pm\i\theta t+\pi vt/4}.
\label{phipmdef}
}
We shall also use the notation
\eq{
\chi^{(\pm)}_i(\theta)
=\phi^{(\pm)}_i(\theta;2-p_i)
+\phi^{(\pm)}_{i+1}(\theta;p_{i+1}-2)
-\phi^{(\pm)}_{i+2}(\theta;p_{i+1}-p_i).
\label{chidef}
}
}
In the expressions below we encounter the integrals
$$
\int^\infty_0 dt\,f(t)
$$
with $f(t)$ having a pole at $t=0$. We shall understand this integral
as~\cite{JKM}
$$
\int_{\cC_0}{dt\over2\pi\i}\,f(t)\log(-t)
$$
with the contour $\cC_0$ going from $+\infty+\i0$ above the real axis, then
around zero, and then below the real axis to $+\infty-\i0$. In fact, we
shall use several simple formulas listed in the
Appendix~\ref{appendix-regint}.

The Fock space is defined by the vacuum $|0\rangle$:%
\footnote{As our goal is calculation of traces, there is no need to
introduce bra-vectors nor to discuss unitarity of the representations.}
\eq{
a_i(t)|0\rangle=0
\text{~~for $t\ge0$}.
}
as the space spanned by the vectors
$$
a_{i_1}(-t_1)\ldots a_{i_n}(-t_n)|0\rangle,
\qquad
t_1,\ldots,t_n>0,
\qquad
n=0,1,2,\ldots.
$$
This defines the evident normal ordering $\lcolon\ldots\rcolon$. In
what follows all traces are understood as traces over this Fock space.

As building blocks for the generators of the algebra~(\ref{HZalgebra})
we introduce the operators
\subeq{\label{VIdef}\Align{
V_i(\theta)
&=\lcolon\exp\left(
\i\phi_{i+1}(\theta;p_{i+1})+\i\phi_{i+2}(\theta;-p_{i+2})
\right)\rcolon,
\label{Vdef}
\\*
I^{(\pm)}_i(\theta)
&=\lcolon\exp(-\i\phibar_i(\theta;p_i)
\pm\i\chi^{(\pm)}_i(\theta))\rcolon,
\label{Idef}
}}
These operators satisfy the following relations:
\subeq{\label{opprods}\Align{
V_i(\theta')V_j(\theta)
&=g_{ij}(\theta-\theta')\lcolon V_i(\theta)V_j(\theta')\rcolon,
\label{VVprod}
\\*
V_i(\theta')I^{(\pm)}_j(\theta)
&=w^{(\pm)}_{ij}(\theta-\theta')\lcolon V_i(\theta')I^{(\pm)}_j(\theta)\rcolon,
\label{VIprod}
\\
I^{(\pm)}_j(\theta')V_i(\theta)
&=w^{(\mp)}_{ij}(\theta-\theta')\lcolon V_i(\theta)I^{(\pm)}_j(\theta')\rcolon,
\label{IVprod}
\\*
I^{(A)}_i(\theta')I^{(B)}_j(\theta)
&=\bg^{(AB)}_{ij}(\theta-\theta')
\lcolon I^{(A)}_i(\theta')I^{(B)}_j(\theta)\rcolon.
\label{IIprod}
}}
The functions $g_{ij}(\theta)$ are defined as follows ($i,j$ are
understood modulo~3):
\subeq{\Align{
g_{ii}(\theta)
&=G^{-1}(p_{i+1},\theta)G^{-1}(p_{i+2},\theta),
\quad
G(p,\theta)
=\exp\int^\infty_0{dt\over t}\,{\sinh^2{\pi t\over2}\cosh{\pi pt\over2}
\over\sinh\pi t\sinh{\pi pt\over2}}e^{-i\theta t},
\\*
g_{ij}(\theta)
&=G^{-1}_1(p_k,\theta)
\quad(i\ne j,~k\ne i,j),
\quad
G_1(p,\theta)
=\exp\int^\infty_0{dt\over t}\,{\sinh^2{\pi t\over2}
\over\sinh\pi t\sinh{\pi pt\over2}}\e^{-\i\theta t}.
}}
The functions $w^{(\pm)}_{ij}(\theta)$ can be expressed in terms of
the gamma-functions:
\subeq{\label{wdef}
\Align{
w^{(+)}_{ii}(\theta)
&=w^{(-)}_{ii}(\theta)=1,
\\*
w^{(+)}_{i-1,i}(\theta)
&=w(p_i,0|\theta),
\qquad
w^{(-)}_{i-1,i}(\theta)
=w(p_i,1|\theta),
\label{wpm01}
\\
w^{(+)}_{i+1,i}(\theta)
&=w^{(-)}_{i+1,i}(\theta)=w(p_i,1/2|\theta),
\label{wpm12}
}
where
\eq{
w(p,z|\theta)
=r_p^{-1}
\>{\Gamma\left({\i\theta\over\pi p}-{1\over2p}+z\right)
\over\Gamma\left({\i\theta\over\pi p}+{1\over2p}+z\right)},
\qquad r_p=\e^{(C_E+\log\pi p)/p}
\label{wpz}
}
}
with $C_E$ being the Euler constant. Note, that all these functions
have one series of poles at the points $\theta=-\i\pi+\i\pi pn$ or
$\theta=-\i\pi+\i\pi p(n+1/2)$ ($n=0,1,2,\ldots$) and one series of
zeros at the points $\theta=\i\pi-\i\pi pn$ or $\theta=\i\pi-\i\pi
p(n+1/2)$.

The functions $\bg^{(AB)}_{ij}(\theta)$ ($A,B=\pm$) read
\subeq{\label{bgdef}
\Align{
\bg^{(-+)}_{ii}(\theta)
&=\bg(p_1,0,0|\theta),
\qquad
\bg^{(--)}_{ii}(\theta)
=\bg^{(++)}_{ii}(\theta)
={\i\theta\over\pi p_i}\bg(p_i,0,1|\theta),
\qquad
\bg^{(+-)}_{ii}(\theta)
=\bg(p_i,1,1|\theta),
\label{bgiidef}
\\
\bg^{(-+)}_{i,i+1}(\theta)
&=\bg^{(+-)}_{i,i+1}(\theta)=1,
\qquad
\bg^{(--)}_{i,i+1}(\theta)
=\bg^{(++)}_{i,i+1}(-\theta)
={\theta-\i\pi(p_{i+1}-2)/2\over\theta-\i\pi p_{i+1}/2},
%\label{bgijdef}
\qquad
\bg^{(AB)}_{i+1,i}(\theta)
=\bg^{(BA)}_{i,i+1}(-\theta)
\label{bgijprop}
}
with
\eq{
\bg(p,z_1,z_2|\theta)
=r_p^2\>
{\Gamma\left({\i\theta\over\pi p}+{1\over p}+z_1\right)
\over\Gamma\left({\i\theta\over\pi p}-{1\over p}+z_2\right)}.
\label{bgfuncdef}
}
}
The functions $\bg^{(AB)}_{i,i\pm1}(\theta)$ are rational and provide
commutativity of the operators $I^{(A)}_i$ and $I^{(B)}_j$ with
$i\ne j$. It will be important later that the residues at the only
poles of the products~$I^{(\pm)}_i(\theta)I^{(\pm)}_{i+1}(\theta')$
satisfy the relation
\eq{
\Res\nolimits_{\theta'=\theta-{i\pi p_{i+1}\over2}}
I_i^{(+)}(\theta)I_{i+1}^{(+)}(\theta')
=-\Res\nolimits_{\theta'=\theta+{i\pi p_{i+1}\over2}}
I_i^{(-)}(\theta)I_{i+1}^{(-)}(\theta')
\label{ResIpIpeqImIm}
}
due to the important identity for the normal products
\eq{
\textstyle
\lcolon I_i^{(+)}(\theta)I_{i+1}^{(+)}
(\theta-{i\pi p_{i+1}\over2})\rcolon
=\lcolon I_i^{(-)}(\theta)I_{i+1}^{(-)}
(\theta+{i\pi p_{i+1}\over2})\rcolon.
\label{IpIpeqImIm}
}

There is a general rule concerning positions of poles, which appear in
the operator products. Let $U_j(\theta)$ be normal ordered
exponentials, like, for example, the operators $V_i(\theta)$ or
$I^{(\pm)}_i(\theta)$. Then the product $U_k(\theta_k)U_j(\theta_j)$
can be reduced to the normal form
$$
U_k(\theta_k)U_j(\theta_j)
=g_{U_kU_j}(\theta_j-\theta_k)
\,\lcolon U_k(\theta_k)U_j(\theta_j)\rcolon.
$$
The set of the poles of the function $g_{U_kU_j}(\theta_j-\theta_k)$ in
the variable $\theta_j$ is bounded in the lower half plane, i.~e.\ the
imaginary parts of all poles are greater than some constant. On the
contrary, the set of the poles in the variable $\theta_k$ is bounded in
the upper half plane. The same is true for any product
$U_N(\theta_N)\ldots U_1(\theta_1)$: the set of poles in the variable
$\theta_j$ arising due to any operator $U_k(\theta_k)$ with $k>j$
($U_k$ is to the left of $U_j$) is bounded in the lower half plane,
while the set of poles related to an operator $U_k(\theta_k)$ with
$k<j$ ($U_k$ is to the right of $U_j$) is bounded in the upper half
plane.

As a result of the relations~(\ref{opprods})--(\ref{bgdef}), the
commutation relations of the operators introduced above are
\subeq{\label{VIicommutations}\Align{
V_i(\theta_1)V_i(\theta_2)
&=-S_{p_{i+1}}(\theta_1-\theta_2)^{++}_{++}
S_{p_{i+2}}(\theta_1-\theta_2)^{++}_{++}V_i(\theta_2)V_i(\theta_1),
\label{VViicommut}
\\*
V_i(\theta_1)V_{i\pm1}(\theta_2)
&={G_1(p_{i+2};\theta_1-\theta_2)
\over G_1(p_{i+2};\theta_2-\theta_1)}
V_{i\pm1}(\theta_2)V_i(\theta_1),
\label{VVijcommut}
\\
V_i(\theta_1)I^{(A)}_{i+1}(\theta_2)
&={\sinh{\theta_2-\theta_1-\i\pi/2\over p}
\over\sinh{\theta_2-\theta_1+\i\pi/2\over p}}
I^{(A)}_{i+1}(\theta_2)V_i(\theta_1),
\label{VIii+1commut}
\\
V_i(\theta_1)I^{(A)}_{i-1}(\theta_2)
&={\cosh{\theta_2-\theta_1-\i\pi/2\over p}
\over\cosh{\theta_2-\theta_1+\i\pi/2\over p}}
I^{(A)}_{i-1}(\theta_2)V_i(\theta_1),
\label{VIii-1commut}
\\
I^{(A)}_i(\theta_1)I^{(B)}_i(\theta_2)
&={\sinh{\theta_2-\theta_1+\i\pi\over p}
\over\sinh{\theta_2-\theta_1-\i\pi\over p}}
I^{(B)}_i(\theta_2)I^{(A)}_i(\theta_1),
\label{IIiicommut}
\\*
I^{(A)}_i(\theta_1)I^{(B)}_{i+1}(\theta_2)
&=I^{(B)}_{i+1}(\theta_2)I^{(A)}_i(\theta_1).
\label{IIijcommut}
}}
Note that the commutation relations are completely independent of the
values $(\pm)$ of the superscripts.

We can see from the relation~(\ref{VViicommut}) that the operator
$V_3(\theta)$ is a good candidate for the operator $Z_{++}(\theta)$.
However, it is known that the ZF algebras corresponding to nondiagonal
$S$ matrices cannot be represented in terms of pure exponentials of
free fields. It is necessary to consider expressions containing
integrals of normal exponents of free fields over the spectral
variable~$\theta$. These integrals first appeared in
CFT~\cite{Dotsenko:1984nm}, where they screened the total charge in the
auxiliary Coulomb gas and were called {\it screening operators}. In
particular, in CFT they are necessary to provide the braiding relations
for the vertex operators, which are CFT analogs of the ZF operators.

In our case the analysis of the commutation relations for the vertex
operators gives us the reasons to guess the following form for the
screening operator:
\eq{
S_i(k,\kappa|\theta)
=c_i\int_{\cC_i}{d\xi\over2\pi\i}\,
(I^{(+)}_i(\xi)\e^\kappa-\i I^{(-)}_i(\xi)\e^{-\kappa})
{\pi\e^{-k\xi}\over\sinh{\xi-\theta-\i\pi/2\over p_i}},
\label{Ssf}
}
with some normalization constants $c_i$, which will be determined
later. The contour $\cC_i$ in this equation goes from $-\infty$ to
$+\infty$ above the pole at the point~$\theta+\i\pi/2$. As for the
poles related to other operators, the contour goes below all poles
arising due to the operators standing to the left of the screening
operator~$S_i$ and above the poles related to the operators standing to
the right of~$S_i$. This is possible due to the remark concerning the
position of the poles, which is formulated above after Eq.~(\ref{wpz}).

In what follows the screening operators appear in the combinations
$V_{i-1}(\theta)S_i(k,\kappa|\theta)$ and $V_{i+1}(\theta+\i\pi
p_i/2)S_i(k,\kappa|\theta)$. In the first product the contour $\cC_i$
goes from $-\infty$ to $+\infty$ above the poles at the points
$\theta+{\i\pi\over2}-\i\pi p_in$ ($n=0,1,2,\ldots$) and below those at
the points $\theta-{\i\pi\over2}+\i\pi p_in$. In the second product it
goes above the poles at $\theta+{\i\pi\over2}-\i\pi p_in$ and below
those at $\theta-{\i\pi\over2}+\i\pi p_i(n+1)$. It is important to note
that there is no real inflection of the contours between the poles at
$\theta+\i\pi/2$ and $\theta-\i\pi/2$ in the first product, because
there is really only one of these two poles in each term: for
$I^{(+)}(\xi)$ a pole may appear at $\theta-\i\pi/2$ and for
$I^{(-)}(\xi)$ at $\theta+\i\pi/2$. In fact, the contour must be chosen
separately for each term of the considered operator product (see
Fig.~\ref{figure-VIcontours}).%
\begin{figure}[t]
$$
\Aligned{
\int{d\xi\over2\pi\i}\,
V_3(\theta)I_1^{(+)}(\xi)
{\pi\e^{-k_1\xi}\over\sinh{\xi-\theta-\i\pi/2\over p_1}}
&\qquad
\vcenter{
\beginpicture
\setcoordinatesystem units <1cm,1cm> point at 0 0
\put{} [lB] at 0 1
\put{} [lB] at 0 -1
\put{} [lB] at -1 0
\put{} [lB] at 2 0
\bfpoint 0 -0.5
\put{$\scriptstyle\theta-{\i\pi\over2}$} [lB] <4pt,-2pt> at 0 -0.5
\setplotsymbol(.)
\setquadratic
\plot -1 0  -0.6 -0.2  -0.2 -0.55  0.2 -0.8
0.6 -0.8  1.0 -0.6  1.4 -0.25  1.6 -0.1  2.0 0 /
\arw 1.98 -0.005 , 2 0
\put{$\scriptstyle\xi$} [lB] <2pt,2pt> at -1 0
\endpicture
}
\\
\int{d\xi\over2\pi\i}\,
V_3(\theta)I_1^{(-)}(\xi)
{\pi\e^{-k_1\xi}\over\sinh{\xi-\theta-\i\pi/2\over p_1}}
&\qquad
\vcenter{
\beginpicture
\setcoordinatesystem units <1cm,1cm> point at 0 0
\put{} [lB] at 0 1
\put{} [lB] at 0 -1
\put{} [lB] at -1 0
\put{} [lB] at 2 0
\bfpoint 0 0.5
\put{$\scriptstyle\theta+{\i\pi\over2}$} [lB] <4pt,-2pt> at 0 0.5
\setplotsymbol(.)
\setquadratic
\plot -1 0  -0.6 0.2  -0.2 0.55  0.2 0.8
0.6 0.8  1.0 0.6  1.4 0.25  1.6 0.1  2.0 0 /
\arw 1.98 0.005 , 2 0
\put{$\scriptstyle\xi$} [lB] <2pt,6pt> at -1 0
\endpicture
}
\\
\int{d\xi\over2\pi}\,
V_3(\theta)I_2^{(\pm)}(\xi)
{\pi\e^{-k_2\xi}\over\cosh{\xi-\theta-\i\pi/2\over p_2}}
&\qquad
\vcenter{
\beginpicture
\setcoordinatesystem units <1cm,1cm> point at 0 0
\put{} [lB] at 0 1.5
\put{} [lB] at 0 -1.5
\put{} [lB] at -1 0
\put{} [lB] at 2 0
\bfpoint 0 1.3
\put{$\scriptstyle\theta+{\i\pi\over2}(p_2-1)$} [lB] <4pt,-2pt> at 0 1.3
\bfpoint 0 -1.3
\put{$\scriptstyle\theta-{\i\pi\over2}(p_2-1)$} [lB] <4pt,-2pt> at 0 -1.3
\setplotsymbol(.)
\ln -1 0 , 2 0
\arw 1.98 0 , 2 0
\put{$\scriptstyle\xi$} [lB] <2pt,3pt> at -1 0
\endpicture
}
}
$$
\caption{Integration contours in the screening operators for different
products that appear in Eqs.~(\ref{HZsf}).}
\label{figure-VIcontours}
\end{figure}

The meaning of the screening operator $S_i$ is to change the
topological number $Q_i$ by $-2$.%
\footnote{The charge $Q_3$ will appear in
Sec.~\ref{sec-identification}, when we will consider other regions of
the parameters of the model~(\ref{sfmodel}).}
Each screening operator $S_i$ depend on its own pair of parameters
$k_i$ and $\kappa_i$. We shall see that, subject to some additional
conditions, their values do not affect the commutation relations of the
vertex operators.

As the $S$ matrix factorizes into a tensor product and the screening
operators change the topological numbers, different screening operators
must commute:
\eq{
[S_i(k_i,\kappa_i|\theta_1),S_j(k_j,\kappa_j|\theta_2)]=0,
\qquad
i\ne j.
\label{SiSjcommut}
}
This condition imposes some relations on the parameters
$k_i$,~$\kappa_i$.

Since the screening currents $I^{(\pm)}_i(\theta)$ commute according to
Eq.~(\ref{IIijcommut}), the only obstacle to commutativity of the screening
operators is the poles of the functions $\bg^{(++)}_{ij}(\theta)$ and
$\bg^{(--)}_{ij}(\theta)$. Due to these poles, we have
\Multline{
[S_1(k_1,\kappa_1|\theta_1),S_2(k_2,\kappa_2|\theta_2)]
=\int{d\xi\over2\pi\i}\,\biggl(
\e^{\kappa_1+\kappa_2-k_1\xi-k_2(\xi-{\i\pi p_2\over2})}
\lcolon I^{(+)}_1(\xi)I^{(+)}_2(\xi-{\textstyle{\i\pi p_2\over2}})\rcolon
\\
-\e^{-\kappa_1-\kappa_2-k_1\xi-k_2(\xi+{\i\pi p_2\over2})}
\lcolon I^{(-)}_1(\xi)I^{(-)}_2(\xi+{\textstyle{\i\pi p_2\over2}})\rcolon
\biggr)
{\pi^3c_1c_2
\over
\sinh{\xi-\theta_1-\i\pi/2\over p_1}\cosh{\xi-\theta_2-\i\pi/2\over p_2}}.
\notag
}
As, according to Eq.~(\ref{IpIpeqImIm}), the normal products in the
right hand side coincide, the commutativity relation (\ref{SiSjcommut})
holds if
$$
\e^{\kappa_1+\kappa_2-k_1\xi-k_2(\xi-{\i\pi p_2\over2})}
=\e^{-\kappa_1-\kappa_2-k_1\xi-k_2(\xi+{\i\pi p_2\over2})}.
$$
To fulfill this equation we can take%
\footnote{Other solutions do not lead to physically different results.}
$$
\textstyle
\kappa_1+\kappa_2=-{\i\pi\over2}p_2k_2,
$$
and, by cyclic permutations,
$$
\textstyle
\kappa_2+\kappa_3=-{\i\pi\over2}p_3k_3,
\qquad
\kappa_1+\kappa_3=-{\i\pi\over2}p_1k_1.
$$
Solving these equations, we get
\eq{
\kappa_i=-{\i\pi\over4}(p_ik_i+p_{i+1}k_{i+1}-p_{i+2}k_{i+2}).
\label{kappadef}
}
In what follows we assume $\kappa_1$, $\kappa_2$, $\kappa_3$ to be the
functions of $k_1$, $k_2$, $k_3$, given by (\ref{kappadef}).

Now we are ready to express the vertex operators and the corner
Hamiltonian in terms of the boson operators $a_i(t)$. To do this we
introduce an auxiliary algebra generated by two elements $\omega$ and
$\rho$ with the relations
\eq{
\omega^2=\rho^2=1,
\quad
\omega\rho=-\rho\omega,
\qquad
\Tr\rho=\Tr\omega=0.
\label{rhoomegaalg}
}
Then the generators of the algebra~(\ref{HZalgebra}) for the
model~(\ref{sfmodel}) can be represented in the form
\subeq{\label{HZsf}\Align{
\span
H=\int^\infty_0dt\,\sum^3_{i=1}
{\sinh\pi t\sinh{\pi p_it\over2}
\over\sinh^2{\pi t\over2}}a_i(-t)a_i(t),
\label{Hsf}
\\
Z_{++}(k_1,k_2,k_3|\theta)
&=\omega V_3(\theta)\e^{(k_1+k_2)\theta/2},
\label{Z++sf}
\\
Z_{-+}(k_1,k_2,k_3|\theta)
&=\omega\rho V_3(\theta)S_1(k_1,\kappa_1|\theta)\e^{(k_1+k_2)\theta/2},
\label{Z-+sf}
\\
Z_{+-}(k_1,k_2,k_3|\theta)
&=-\omega\rho V_3(\theta)
S_2(k_2,\kappa_2|\theta-{\textstyle{\i\pi p_2\over2}})
\e^{(k_1+k_2)\theta/2},
\label{Z+-sf}
\\
Z_{--}(k_1,k_2,k_3|\theta)
&=-\omega V_3(\theta)
S_1(k_1,\kappa_1|\theta)
S_2(k_2,\kappa_2|\theta-{\textstyle{\i\pi p_2\over2}})
\e^{(k_1+k_2)\theta/2}.
\label{Z--sf}
}}
These operators satisfy the commutation relations~(\ref{Zcommut})
and~(\ref{HZcommut}). In addition, if the normalization constants are
given by
\eq{
c_i=-{\e^{2(C_E+\log\pi p_i)/p_i}\over\pi^{3/2}}\,
{\Gamma(1+1/p_i)\over\Gamma(-1/p_i)}\,
G(p_i,-\i\pi),
\label{cidef}
}
they satisfy the normalization condition~(\ref{Znorm}). The proofs of
the commutation relations and normalization condition is rather
standard and are presented in Appendix~\ref{appendix-commut-gen}.

The expressions for the operators~(\ref{HZsf}) define a free field
realization of the algebra~(\ref{HZalgebra}) for our model. This
realization reduces the problem of construction of the integral
representation for form factors to application of the Wick theorem. For
arbitrary free field exponents (e.~g.\ $V(\theta)$,
$I^{(\pm)}_i(\theta)$) of the form
\eq{\Aligned{
U_j(\theta)=\lcolon\e^{\phi_j(\theta)}\rcolon
=\e^{\phi^-_j(\theta)}\e^{\phi^+_j(\theta)},
\qquad
\phi^\pm_j(\theta)
&=\sum_i\int^\infty_0dt\,A^i_j(\pm t)a_i(\pm t)\e^{\pm\i\theta t},
\\
\phi_j(\theta)
&=\phi^+_j(\theta)+\phi^-_j(\theta),
}\label{Udef}
}
the Wick theorem gives
\eq{
\llangle U_1(\theta_1)\ldots U_n(\theta_n)\rrangle
=\prod^n_{j=1}C_{U_j}\prod_{j<k}G_{U_jU_k}(\theta_k-\theta_j)
\label{TraceFormula}
}
with
\subeq{\label{CGdef}\Align{
\relax\span
\log C_{U_j}=\llangle\phi^-_j(0)\phi^+_j(0)\rrangle,
\label{Cjdef}
\\*
\span
\log G_{U_jU_k}(\theta)=\llangle\phi_j(0)\phi_k(\theta)\rrangle.
\label{Gjkdef}
}}
The traces (\ref{CGdef}) can be easily calculated. The values of
$C_{U_j}$, $G_{U_jU_k}(u)$ for the operators, introduced in the paper,
are listed in the Appendix~\ref{appendix-traces}. The
expressions~(\ref{sfformfactors})--(\ref{barGvthetadef}) are obtained
from these formulas after the substitution $I_1^{(\pm)}(\xi)\to
I_1^{(\pm)}(\xi\mp\i\pi p_1/2)$. These shifts of the integration
variables allow one to deform two first contours depicted in
Fig.~\ref{figure-VIcontours} to the single straight contour like the
third one in Fig.~\ref{figure-VIcontours}.

We have a three-parametric family of form factors labeled by $k_1$,
$k_2$,~$k_3$. Now we have to identify them with a three-parametric
family of local operators in the Lagrangian formulation. It will be
done in the next section.

%%%%%%%%%%%%%%%%%%%%%%%%%%%%%%%%%%%%%%%%%%%%%%%%%%%%%%%%%%%%%%%%%%%%%%%%%

\section{Identification of form factors}
\label{sec-identification}

Due to the formal symmetry of the action~(\ref{sfmodel}) with respect
to the substitutions
$(\varphi_i,\alpha_i)\leftrightarrow(\varphi_j,\alpha_j)$, the model is
unitary in three regions:
\eq{
\Aligned{
\rI_1:
&\quad p_1<0,\quad p_2,p_3>0;
\\
\rI_2:
&\quad p_2<0,\quad p_1,p_3>0;
\\
\rI_3:
&\quad p_3<0,\quad p_1,p_2>0.
}\label{I1I2I3}
}
What we considered above, while considering the regime~I, was the
region~$\rI_3$.

In the region~$\rI_i$ the operators corresponding to the
fundamental particle is defined in terms of the operator $V_i$. Namely,
we can introduce the vertex operators (recall that $i\in\Z_3$)
\subeq{\label{Zisf}\Align{
Z^i_{++}(k_1,k_2,k_3|\theta)
&=\omega V_i(\theta)\e^{(k_{i+1}+k_{i+2})\theta/2},
\label{Zippsf}
\\
Z^i_{-+}(k_1,k_2,k_3|\theta)
&=\omega\rho V_i(\theta)
S_{i+1}(k_{i+1},\kappa_{i+1}|\theta)
\e^{(k_{i+1}+k_{i+2})\theta/2},
\label{Zimpsf}
\\
Z^i_{+-}(k_1,k_2,k_3|\theta)
&=-\omega\rho V_i(\theta)
S_{i+2}(k_{i+2},\kappa_{i+2}|\theta-{\textstyle{\i\pi p_{i+2}\over2}})
\e^{(k_{i+1}+k_{i+2})\theta/2},
\label{Zipmsf}
\\
Z^i_{--}(k_1,k_2,k_3|\theta)
&=-\omega V_i(\theta)
S_{i+1}(k_{i+1},\kappa_{i+1}|\theta)
S_{i+2}(k_{i+2},\kappa_{i+2}|\theta-{\textstyle{\i\pi p_{i+2}\over{i+2}}})
\e^{(k_{i+1}+k_{i+2})\theta/2}.
\label{Zimmsf}
}}
Evidently, in the previous section we constructed the generators
$Z^3_{\ve\ve'}(k_1,k_2,k_3|\theta)$. The vertex operators
$Z^i_{\ve\ve'}(\theta)$ for a given $i$ satisfy the
relations~(\ref{HZalgebra}) with the $S$ matrix
$-S_{p_{i+1}}(\theta)\otimes S_{p_{i+2}}(\theta)$. The topological
charges in these regions are given by the same formula~(\ref{Q1Q2}) for
$Q_i$, but without the limitation to $i=1,2$.

We have a three-parametric family of form factors. We expect that they
are in a one-to-one correspondence with the exponential
operators~(\ref{Oa1a2a3}). Our main conjecture is that the
correspondence between $a_1,a_2,a_3$ and $k_1,k_2,k_3$ analytically
depends on the parameters $p_1$, $p_2$, $p_3$ for any values of these
parameters including all three regions~$\rI_1$, $\rI_2$, $\rI_3$. This
analyticity conjecture becomes more natural, if we consider the
intermediate region of parameters, corresponding to the regime~II, where
all three families of vertex operators coexist (see
Appendix~\ref{appendix-regimeII}).

Now we want to exploit the mutual locality indices, introduced in
Eq.~(\ref{mutuallocality}). For this purpose we need operators that
create the fundamental particles $z^i_{\ve\ve'}$. Such operators are
given by
$$
\cV^i_{\ve\ve'}(x)
=\cO^i_{\ve\ve'}(x)
\exp\left({\i\ve\tilde\varphi_{i+1}(x)\over4\alpha_{i+1}}
+{\i\ve'\tilde\varphi_{i+2}(x)\over4\alpha_{i+2}}\right)
$$
with $\tilde\varphi_i(x)$ being the dual field to $\varphi_i(x)$,
$\d^\mu\tilde\varphi_i(x)=\ve^{\mu\nu}\d_\nu\varphi_i(x)$ and
$\cO^i_{\ve\ve'}(x)$ being an operator mutually local with the
operators $\varphi_i(x)$. Indeed, consider the commutation relations
$[Q_j,\cV^i_{\ve\ve'}]$ in the Euclidean plane. We have
$$
[Q_j,\cV^i_{\ve\ve'}(x)]
=-{\alpha_j\over\pi}\oint dy^\mu\,\d_\mu\varphi_j(y)
\,\cV^i_{\ve\ve'}(x).
$$
The last integral can be taken over a small circle around the point
$x$, where we can neglect the interaction term in Eq.~(\ref{sfmodel})
and consider the model as that of three free bosons. It can be easily
taken using the complex variables $z=x^1+\i x^2$, $\bar z=x^1-\i x^2$
with the result
$$
[Q_{i+1},\cV^i_{\ve\ve'}(x)]=\ve\cV^i_{\ve\ve'}(x),
\qquad
[Q_{i+2},\cV^i_{\ve\ve'}(x)]=\ve'\cV^i_{\ve\ve'}(x).
$$
It means that this operator generally creates from the vacuum all states
with topological charges $Q_{i+1}=\ve$, $Q_{i+2}=\ve'$ including the
one-particle ones.

The mutual locality index of the operator $O_{a_1a_2a_3}(x)$ with the
operator $\cV^i_{\ve\ve'}(x)$ is equal to
$$
\e^{2\pi\i\Omega^i_{\ve\ve'}}
=\e^{\i\pi(\ve a_{i+1}/\alpha_{i+1}+\ve'a_{i+2}/\alpha_{i+2})}.
$$
On the other hand, using Eqs.~(\ref{HZcommut}), it is possible to
calculate the mutual locality index of the operator
$\cV^i_{\ve\ve'}(x)$ and the operator that corresponds to the form
factors defined by the generators~(\ref{Zisf}) with given $k_1$, $k_2$,
$k_3$:
$$
\e^{2\pi\i\Omega^i_{\ve\ve'}}
=\e^{\i\pi(\ve k_{i+1}+\ve'k_{i+2})}.
$$

Hence, we conjecture that the form factors of the operator
$O_{a_1a_2a_3}(x)$ in the region $\rI_3$ are given by the traces of
the operators~(\ref{HZsf}) with
\eq{
k_i={a_i\over\alpha_i},
\qquad
\kappa_i=-{\i\pi\over4}
\left(
p_i{a_i\over\alpha_i}
+p_{i+1}{a_{i+1}\over\alpha_{i+1}}-p_{i+2}{a_{i+2}\over\alpha_{i+2}}
\right).
\label{kdefsf}
}

%%%%%%%%%%%%%%%%%%%%%%%%%%%%%%%%%%%%%%%%%%%%%%%%%%%%%%%%%%%%%%%%%%%%%%%%%

\section{Sausage Model}
\label{sec-sausage}

In the limit $\alpha_2\to0$, the action~(\ref{sfmodel}) splits into two
parts
\eq{
\cS[\varphi_1,\varphi_2,\varphi_3]
=\cS_{\rm SM}[\varphi_1,\varphi_3]
+\int d^2x\,{(\d_\mu\varphi_2)^2-(2\mu)^2\varphi_2^2\over8\pi},
\label{p20splitting}
}
where $\cS_{\rm SM}$ is the action of the sausage model in the dual
representation:
\eq{
\cS_{\rm SM}[\varphi,\chi]=\int d^2x\left(
{(\d_\mu\varphi)^2+(\d_\mu\chi)^2
\over8\pi}
+{2\mu\over\pi}\cos\alpha\varphi\cosh\beta\chi
\right),
\qquad
\alpha^2-\beta^2=1
\label{smodel}
}
with $\alpha=\alpha_1$. The mass term for the field $\varphi_2(x)$ in
Eq.~(\ref{p20splitting}) appears in the limit $p_2\to0$ due to the
quantum corrections (see Ref.~\cite{Fateev96} for details).

On the level of the factorized scattering theory the limit $p_2\to0$
must be performed as follows. According to Eq.~(\ref{regimeImass}),
(\ref{boundstatemass}) the mass of the fundamental particles $m$ tends
to infinity, while the masses of bound states remain finite. Hence, the
fundamental particles decouple. The mass of the $n$th bound state is
$2\mu n$, which means that $n$th bound state is simply $n$ the first
bound state particles with the same rapidity. It means that the model in
this limit is completely described by the first bound state particles.

Consider the first bound state of the particles $z_{\ve_1+}$ and
$z_{\ve_2-}$. This bound state forms a quadruplet. Its scattering
matrix is given by the diagram
$$
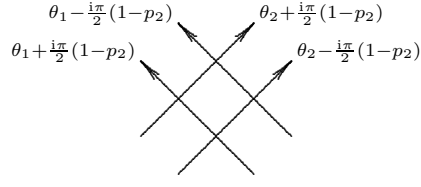
\text{
\beginpicture
\setcoordinatesystem units <1cm,1cm> point at 0 0
\ar 1.5 0.0 , 0.0 1.5
\ar 2.0 0.5 , 0.5 2.0
\ar 0.0 0.5 , 1.5 2.0
\ar 0.5 0.0 , 2.0 1.5
\put{$\scriptstyle\theta_1+{\i\pi\over2}(1-p_2)$} [Br] <-2pt,2pt> at 0.0 1.5
\put{$\scriptstyle\theta_1-{\i\pi\over2}(1-p_2)$} [Br] <-2pt,2pt> at 0.5 2.0
\put{$\scriptstyle\theta_2+{\i\pi\over2}(1-p_2)$} [Bl] <2pt,2pt> at 1.5 2.0
\put{$\scriptstyle\theta_2-{\i\pi\over2}(1-p_2)$} [Bl] <2pt,2pt> at 2.0 1.5
\put{} [Bc] at 0.0 0.0
\endpicture
}
$$
The values of the rapidities are written near the arrows. For $p_2=0$
the well known fusion phenomenon takes place. The quadruplet
$V^2\otimes V^2$ splits into a triplet $V^3$ and a singlet~$V^1$. The
$S$ matrix acts on the spaces $V^1\otimes V^1$, $V^1\otimes V^3$, and
$V^3\otimes V^1$ trivially and on the space $V^3\otimes V^3$ as a
nontrivial $S$ matrix. It means that the singlet $Z'_0(\theta)$
describes a free boson particle $\varphi_2(x)$ and can be decoupled.
The remaining triplet $Z_I(\theta)$ (where the subscript $I=+,0,-$
labels the particles with the charge $Q_1=2,0,-2$ respectively)
corresponds to particles with the mass $M$ related to the parameter
$\mu$ in the action as
\eq{
M=2\mu.
\label{mass-sausage}
}
Scattering of these particles is described by the $S$ matrix of the
sausage model $SM_\lambda$~\cite{Fateev:1992tk} with
\eq{
\lambda={1\over p}={1\over2\alpha^2}.
}
The $S$ matrix of the sausage model
has the form~\cite{Zamolodchikov:ku,Fateev:1992tk}
\eq{
\Aligned{
S(\theta)^{++}_{++}
&={\sinh{\theta-\i\pi\over p}\over\sinh{\theta+\i\pi\over p}},
\\
S(\theta)^{0+}_{+0}
&=-{\i\sin{2\pi\over p}\over\sinh{\theta-2\pi\i\over p}}
S(\theta)^{++}_{++},
&\qquad
S(\theta)^{+0}_{+0}
&={\sinh{\theta\over p}\over\sinh{\theta-2\pi\i\over p}}
S(\theta)^{++}_{++},
\\
S(\theta)^{-+}_{+-}
&=-{\sin{\pi\over p}\sin{2\pi\over p}
\over\sinh{\theta-\i\pi\over p}\sinh{\theta-2\pi\i\over p}}
S(\theta)^{++}_{++},
&\qquad
S(\theta)^{+-}_{+-}
&={\sinh{\theta\over p}\sinh{\theta+\i\pi\over p}
\over\sinh{\theta-\i\pi\over p}\sinh{\theta-2\pi\i\over p}}
S(\theta)^{++}_{++},
\\
S(\theta)^{\0\0}_{+-}
&={\i\sin{2\pi\over p}\sinh{\theta\over p}
\over\sinh{\theta-\i\pi\over p}\sinh{\theta-2\pi\i\over p}}
S(\theta)^{++}_{++},
&\qquad
S(\theta)^{00}_{00}
&=S(\theta)^{+0}_{+0}+S(\theta)^{-+}_{+-},
\\
\span\span
S(\theta)^{-I'_1,-I'_2}_{-I_1,-I_2}
=S(\theta)^{I_1I_2}_{I'_1I'_2}
=S(\theta)^{I'_1I'_2}_{I_1I_2},
}\label{Smatrix-sausage}
}
and the charge conjugation matrix is
\eq{
C_{IJ}=\delta_{-I,J}.
}

To obtain the generators of the algebra~(\ref{HZalgebra}) for the
sausage model we have to consider the first bound state operators
$B^{2,1}_\nu(\theta)$ ($\nu=+,\rS,\rA,-$) for $p_2<1$ and then take the
limit $p_2\to0$. The operators $B^{2,n}_\nu(\theta)$
can be obtained from the operators $B^{1,n}_\nu(\theta)$ defined in
Eq.~(\ref{boundstateVOs}) by the substitutions $p_1\leftrightarrow p_2$,
$u_{1,n}\to u_{2,n}$ and $Z_{\ve\ve'}(\theta)\to Z_{\ve'\ve}(\theta)$.

In the limit $p_2\to0$ we easily obtain
\eq{
\Gamma_{2,1}\to{2\over\sqrt{p_1\sin{\pi\over p_1}}},
\qquad
K^{2,1}_\rS
\sim2p_2^{-1/2}\sqrt{{p_1\over\pi}\sin{\pi\over p_1}},
\qquad
K^{2,1}_\rA
\to\sqrt{2\cos{\pi\over p_1}},
\label{p20limitcoeffs}
}
We see that the coefficients at $B^{2,1}_+(\theta)$,
$B^{2,1}_\rA(\theta)$, and $B^{2,1}_-(\theta)$ in equations analogous to
Eq.~(\ref{boundstateVOs}) remain finite. These three
operators form the triplet. The coefficients at the remaining operator
$B^{2,1}_\rS(\theta)$ is singular. This is just the singlet operator.

We note that the contributions of the oscillators $a_1(t)$ and $a_3(t)$
to the vertex operators $B^{2,1}_\rS$ corresponding to the singlet
particle vanish, and this operator can be written as
\eq{
B^{2,1}_\rS(\theta)
=Z'_0(a_2|\theta)=\int^\infty_{-\infty}dt\,\ta(t)\e^{\i\theta t}
+\i\sqrt{2\pi}\,a_2
\label{Z0prime}
}
with the parameter $a_2$ defined in Eq.~(\ref{Oa1a2a3}) and the
operators $\ta(t)$ related with the operators $a_2(t)$ as
\eq{
\ta(t)=-{\sinh\pi t\over\sinh{\pi t\over2}}
\sqrt{2\sinh{\pi p_2t\over2}\over t}\,a_2(t),
\qquad
[\ta(t),\ta(t')]=2\sinh\pi t\>\delta(t+t').
\label{a2tilde}
}

The corresponding traces consisting of the operators $Z'_0(\theta)$ are
given by
\Multline{
\llangle Z'_0(\vartheta_m+\i\pi/2)\ldots Z'_0(\vartheta_1+\i\pi/2)
Z'_0(\theta_1-\i\pi/2)\ldots Z'_0(\theta_n-\i\pi/2)\rrangle
\\
=(\i\sqrt{2\pi}\,a_2)^{m+n}+(\text{$\delta$-functional terms}).
\label{FreeFieldTrace}
}
They reproduce the form factors of a free field. Indeed, for a free
boson field with the action
\eq{
\cS[\varphi]=\int {d^2x\over8\pi}\,
\left((\d_\mu\varphi)^2-M^2\varphi^2\right)
\label{FreeFieldAction}
}
we have
\Align{
\langle\vartheta_1,\ldots,\vartheta_m|
e^{\i a\varphi(0)}|\theta_1,\ldots,\theta_n\rangle
&=\sum^{\min(m,n)}_{k=0}(\i\sqrt{2\pi}\,a)^{m+n-2k}
\sum_{1\le i_1<\ldots<i_k\le m\atop 1\le j_1<\ldots<j_k\le n}
\prod^k_{l=1}2\pi\delta(\vartheta_{i_l}-\theta_{j_l})
\notag
\\
&=(\i\sqrt{2\pi}\,a)^{m+n}+(\text{$\delta$-functional terms}).
\label{FreeFieldFormFactors}
}
It is easy to see that the $\delta$-functional terms (which correspond
to nonconnected diagrams) in Eq.~(\ref{FreeFieldTrace}) are the same as
in Eq.~(\ref{FreeFieldFormFactors}).

This provides a check for our construction. The consideration in
Sec.~\ref{sec-identification} does not distinguish between primary
(exponential) fields~(\ref{Oa1a2a3}) and zero spin linear combinations
of their descendants~(\ref{Oa1a2a3descendants}). The result
(\ref{FreeFieldTrace}), as well as other free field cases described
below, provide an evidence in favor of the identification of the form
factors just with the primary operators.

In the triplet operators $B^{2,1}_\nu(\theta)$ ($\nu=+,\rA,-$) the
contribution of the oscillators $a_2(t)$ vanishes, and we obtain the
generators of the algebra~(\ref{HZalgebra}) for the sausage model in
terms of two families of boson operators $a(t)=a_1(t)$ and
$b(t)=a_3(t)$.

Namely, let
\subeq{\label{absausage}\Align{
[a(t),a(t')]
&=t{\sinh^2{\pi t\over2}\over\sinh\pi t\sinh{\pi pt\over2}}\delta(t+t'),
\label{asausage}
\\
[b(t),b(t')]
&=t{\sinh^2{\pi t\over2}\over\sinh\pi t\sinh{\pi(2-p)t\over2}}\delta(t+t').
\label{bsausage}
}}
The fields associated to the operators $a(t)$ according to the
rules~(\ref{allphidef}) will be denoted as $\phi$, while those
associated to the operators $b(t)$ will be denoted as $\psi$. Now we
introduce some building blocks for the explicit representation of the
triplet vertex operators. Let
\subeq{\label{VIpmsausage}\Align{
V^{(\pm)}(\theta)
&=\lcolon\exp\left(
\i\phibar(\theta;p)\mp\i(\phi^{(\pm)}(\theta;2-p)
-\psi^{(\pm)}(\theta;-p))
\right)\rcolon,
\label{Vpmsausage}
\\
I^{(\pm)}(\theta)
&=\lcolon\exp\left(
-\i\phibar(\theta;p)\pm\i(\phi^{(\pm)}(\theta;2-p)
-\psi^{(\pm)}(\theta;-p))
\right)\rcolon.
\label{Ipmsausage}
}}
The operator products for these operators are given by
\subeq{\label{opprods-sausage}\Align{
V^{(A)}(\theta')V^{(B)}(\theta)
&=\bg^{(AB)}(\theta-\theta')\,
\lcolon V^{(A)}(\theta')V^{(B)}(\theta)\rcolon,
\label{VVprod-sausage}
\\*
I^{(A)}(\theta')V^{(B)}(\theta)
&=(\bg^{(AB)}(\theta-\theta'))^{-1}\,
\lcolon I^{(A)}(\theta')V^{(B)}(\theta)\rcolon,
\label{IVprod-sausage}
\\
V^{(A)}(\theta')I^{(B)}(\theta)
&=(\bg^{(AB)}(\theta-\theta'))^{-1}\,
\lcolon V^{(A)}(\theta')I^{(B)}(\theta)\rcolon,
\label{VIprod-sausage}
\\*
I^{(A)}(\theta')I^{(B)}(\theta)
&=\bg^{(AB)}(\theta-\theta')\,
\lcolon I^{(A)}(\theta')I^{(B)}(\theta)\rcolon,
\label{IIprod-sausage}
}}
with
$$
\bg^{(AB)}(\theta)=\bg^{(AB)}_{11}(\theta)|_{p_1\to p}.
$$

Then we define the `bare' vertex and screening operators as
\Align{
V(k,\kappa|\theta)
&=\left(
V^{(+)}(\theta)\e^{-\kappa}-\i V^{(-)}(\theta)\e^\kappa
\right)\e^{k\theta},
\label{Vsausage}
\\*
S_\pm(k,\kappa|\theta)
&=\int_{\cC_\pm}{d\xi\over2\pi\i}\left(
I^{(+)}(\xi)\e^\kappa-\i I^{(-)}(\xi)\e^{-\kappa}
\right)
{\pi\e^{-k\xi}\over\sinh{\xi-\theta\pm\i\pi\over p}}.
\label{Ssausage}
}
The contour $\cC_+$ ($\cC_-$) goes below (above) the pole at the point
$\theta-\i\pi$ ($\theta+\i\pi$).

The screening operators appear in the products
$S_+(k,\kappa|\theta)V^{(\pm)}(\theta)$ and
$V^{(\pm)}(\theta)S_-(k,\kappa|\theta)$. The products
$V^{(\pm)}(\theta)I^{(\pm)}(\xi)$ possess a simple pole at the point
$\xi=\theta$. Taking this fact into account we should carefully define
the prescription for the integration contours in the screening
operators $S_\pm(k,\kappa|\theta)$. According to the general rule,
since the screening operator $S_+$ ($S_-$) is placed to the left
(right) of the operator $V^{(\pm)}(\theta)$, the contour $\cC_+$
($\cC_-$) goes above (below) the point $\theta$. Besides, both of them
go below the poles at the points $\theta-\i\pi+\i\pi pn$
($n=0,1,2,\ldots$) and above the poles at $\theta+\i\pi-\i\pi pn$.
Similarly to the case of the operators $S_i$, there are no inflections
in the contours, because the poles
$\xi=\theta,\theta+\i\pi,\theta-\i\pi$ never appear in the same product
simultaneously.

We can find from Eqs.~(\ref{IVprod-sausage}), (\ref{VIprod-sausage})
that
\eq{
V^{(A)}(\theta)I^{(B)}(\xi){1\over\sinh{\xi-\theta-\i\pi\over p}}
=I^{(B)}(\xi)V^{(A)}(\theta){1\over\sinh{\xi-\theta+\i\pi\over p}}.
\label{VIcommut}
}
Using this relation we obtain
\eq{
S_+(k,\kappa|\theta)V(k,\kappa|\theta)
=V(k,\kappa|\theta)S_-(k,\kappa|\theta),
\label{VSequal}
}
The pole at $\xi=\theta$ is cancelled due to the identity
$g^{(++)}(\theta)=g^{(--)}(\theta)$. But for the product of the
operator $V(k,\kappa|\theta)$ with two screening operators the pole
contributions do not cancel. Namely,
\eq{
\Aligned{
S_+(k,\kappa|\theta)\Bigl(S_+(k,\kappa|\theta)V(k,\kappa|\theta)
-V(k,\kappa|\theta)S_-(k,\kappa|\theta)\Bigr)
&=\tilde c\e^{\kappa-k(\theta-\pi\i)}I^{(+)}(\theta-\pi\i),
\\
\Bigl(S_+(k,\kappa|\theta)V(k,\kappa|\theta)
-V(k,\kappa|\theta)S_-(k,\kappa|\theta)\Bigr)S_-(k,\kappa|\theta)
&=-\i\tilde c\e^{-\kappa-k(\theta+\pi\i)}I^{(-)}(\theta+\pi\i),
\\
\span
\tilde c=r_p^{-2}\pi^2\Gamma^2(-1/p).
}\label{VSSdiffs}
}
These relations take place due to pinching the contours of two
screening operators between poles. The situation is analyzed in detail
in Appendix~\ref{appendix-sausage}.

Now we are in position to introduce the generators of the
algebra~(\ref{HZalgebra}) for the sausage model. They read
\eq{
B^{2,1}_+(\theta)=c_1^{-1}Z_+(k,\kappa|\theta),
\qquad
B^{2,1}_\rA(\theta)=-Z_0(k,\kappa|\theta),
\qquad
B^{2,1}_-(\theta)=c_1Z_-(k,\kappa|\theta),
\label{p20BZrelation}
}
where the constant $c_1$ is defined in Eq.~(\ref{cidef}) and the
renormalized generators are
\subeq{\label{HZsausage}
\eq{
H=\int^\infty_0dt\,
\biggl(
{\sinh\pi t\sinh{\pi pt\over2}\over\sinh^2{\pi t\over2}}a(-t)a(t)
+{\sinh\pi t\sinh{\pi(2-p)t\over2}\over\sinh^2{\pi t\over2}}b(-t)b(t)
\biggr),
\label{Hcornersausage}
}
\interdisplay
\Align{
Z_+(k,\kappa|\theta)
&=c\rho V(k,\kappa|\theta),
\label{Z+sausage}
\\*
Z_0(k,\kappa|\theta)
&=-c\sqrt{2\cos\ts{\pi\over p}}\,
V(k,\kappa|\theta)S_-(k,\kappa|\theta),
\label{Z0sausage}
\\*
Z_-(k,\kappa|\theta)
&=-c\rho S_+(k,\kappa|\theta)V(k,\kappa|\theta)S_-(k,\kappa|\theta).
\label{Z-sausage}
}}
As we have mentioned above, the definition of the bound state vertex
operators is not unique. The operators $Z_I(k,\kappa|\theta)$
($I=+,0,-$) can be considered as vertex operators for the sausage
model, because the transformation~(\ref{p20BZrelation}) respects the
normalization condition~(\ref{Znorm}). The guidelines for checking the
ZF algebra commutation relations in this case are given in
Appendix~\ref{appendix-sausage}.

The constants $k$, $\kappa$ are related to the parameters of the
exponential operators. For the operator
\eq{
O_{ab}(x)=\e^{\i a\varphi(x)+b\chi(x)}
\label{Oabdef}
}
we have
\eq{
k={a\over\alpha},
\qquad
\kappa=-{\i\pi\over4}\left(
p{a\over\alpha}+(p-2){b\over\beta}
\right).
\label{kdefsausage}
}
The normalization constant $c$ in Eq.~(\ref{HZsausage}) is given by
\eq{
c={\e^{2(C_E+\log\pi p)/p}\over(\pi p)^{3/2}}
{\Gamma^{1/2}(1/p)
\over\Gamma^{3/2}(1-1/p)}.
\label{cdef}
}

\begin{figure}[t]
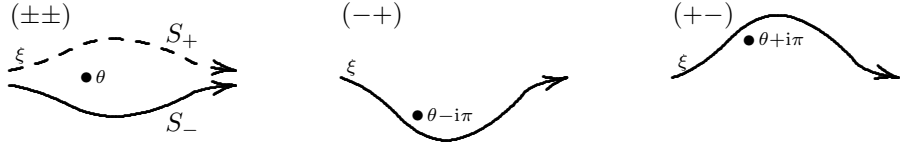

$$
\vcenter{
\beginpicture
\setcoordinatesystem units <1cm,1cm> point at 0 0
\put{$(\pm\pm)$} [lt] at -1 1
\put{} [lB] at 0 1
\put{} [lB] at 0 -1
\put{} [lB] at -1 0
\put{} [lB] at 2 0
\bfpoint 0 0
\put{$\scriptstyle\theta$} [lB] <4pt,-2pt> at 0 0
\setplotsymbol(.)
\setquadratic
\setdashes
\plot -1 0.1  -0.6 0.2  -0.2 0.37  0.2 0.5
0.6 0.5  1.0 0.4  1.4 0.225  1.6 0.15  2.0 0.1 /
\setsolid
\arw 1.98 0.1025 , 2 0.1
\put {$S_+$} [lB] <2pt,2pt> at 1.0 0.4
\put{$\scriptstyle\xi$} [lB] <2pt,4pt> at -1 0.1
\plot -1 -0.1  -0.6 -0.2  -0.2 -0.37  0.2 -0.5
0.6 -0.5  1.0 -0.4  1.4 -0.225  1.6 -0.15  2.0 -0.1 /
\arw 1.98 -0.1025 , 2 -0.1
\put {$S_-$} [lt] <2pt,-2pt> at 1.0 -0.4
\endpicture
}
\qquad\qquad
\vcenter{
\beginpicture
\setcoordinatesystem units <1cm,1cm> point at 0 0
\put{$(-+)$} [lt] at -1 1
\put{} [lB] at 0 1
\put{} [lB] at 0 -1
\put{} [lB] at -1 0
\put{} [lB] at 2 0
\bfpoint 0 -0.5
\put{$\scriptstyle\theta-\i\pi$} [lB] <4pt,-2pt> at 0 -0.5
\setplotsymbol(.)
\setquadratic
\plot -1 0  -0.6 -0.2  -0.2 -0.55  0.2 -0.8
0.6 -0.8  1.0 -0.6  1.4 -0.25  1.6 -0.1  2.0 0 /
\arw 1.98 -0.005 , 2 0
\put{$\scriptstyle\xi$} [lB] <2pt,2pt> at -1 0
\endpicture
}
\qquad\qquad
\vcenter{
\beginpicture
\setcoordinatesystem units <1cm,1cm> point at 0 0
\put{$(+-)$} [lt] at -1 1
\put{} [lB] at 0 1
\put{} [lB] at 0 -1
\put{} [lB] at -1 0
\put{} [lB] at 2 0
\bfpoint 0 0.5
\put{$\scriptstyle\theta+\i\pi$} [lB] <4pt,-2pt> at 0 0.5
\setplotsymbol(.)
\setquadratic
\plot -1 0  -0.6 0.2  -0.2 0.55  0.2 0.8
0.6 0.8  1.0 0.6  1.4 0.25  1.6 0.1  2.0 0 /
\arw 1.98 0.005 , 2 0
\put{$\scriptstyle\xi$} [lB] <2pt,4pt> at -1 0
\endpicture
}
$$
\caption{Integration contours in the terms~(\ref{Z0terms}) consisting
the operator $Z_0(\theta)$. The pictures are labeled by values of the
pair~$(AB)$. On the left picture the solid line corresponds to the
screening operator~$S_-$, while the dashed one to~$S_+$.}
\label{figure-VIcontours-sausage}
\end{figure}

\begin{figure}[t]
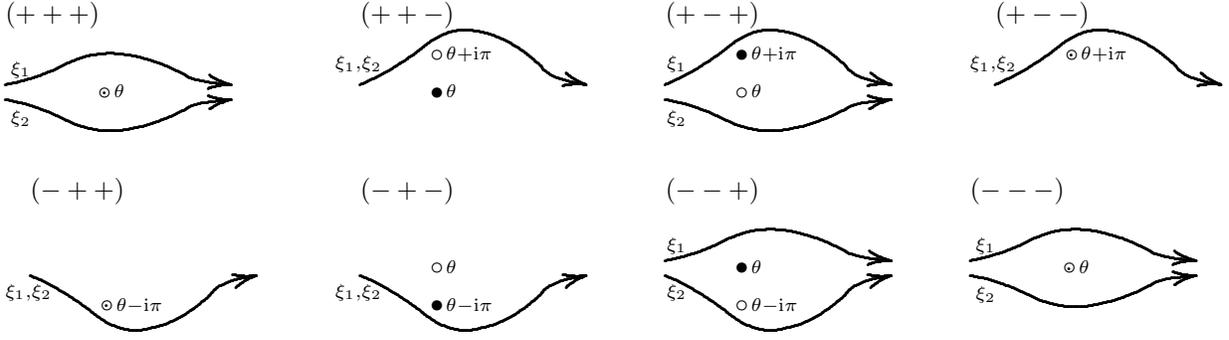

$$
\Aligned{
&\vcenter{
\beginpicture
\setcoordinatesystem units <1cm,1cm> point at 0 0
\put{$(+++)$} [lt] at -1 1.2
\put{} [lB] at 0 1
\put{} [lB] at 0 -1
\put{} [lB] at -1 0
\put{} [lB] at 2 0
\point 0.3 0
\vdpoint 0.3 0
\put {$\scriptstyle\theta$} [lB] <4pt,-2pt> at 0.3 0
\setquadratic
\setplotsymbol(.)
\plot -1 0.1  -0.6 0.2  -0.2 0.37  0.2 0.5
0.6 0.5  1.0 0.4  1.4 0.225  1.6 0.15  2.0 0.1 /
\arw 1.98 0.1025 , 2 0.1
\put{$\scriptstyle\xi_1$} [lB] <2pt,4pt> at -1 0.1
\plot -1 -0.1  -0.6 -0.2  -0.2 -0.37  0.2 -0.5
0.6 -0.5  1.0 -0.4  1.4 -0.225  1.6 -0.15  2.0 -0.1 /
\arw 1.98 -0.1025 , 2 -0.1
\put{$\scriptstyle\xi_2$} [lt] <2pt,-4pt> at -1 -0.1
\endpicture
}
&\qquad
&\vcenter{
\beginpicture
\setcoordinatesystem units <1cm,1cm> point at 0 0
\put{$(++-)$} [lt] at -1 1.2
\put{} [lB] at 0 1
\put{} [lB] at 0 -1
\put{} [lB] at -1 0
\put{} [lB] at 2 0
\vdpoint 0 0.5
\put {$\scriptstyle\theta+\i\pi$} [lB] <4pt,-2pt> at 0 0.5
\bfpoint 0 0
\put {$\scriptstyle\theta$} [lB] <4pt,-2pt> at 0 0
\setquadratic
\setplotsymbol(.)
\plot -1 0.1  -0.6 0.3  -0.2 0.57  0.2 0.8
0.6 0.8  1.0 0.65  1.4 0.35  1.6 0.2  2.0 0.1 /
\arw 1.98 0.105 , 2 0.1
\put{$\scriptstyle\xi_1,\xi_2$} [rB] <8pt,6pt> at -1 0.1
\endpicture
}
&\qquad
&\vcenter{
\beginpicture
\setcoordinatesystem units <1cm,1cm> point at 0 0
\put{$(+-+)$} [lt] at -1 1.2
\put{} [lB] at 0 1
\put{} [lB] at 0 -1
\put{} [lB] at -1 0
\put{} [lB] at 2 0
\bfpoint 0 0.5
\put{$\scriptstyle\theta+\i\pi$} [lB] <4pt,-2pt> at 0 0.5
\vdpoint 0 0
\put{$\scriptstyle\theta$} [lB] <4pt,-2pt> at 0 0
\setquadratic
\setplotsymbol(.)
\plot -1 0.1  -0.6 0.3  -0.2 0.57  0.2 0.8
0.6 0.8  1.0 0.65  1.4 0.35  1.6 0.2  2.0 0.1 /
\arw 1.98 0.105 , 2 0.1
\put{$\scriptstyle\xi_1$} [rB] <8pt,6pt> at -1 0.1
\plot -1 -0.1  -0.6 -0.2  -0.2 -0.37  0.2 -0.5
0.6 -0.5  1.0 -0.4  1.4 -0.225  1.6 -0.15  2.0 -0.1 /
\arw 1.98 -0.1025 , 2 -0.1
\put{$\scriptstyle\xi_2$} [rt] <8pt,-4pt> at -1 -0.1
\endpicture
}
&\qquad
&\vcenter{
\beginpicture
\setcoordinatesystem units <1cm,1cm> point at 0 0
\put{$(+--)$} [lt] at -1 1.2
\put{} [lB] at 0 1
\put{} [lB] at 0 -1
\put{} [lB] at -1 0
\put{} [lB] at 2 0
\point 0 0.5
\vdpoint 0 0.5
\put{$\scriptstyle\theta+\i\pi$} [lB] <4pt,-2pt> at 0 0.5
\setquadratic
\setplotsymbol(.)
\plot -1 0.1  -0.6 0.3  -0.2 0.57  0.2 0.8
0.6 0.8  1.0 0.65  1.4 0.35  1.6 0.2  2.0 0.1 /
\arw 1.98 0.105 , 2 0.1
\put{$\scriptstyle\xi_1,\xi_2$} [rB] <8pt,6pt> at -1 0.1
\endpicture
}
\\
&\vcenter{
\beginpicture
\setcoordinatesystem units <1cm,1cm> point at 0 0
\put{$(-++)$} [lt] at -1 1.2
\put{} [lB] at 0 1
\put{} [lB] at 0 -1
\put{} [lB] at -1 0
\put{} [lB] at 2 0
\point 0 -0.5
\vdpoint 0 -0.5
\put {$\scriptstyle\theta-\i\pi$} [lB] <4pt,-2pt> at 0 -0.5
\setquadratic
\setplotsymbol(.)
\plot -1 -0.1  -0.6 -0.3  -0.2 -0.57  0.2 -0.8
0.6 -0.8  1.0 -0.65  1.4 -0.35  1.6 -0.2  2.0 -0.1 /
\arw 1.98 -0.105 , 2 -0.1
\put{$\scriptstyle\xi_1,\xi_2$} [rB] <8pt,-8pt> at -1 -0.1
\endpicture
}
&\qquad
&\vcenter{
\beginpicture
\setcoordinatesystem units <1cm,1cm> point at 0 0
\put{$(-+-)$} [lt] at -1 1.2
\put{} [lB] at 0 1
\put{} [lB] at 0 -1
\put{} [lB] at -1 0
\put{} [lB] at 2 0
\vdpoint 0 0
\put{$\scriptstyle\theta$} [lB] <4pt,-2pt> at 0 0
\bfpoint 0 -0.5
\put{$\scriptstyle\theta-\i\pi$} [lB] <4pt,-2pt> at 0 -0.5
\setquadratic
\setplotsymbol(.)
\plot -1 -0.1  -0.6 -0.3  -0.2 -0.57  0.2 -0.8
0.6 -0.8  1.0 -0.65  1.4 -0.35  1.6 -0.2  2.0 -0.1 /
\arw 1.98 -0.105 , 2 -0.1
\put{$\scriptstyle\xi_1,\xi_2$} [rB] <8pt,-8pt> at -1 -0.1
\endpicture
}
&\qquad
&\vcenter{
\beginpicture
\setcoordinatesystem units <1cm,1cm> point at 0 0
\put{$(--+)$} [lt] at -1 1.2
\put{} [lB] at 0 1
\put{} [lB] at 0 -1
\put{} [lB] at -1 0
\put{} [lB] at 2 0
\bfpoint 0 0
\put{$\scriptstyle\theta$} [lB] <4pt,-2pt> at 0 0
\vdpoint 0 -0.5
\put{$\scriptstyle\theta-\i\pi$} [lB] <4pt,-2pt> at 0 -0.5
\setquadratic
\setplotsymbol(.)
\plot -1 0.1  -0.6 0.2  -0.2 0.37  0.2 0.5
0.6 0.5  1.0 0.4  1.4 0.225  1.6 0.15  2.0 0.1 /
\arw 1.98 0.1025 , 2 0.1
\put{$\scriptstyle\xi_1$} [rB] <8pt,4pt> at -1 0.1
\plot -1 -0.1  -0.6 -0.3  -0.2 -0.57  0.2 -0.8
0.6 -0.8  1.0 -0.65  1.4 -0.35  1.6 -0.2  2.0 -0.1 /
\arw 1.98 -0.105 , 2 -0.1
\put{$\scriptstyle\xi_2$} [rB] <8pt,-8pt> at -1 -0.1
\endpicture
}
&\qquad
&\vcenter{
\beginpicture
\setcoordinatesystem units <1cm,1cm> point at 0 0
\put{$(---)$} [lt] at -1 1.2
\put{} [lB] at 0 1
\put{} [lB] at 0 -1
\put{} [lB] at -1 0
\put{} [lB] at 2 0
\point 0.3 0
\vdpoint 0.3 0
\put {$\scriptstyle\theta$} [lB] <4pt,-2pt> at 0.3 0
\setquadratic
\setplotsymbol(.)
\plot -1 0.1  -0.6 0.2  -0.2 0.37  0.2 0.5
0.6 0.5  1.0 0.4  1.4 0.225  1.6 0.15  2.0 0.1 /
\arw 1.98 0.1025 , 2 0.1
\put{$\scriptstyle\xi_1$} [lB] <2pt,4pt> at -1 0.1
\plot -1 -0.1  -0.6 -0.2  -0.2 -0.37  0.2 -0.5
0.6 -0.5  1.0 -0.4  1.4 -0.225  1.6 -0.15  2.0 -0.1 /
\arw 1.98 -0.1025 , 2 -0.1
\put{$\scriptstyle\xi_2$} [lt] <2pt,-4pt> at -1 -0.1
\endpicture
}
}
$$
\caption{Integration contours in the terms~(\ref{Z-terms}) consisting
the operator~$Z_-(\theta)$. The pictures are labeled by values of the
triple~$(ABC)$. The bold points ($\bullet$) correspond to poles in the
variable $\xi_1$, while the circles ($\circ$) correspond to poles in
$\xi_2$. A circle with a dot in the center ($\rlap{$\kern
1pt\cdot$}\circ$) means a pole for both variables.}
\label{figure-VIIcontours-sausage}
\end{figure}

Now we make the integration contours for the operators $Z_0$ and $Z_-$
more obvious. The operator $Z_0$ contains the terms of the form
\eq{
\const\times
\int{d\xi\over2\pi\i}\,V^{(A)}(\theta)I^{(B)}(\xi)
{\pi\e^{k(\theta-\xi)}\over\sinh{\xi-\theta-\i\pi\over p}},
\qquad
A,B=\pm.
\label{Z0terms}
}
The integrations contours for these terms labeled by values of the pair
$(AB)$ are shown in Fig.~\ref{figure-VIcontours-sausage}.%
The operator $Z_-$ consists of the terms
\eq{
\const\times
\int_{C_+}{d\xi_1\over2\pi\i}\int_{C_-}{d\xi_2\over2\pi\i}\,
V^{(A)}(\theta)I^{(B)}(\xi_1)I^{(C)}(\xi_2)
{\pi\e^{k(\theta-\xi_1)}\over\sinh{\xi_1-\theta-\i\pi\over p}}
{\pi\e^{-k\xi_2}\over\sinh{\xi_2-\theta-\i\pi\over p}},
\qquad
A,B,C=\pm.
\label{Z-terms}
}
The integration contours for these terms labeled by values of the
triples $(ABC)$ are presented in Fig.~\ref{figure-VIIcontours-sausage}.%
Note that the contour for the integration variable $\xi_2$ is always
below the contour for $\xi_1$, which makes it possible to avoid the
pole at the point $\xi_2=\xi_1+\i\pi$. It is interesting to note that
for the products $S_+S_+V$ and $VS_-S_-$ the contour for $\xi_2$ is
above the contour for $\xi_1$ in the case $B=+$, $C=-$ only, where
there is no such pole. This observation makes these operator products,
which enter Eq.~(\ref{VSSdiffs}), well defined.

Consider now the free particle point $p=2$. At this point the action
takes the form
\eq{
\cS_{\rm SM}=\int d^2x\biggl(
{(\d_\mu\varphi)^2\over8\pi}+{M\over\pi}\cos\varphi
+{(\d_\mu\chi)^2-M^2\chi^2\over8\pi}\biggr),
\label{SMffaction}
}
where the term $-M^2\chi^2/8\pi$ in the Lagrangian appears due to the
quantum corrections, which make the action to be well defined. Hence,
it is possible to compare the result with the known results for the
sine-Gordon model at the free fermion point.

At the point $p=2$ we can derive from the $S$ matrix
(\ref{Smatrix-sausage}) that
\eq{
\Aligned{\relax
Z_\pm(\theta_1)Z_\pm(\theta_2)
&=-Z_\pm(\theta_2)Z_\pm(\theta_1),
&\qquad
Z_+(\theta_1)Z_-(\theta_2)
&=-Z_-(\theta_2)Z_+(\theta_1),
\\
Z_0(\theta_1)Z_0(\theta_2)
&=Z_0(\theta_2)Z_0(\theta_1),
&\qquad
Z_\pm(\theta_1)Z_0(\theta_2)
&=Z_0(\theta_2)Z_\pm(\theta_1).
}
}
It means that the operators $Z_\pm(\theta)$ describe free fermions while
the operator $Z_0(\theta)$ describes a free boson.

Despite of simplicity of commutation relations, the limit $p\to2$ in
the free field construction is rather complicated. First of all, in
this limit the operators
\eq{
V^{(\pm)}(\theta)
=\lcolon\e^{\i\phibar(\theta\pm\i\pi/2)}\rcolon,
\qquad
I^{(\pm)}(\theta)
=\lcolon\e^{-\i\phibar(\theta\pm\i\pi/2)}\rcolon
\label{VI-FPL}
}
are all mutually anticommuting. This expression determines the limit
for~$Z_+(\theta)$. The situation with $Z_0(\theta)$ and $Z_-(\theta)$ is
more complicated.

We start with the operator
$$
Z_0(k,\kappa|\theta)
\simeq-{\e^{C_E}(p-2)^{1/2}\over2\sqrt\pi}
\sum_{A,B=\pm}(-\i)^{A+B}J^{(AB)},
\qquad
J^{(AB)}
=\int{d\xi\over2\pi\i}\,V^{(A)}(\theta)I^{(B)}(\xi)
{\pi\e^{k(\theta-\xi)-A\kappa+B\kappa}
\over\sinh{\xi-\theta-\i\pi\over p}}.
$$
In the limit $p\to2$ the integration contour in $J^{(+-)}$ is pinched
between the poles at $\xi=\theta+\i\pi$ and $\xi=\theta+\i\pi(p-1)$.
The contour for the product in $J^{(-+)}$ is pinched between the poles
at $\xi=\theta-\i\pi$ and $\xi=\theta-\i\pi(p-1)$. This results in the
following identities
$$
\Aligned{
J^{(+-)}
&\simeq
-{2\e^{-C_E}\over p-2}e^{-\i\pi k-2\kappa}
\lcolon V^{(+)}(\theta)I^{(-)}(\theta+\i\pi)\rcolon
+\text{finite integral},
\\
J^{(-+)}
&\simeq
{2\e^{-C_E}\over p-2}e^{-\i\pi k+2\kappa}
\lcolon V^{(-)}(\theta)I^{(+)}(\theta-\i\pi)\rcolon
+\text{finite integral}.
}
$$
The sum of these two terms is equal to
$$
J^{(+-)}+J^{(-+)}
\simeq-{2\e^{-C_E}\over p-2}
\left(\e^{-\i\pi k-2\kappa}\,
\lcolon\e^{\i\psi^{(+)}(\theta;-2)+\i\psi^{(-)}(\theta+\i\pi;-2)}
\rcolon
-\e^{\i\pi k+2\kappa}\,
\lcolon\e^{-\i\psi^{(-)}(\theta;-2)-\i\psi^{(+)}(\theta-\i\pi;-2)}
\rcolon
\right).
$$
It follows from Eqs.~(\ref{kdefsausage}) that $\i\pi
k+2\kappa\sim(p-2)^{1/2}$ and from Eq.~(\ref{phipmdef}) that
$\psi^{(\pm)}(\theta)\sim(p-2)^{1/2}$. Hence, the sum
$J^{(+-)}+J^{(-+)}$ is proportional to $(p-2)^{-1/2}$, which is much
greater than the finite contributions of $J^{(++)}$ and $J^{(--)}$. For
this reason, we introduce the operators
\eq{
\tb(t)
=\i{\sinh\pi t\over\sinh{\pi t\over2}}
\sqrt{2\sinh{\pi(p-2)t\over2}\over t}\,b(t),
\qquad
[\tb(t),\tb(t')]=2\sinh\pi t\>\delta(t+t').
\label{btilde}
}
Finally, we obtain that the operator $Z_0(\theta)$ depends on the
parameter $b$ defined in Eq.~(\ref{Oabdef}) only and is given by
\eq{
Z_0(b|\theta)=\int^\infty_{-\infty}dt\,\tb(t)\e^{\i\theta t}
+\sqrt{2\pi}\,b.
\label{Z0-FPL}
}
Comparing this with (\ref{Z0prime}) and (\ref{FreeFieldTrace}) we see
that this correctly reproduces the form factors of the
operator~$\e^{b\chi}$.

Now we proceed with the operator $Z_-(\theta)$. The integration
contours are again pinched between the poles at $\xi_i=\theta\pm\i\pi$
and $\xi_i=\theta\pm\i\pi(p-1)$ ($i=1,2$). This pinching leads to
taking out one integration. The remaining integration contour goes
above or below the point $\theta=0$ depending on the kind of the
remaining screening operator, $S_+$ or $S_-$. In the limit $p\to2$ the
sum of all terms reduces to a residue at $\xi=\theta$. The answer for
the fermion subsystem turns out to be
\eq{
\Aligned{
Z_+(a|\theta)
&=\rho c\e^{a\theta}\bigl(\lcolon\e^{\i\phibar(\theta+\i\pi/2)}\rcolon\,
\e^{\i\pi a/2}
-\i\,\lcolon\e^{\i\phibar(\theta-\i\pi/2)}\rcolon\,
\e^{-\i\pi a/2}\bigr),
\\
Z_-(a|\theta)
&=-\rho c^{-1}\e^{C_E}\e^{-a\theta}
\bigl(\lcolon\e^{-\i\phibar(\theta-\i\pi/2)}\rcolon\,
\e^{\i\pi a/2}
+\i\,\lcolon\e^{-\i\phibar(\theta+\i\pi/2)}\rcolon\,
\e^{-\i\pi a/2}\bigr)
}\label{Z+--FPL}
}
with $c=2^{-1/2}\pi^{-1}\e^{C_E}$ for $p=2$.

This expression can be compared to the known results for the free
fermion point of the sine-Gordon model. Though it differs from the known
representation~\cite{Lukyanov:1997bp} for the vertex operators of the
sine-Gordon model at the free fermion point, it leads to the correct
form factors. Indeed, the straightforward calculation of the trace
gives
\Multline{
\llangle Z_+(\vartheta_n)\ldots Z_+(\vartheta_1)
Z_-(\theta_n)\ldots Z_-(\theta_1)\rrangle
\\*
={1\over(2\i)^n}
\sum_{\ve_1,\ldots,\ve_n=\pm1\atop\delta_1,\ldots,\delta_n=\pm1}
\i^{{1\over2}\sum_i(\ve_i+\delta_i)}
\e^{a\sum_i\left(\vartheta_i-\theta_i
+{\i\pi\over2}(\ve_i-\delta_i)\right)}
{\prod_{i<j}\sinh{\vartheta_i-\vartheta_j+{\i\pi\over2}(\ve_i-\ve_j)\over2}
\sinh{\theta_i-\theta_j+{\i\pi\over2}(\delta_i-\delta_j)\over2}
\over
\prod_{i,j}\sinh{\vartheta_i-\theta_j+{\i\pi\over2}(\ve_i-\delta_j)\over2}}.
}
It can be proved that this expression is equal to the known
one~\cite{SJM-IV,Bernard:1994re}
\eq{
\llangle Z_+(\vartheta_n)\ldots Z_+(\vartheta_1)
Z_-(\theta_n)\ldots Z_-(\theta_1)\rrangle
=(-)^{n(n+1)/2}\e^{a\sum_i(\vartheta_i-\theta_j)} (\i\sin\pi a)^n
{\prod_{i<j}\sinh{\vartheta_i-\vartheta_j\over2}\sinh{\theta_i-\theta_j\over2}
\over\prod_{i,j}\cosh{\vartheta_i-\theta_j\over2}}.
\label{FreeFermionFormFactors}
}

%%%%%%%%%%%%%%%%%%%%%%%%%%%%%%%%%%%%%%%%%%%%%%%%%%%%%%%%%%%%%%%%%%%%%%%%

\section{Cosine-cosine model: $p_1+p_2=2$ case}
\label{sec-p3zero}

The limit $p_1+p_2\to2$ ($p_3\to0$) is not singular for the operators
corresponding to fundamental particles. Nevertheless, the first bound
state splits again into a triplet and a singlet.

Let
$$
a(t)=a_2(t),
\qquad
b(t)=a_1(t),
\qquad
p=p_2.
$$
Without loss of generality we shall assume that
\eq{
1\le p\le2.
\label{pregion}
}
In the limit $p_3\to0$ they satisfy the commutation
relations~(\ref{absausage}). The corner Hamiltonian is given by
(\ref{Hcornersausage}). As in the last section, we denote by $\phi$ and
$\psi$ the fields associated with the oscillators $a(t)$ and $b(t)$
according to Eqs.~(\ref{allphidef}) respectively. Then
\subeq{\label{p30VI}\Align{
V_3(\theta)
&=\lcolon\exp(\i\phi(\theta;-p)+\i\psi(\theta;2-p))\rcolon,
\label{p30V3}
\\
I_1^{(\pm)}(\theta)
&=\lcolon\exp\left(
\i\psi(\theta;2-p)\pm\i(\phi^{(\pm)}(\theta;p-2)+\psi^{(\pm)}(\theta;p))
\right)\rcolon,
\label{p30I1}
\\
I_2^{(\pm)}(\theta)
&=\lcolon\exp\left(
\i\phi(\theta;p)\pm\i(\phi^{(\pm)}(\theta;2-p)-\psi^{(\pm)}(\theta;-p))
\right)\rcolon.
\label{p30I2}
}}
The fundamental particles are given by Eqs.~(\ref{Z++sf}--\ref{Z--sf})
with the operators $V_3$, $I^{(\pm)}_i$ given by Eq.~(\ref{p30VI}).
According to Eq.~(\ref{regimeImass}) the mass of the fundamental
particles is related with the parameter $\mu$ in the action as
\eq{
m={\mu\over\sin{\pi p\over2}}={\mu\over\sin{\pi(2-p)\over2}}.
\label{p30kinkmass}
}
Beside the fundamental particles there is a set of bound states with the
masses
\eq{
M_n=2\mu{\sin{\pi(2-p)n\over2}\over\sin{\pi(2-p)\over2}},
\qquad
n=1,2,\ldots,
\quad
n(2-p)<1,
\label{p30boundstatemasses}
}
which form quadruplets as in general case.

Consider the quadruplet of the mass $M_1=2\mu$. It is instructive to
consider the bound state vertex operators
$B^{1,1}_\nu$ ($\nu=+,\rS,\rA,-$) defined in Eq.~(\ref{boundstateVOs})
for small but nonzero $p_3$. In this case we have
\eq{
\Gamma_{1,1}\to\sqrt{p_2\sin{\ts{\pi\over p_2}}},
\qquad
K^{1,1}_\rS\to\sqrt{-2\over\cos{\pi\over p_2}},
\qquad
K^{1,1}_\rA
\sim|p_3|^{1/2}\sqrt{\pi\over p_2\sin{\pi\over p_2}}.
\label{p30limitcoeffs}
}
The operators $B^{1,1}_+(\theta)$, $B^{1,1}_\rS(\theta)$, and
$B^{1,1}_-(\theta)$ correspond to the triplet. The coefficients at
these operators in Eq.~(\ref{boundstateVOs}) are finite. The operator
$B^{1,1}_\rA(\theta)$ corresponds to the singlet. It describes the field
$\varphi_3(x)$ that decouples and becomes a free massive field in the
limit $p_3\to0$. The operator $B^{1,1}_\rA(\theta)$ explicitly commutes
with the operators $Z_{\ve\ve'}(\theta)$. It means that the
corresponding particle does not interact with the fundamental ones.
Besides, it is not a bound state of the fundamental particles, because
$K^{1,1}_\rA\to0$ as $p_3\to0$.

It is remarkable that the operators for the first bound state particle
can be written in terms of the vertex operators used for the sausage
model (see Appendix~\ref{appendix-p3zero-bs}), if we assume that
\eq{
a(t)=a_2(t),
\qquad
b(t)=a_1(t),
\qquad
\sqrt{2\cos\ts{\pi\over p}}\to-\i\sqrt{2|\cos\ts{\pi\over p}|}
}
in the definition of the triplet operators~(\ref{HZsausage}) and
\eq{
\ta(t)={\sinh\pi t\over\sinh{\pi t\over2}}
\sqrt{{2\sinh{\pi|p_3|t\over2}\over t}}\,a_3(t)
\label{a3tilde}
}
in the definition of the singlet operator~(\ref{Z0prime}). Namely,
\eq{
\Aligned{
B^{1,1}_+(\theta)=\i c_2^{-1}Z_+(k_2,\kappa_2|\theta),
\qquad
B^{1,1}_\rS(\theta)
&=Z_0(k_2,\kappa_2|\theta),
\qquad
B^{1,1}_-(\theta)=-\i c_2Z_-(k_2,\kappa_2|\theta),
\\
B^{1,1}_\rA(\theta)
&=Z'_0(a_3|\theta),
}\label{p30BZrelation}
}
where the constant $c_2$ is defined by Eq.~(\ref{cidef}). The scattering
matrix of the triplet is the analytical continuation of the sausage $S$
matrix~(\ref{Smatrix-sausage}) to the region $p<2$.

The fundamental particles, the first triplet bound state and the higher
($n\ge2$) quadruplets form the particle contents of the cosine-cosine
model with the action%
\footnote{This is a corrected (integrable) version of the model proposed
in Ref.~\cite{Bukhvostov:1980sn}.}
\eq{
\cS_{\rm CC}[\varphi_1,\varphi_2]=\int d^2x\left(
{(\d_\mu\varphi_1)^2+(\d_\mu\varphi_2)^2
\over8\pi}
+{2\mu\over\pi}
\cos\alpha_1\varphi_1\cos\alpha_2\varphi_2
\right),
\qquad
\alpha_1^2+\alpha_2^2=1.
\label{p30model}
}

Note that the form factors for the first bound state of this model can
be analytically continued to the sausage model region $p>2$ by taking
$\alpha_2=\alpha$, $\alpha_1=-\i\beta$. The situation is very similar
to that of the sine-Gordon and sinh-Gordon model, where the form
factors in the sinh-Gordon model are the analytic continuation of the
form factors of the sine-Gordon model that contain the first bound
state particles only.

Note, that the $S$ matrix of the fundamental quadruplet
$Z_{\ve_1\ve_2}(\theta)$ and the triplet $B^{1,1}_\nu(\theta)$
($\nu=+,\rS,-$) is rather simple. Namely, let
\eq{
Z_{\ve\ve'}(\theta_1)B^{1,1}_\nu(\theta_2)
=\sum_{\scriptstyle\ve'_1=\pm\atop\scriptstyle\nu_1=+,\rS,-}
S^{(1)}(\theta_1-\theta_2)^{\ve'_1\nu_1}_{\ve'\,\nu}
B^{1,1}_{\nu_1}(\theta_2)Z_{\ve\ve'_1}(\theta_1).
\label{Zve1ve2ZIalg}
}
Then, the nonzero matrix elements of the $S^{(1)}(\theta)$ matrix are
given by
\eq{
\Aligned{
S^{(1)}(\theta)^{++}_{++}
&=S^{(1)}(\theta)^{--}_{--}
=-{\cosh{\theta-\i\pi/2\over p}\over\cosh{\theta+\i\pi/2\over p}},
\qquad
S^{(1)}(\theta)^{+-}_{+-}
=S^{(1)}(\theta)^{-+}_{-+}
=-{\cosh{\theta-\i\pi/2\over p}\over\cosh{\theta-3\i\pi/2\over p}},
\\
S^{(1)}(\theta)^{+\,\rS}_{+\,\rS}
&=S^{(1)}(\theta)^{-\,\rS}_{-\,\rS}
={\cosh^2{\theta-\i\pi/2\over p}
\over\cosh{\theta-3\i\pi/2\over p}\cosh{\theta+\i\pi/2\over p}},
\\
S^{(1)}(\theta)^{-\,\rS}_{+-}
&=S^{(1)}(\theta)^{-+}_{+\,\rS}
=S^{(1)}(\theta)^{+\,\rS}_{-+}
=S^{(1)}(\theta)^{+-}_{-\,\rS}
=-\sqrt{\ts2|\cos{\pi\over p}|}
{\sin{\pi\over p}\cosh{\theta-\i\pi/2\over p}
\over\cosh{\theta-3\i\pi/2\over p}\cosh{\theta+\i\pi/2\over p}}.
}\label{Zve1ve2ZISmatrix}
}

Consider now the case $p_1=p_2=1$. At this point the action can be
rewritten as a sum of two sine-Gordon models at the free fermion point:
$$
\cS_{\rm CC}[\varphi_1,\varphi_2]=\sum^2_{i=1}\int d^2x\biggl(
{(\d_\mu\chi_i)^2\over8\pi}+{\mu\over\pi}\cos\chi_i\biggr),
\qquad
\chi_{1,2}(x)={\varphi_1(x)\pm\varphi_2(x)\over\sqrt2}.
$$
This free fermion point provides one more check of the construction. In
the limit $p_1\to1$ the integration contours in $Z_{\ve\ve'}(\theta)$
are pinched and it is possible to take all integrals explicitly. Let
\eq{
a_\pm(t)=a_1(t)\e^{\pm\pi|t|/4}\pm a_2(t)\e^{\mp\pi|t|/4},
\qquad
[a_\pm(t),a_\pm(t')]=t\delta(t+t'),
\qquad
[a_+(t),a_-(t)]=0.
\label{apmdef}
}
Define
\eq{
\phi_\pm(\theta)
=\int^\infty_{-\infty}{dt\over t}\,a_\pm(t)\e^{\i\theta t}.
\label{ffphipmdef}
}
Then
\subeq{\Align{
Z_{++}(k_1,k_2|\theta)
&=\omega\e^{k_+\theta}\,\lcolon\e^{\i\phi_+(\theta)}\rcolon,
\label{Z++ff}
\\*
Z_{-+}(k_1,k_2|\theta)
&=\omega\rho\e^{C_E/2}\e^{-k_-\theta}
\left(
\e^{\i\pi k_-/2}\,\lcolon\e^{-\i\phi_-(\theta-{\i\pi\over2})}\rcolon
+\i\e^{-\i\pi k_-/2}\,\lcolon\e^{-\i\phi_-(\theta+{\i\pi\over2})}\rcolon
\right),
\label{Z-+ff}
\\
Z_{+-}(k_1,k_2|\theta)
&=\omega\rho\e^{C_E/2}\e^{k_-\theta}
\left(
\e^{\i\pi k_-/2}\,\lcolon\e^{\i\phi_-(\theta+{\i\pi\over2})}\rcolon
-\i\e^{-\i\pi k_-/2}\,\lcolon\e^{\i\phi_-(\theta-{\i\pi\over2})}\rcolon
\right),
\label{Z+-ff}
\\*
Z_{--}(k_1,k_2|\theta)
&=-\omega\e^{C_E}\e^{-k_+\theta}
\left(
\e^{-\i\pi k_+}\,\lcolon\e^{-\i\phi_+(\theta+\i\pi)}\rcolon
+\e^{\i\pi k_+}\,\lcolon\e^{-\i\phi_+(\theta-\i\pi)}\rcolon
\right),
\label{Z--ff}
}}
where
\eq{
k_\pm={k_1\pm k_2\over2}={a_1\pm a_2\over\sqrt2}
}
and the operators give the form factors of the exponential fields
\eq{
O_{a_1a_2}(x)=\e^{\i a_1\varphi_1(x)+\i a_2\varphi_2(x)}
=\e^{\i k_+\chi_1(x)+\i k_-\chi_2(x)}.
}
The vertex operators anticommute:
\eq{
Z_{\vep_1\ve'_1}(\theta_1)Z_{\vep_2\ve'_2}(\theta_2)
=-Z_{\vep_2\ve'_2}(\theta_2)Z_{\vep_1\ve'_1}(\theta_1),
\qquad\text{for $p_1=p_2=1$}.
\label{peq1ZZcommut}
}
The pair $Z_{++}(\theta)$, $Z_{--}(\theta)$ describes the Dirac fermion
related to the field $\chi_1$, while the pair $Z_{-+}(\theta)$,
$Z_{+-}(\theta)$ describes that related to $\chi_2$. The representation
(\ref{Z++ff}), (\ref{Z--ff}) coincides with that described
in~\cite{Lukyanov:1997bp} for the field $\e^{\i a\varphi}$ (see
Eq.~(\ref{SMffaction}) for normalizations) with $a=k_+$, while the
representation (\ref{Z-+ff}), (\ref{Z+-ff}) gives the representation
(\ref{Z+--FPL}) with $a=k_-$. Both of them provide the right hand side
of Eq.~(\ref{FreeFermionFormFactors}) with $a=k_\pm$ as the final
result for form factors.

%%%%%%%%%%%%%%%%%%%%%%%%%%%%%%%%%%%%%%%%%%%%%%%%%%%%%%%%%%%%%%%%%%%%%%%%

\section{Conclusion}

The form factors of the exponential operators~(\ref{Oa1a2b}) in the
model with the action~(\ref{sfmodel}) in the unitary regime can be
expressed as traces of the vertex operators, which are realized in
terms of three sets of free boson operators. This bosonization
procedure provides an algorithm for construction of an integral
representation for form factors. For the fundamental particles a
general explicit integral
formula~(\ref{sfformfactors})--(\ref{barGvthetadef}) for any
$N$-particle form factor has been found.

There are two important limiting cases. The first one, $p_2\to0$, gives
the sausage model in the dual representation. The free field
representation for the vertex operators is constructed by the limit in
the free field representation for the first bound state of two
fundamental particles. The second limiting case, $p_1+p_2\to2$ gives
the known cosine-cosine model, that contains the fundamental particles
as well as their bound states. The first bound state forms a triplet
and the respective vertex operators can be considered as analytic
continuation of the vertex operators of the sausage model after the
substitution $p_1\to p_2$, $p_2\to p_3$, $p_3\to p_1$.

There are three interesting problems, which we did not solve in this
paper. The first one is to compute analytically the integrals for
two-particle form factors of some special exponential operators. By the
analogy to the sine-Gordon theory, it is expected to be possible to
compute them for $a_i={n\over2}\alpha_i$ ($n\in\Z$). In this paper the
integral was only computed for the one-particle form factor in the
sausage model (see Appendix~\ref{appendix-traces}).

Another task is to obtain a free field realization of the vertex
operators for restricted models like the parafermion sine-Gordon or the
coset models, which are related the family of models under
consideration by the quantum group reduction. It is expected that it
can be achieved by an appropriate change of the screening
operators~\cite{Lukyanov:1996qs} together with, probably, some Felder
type resolution.

The third and the most important problem is related with the
prescription for calculation of the form factors of the descendent
operators~(\ref{Oa1a2a3descendants}) and of the nonlocal with respect
to the fields $\varphi_i(x)$ operators. These problems will be
addressed in forthcoming papers.

%%%%%%%%%%%%%%%%%%%%%%%%%%%%%%%%%%%%%%%%%%%%%%%%%%%%%%%%%%%%%%%%%%%%%%%%

\section*{Acknowledgments}

The authors are grateful to Ya.~Pugai for stimulating discussions. The
work was supported, in part, by EU under the contract
HPRN--CT--2002--00325, by INTAS under the grant INTAS--OPEN--00--00055,
by Russian Foundation for Basic Research under the grant RFBR
02--01--01015, and by Russian Ministry of Science and Technology under
the Scientific Schools grant 2044.2003.2. M.~L. is indebted to
P.~Forgacs and M.~Niedermaier for organizing his visit to Universit\'e
de Tours and Universit\'e Montpellier~II during the autumn of 2002,
where this work was started, and to CNRS for support of this visit.

%%%%%%%%%%%%%%%%%%%%%%%%%%%%%%%%%%%%%%%%%%%%%%%%%%%%%%%%%%%%%%%%%%%%%%%%

\Appendix

%%%%%%%%%%%%%%%%%%%%%%%%%%%%%%%%%%%%%%%%%%%%%%%%%%%%%%%%%%%%%%%%%%%%%%%%

\section{Regularized $t$-integration}
\label{appendix-regint}

The Jimbo--Konno--Miwa regularization rule, described in
Sec.~\ref{sec-bosonization}, leads to the following simple formulas for
regularized integrals:
\Align{
\exp\int^\infty_0{dt\over t}\,\e^{-zt}
&={\e^{-C_E}\over z},
\\*
\exp\int^\infty_0{dt\over t}\,{\e^{-zt}-1\over\e^t-1}
&=\e^{C_Ez}\Gamma(z+1),
\\
\exp\int^\infty_0{dt\over t}\,{1\over\e^t-1}
&={1\over\sqrt{2\pi}}\e^{C_E/2},
\\
\int^\infty_0 dt\,f(at)
&={1\over a}\int^\infty_0 dt\,f(t)
-\Res_{t=0}f(at)\cdot\log a.
}
Here $C_E$ is the Euler constant.

%%%%%%%%%%%%%%%%%%%%%%%%%%%%%%%%%%%%%%%%%%%%%%%%%%%%%%%%%%%%%%%%%%%%%%%%

\section{List of trace functions}
\label{appendix-traces}

Here we list formulas for the constants $C_{U_j}$ and functions
$G_{U_jU_k}(\theta)$ defined in Eqs.~(\ref{Cjdef}),~(\ref{Gjkdef}),
which determine the traces according to Eq.~(\ref{TraceFormula}), for
the operators $V_i(\theta)$, $I^{(\pm)}_i(\theta)$ defined in
Eqs.~(\ref{VIdef}) and $V^{(\pm)}(\theta)$, $I^{(\pm)}(\theta)$ defined
in Eqs.~(\ref{VIpmsausage}). These calculations are performed
straightforwardly by application of the formula (we recall that
subscripts $i$, $j$ are defined modulo~3):
$$
\llangle a_i(t)a_j(t')\rrangle
=t{\sinh^2{\pi t\over2}\over\sinh\pi t\sinh{\pi p_it\over2}}
{1\over1-\e^{-2\pi t}}\,\delta_{ij}\delta(t+t').
$$
For the constants we have
\Align{
C_{V_i}
&=C_i
=\exp\left(-\int^\infty_0{dt\over t}\,
{e^{-\pi t}\sinh^2{\pi t\over2}\sinh{\pi(p_{i+1}+p_{i+2})t\over2}\over
2\sinh^2\pi t\sinh{\pi p_{i+1}t\over2}\sinh{\pi p_{i+2}t\over2}}\right),
\\
C_{I^{(\pm)}_i}
&=\bar C_i
=\e^{-(C_E+\log\pi p_i)/p_i}
{\pi p_i^{-1/2}\over\Gamma(1+1/p_i)},
\\
C_{V^{(\pm)}}
&=C_{I^{(\pm)}}=\bar C=\bar C_1|_{p_1\to p}.
}
For the functions $G_{jk}$ the result is
\Gather{
G_{V_iV_j}(\theta)=G_{ij}(\theta),
\\
G_{V_iI^{(\pm)}_j}(\theta)
=G_{I^{(\mp)}_jV_i}(\theta)
=W^{(\pm)}_{ij}(\theta),
\\
G_{I^{(A)}_iI^{(B)}_j}(\theta)
=\bar G^{(AB)}_{ij}(\theta),
\\
G_{V^{(A)}V^{(B)}}(\theta)
=G_{I^{(A)}I^{(B)}}(\theta)
=\bar G^{(AB)}(\theta)
=\bar G^{(AB)}_{11}(\theta)|_{p_1\to p},
\\
G_{V^{(A)}I^{(B)}}(\theta)
=G_{I^{(A)}V^{(B)}}(\theta)
=\bar W^{(AB)}(\theta)
=(\bar G^{(AB)}(\theta))^{-1}.
}
The functions $G_{ij}(\theta)$ are rather complicated:
\subeq{\Align{
G_{ii}(\theta)
&=\exp\left(-\int^\infty_0{dt\over t}\,
{\sinh^2{\pi t\over2}\sinh{\pi(p_{i+1}+p_{i+2})t\over2}\cos(\theta+\i\pi)t
\over\sinh^2\pi t\sinh{\pi p_{i+1}t\over2}\sinh{\pi p_{i+2}t\over2}}\right),
\\
G_{ij}(\theta)
&=\exp\left(-\int^\infty_0{dt\over t}\,
{\sinh^2{\pi t\over2}\cos(\theta+\i\pi)t
\over\sinh^2\pi t\sinh{\pi p_kt\over2}}\right)
\qquad
(i\ne j,\ k\ne i,j),
}}
but they never appear in the integrands, i.~e.\ their arguments never
contain integration variables. The functions $W^{(\pm)}_{ij}(\theta)$
can be expressed in terms of a single function $W(p;\theta)$:
\subeq{\Align{
W^{(\pm)}_{ii}(\theta)
&=1,
\\*
W^{(\pm)}_{i-1,i}(\theta)
&=W(p_i;\theta\pm\ipihalf p_i),
\\
W^{(\pm)}_{i+1,i}(\theta)
&=W(p_i;\theta),
\\*
W(p;\theta)
&=\exp\int^\infty_0{dt\over t}\,
{\sinh{\pi t\over2}\cos(\theta+\i\pi)t\over\sinh\pi t\sinh{\pi pt\over2}}.
}}
The function $W(p;\theta)$ possesses poles at the points
$\theta={\i\pi\over2}(p-1+2Mp)+2\pi\i N$ ($M,N=0,1,2,\ldots$) and
$\theta=-{\i\pi\over2}(p-1+2Mp)-2\pi\i(N+1)$, and satisfies the relations
\Align{
&W(p;\theta)=W(p;-\theta-2\pi\i),
\notag
\\*
&W(p;\theta-\i\pi)W(p;\theta)={1\over2\cosh{\theta+\i\pi/2\over p}},
\qquad
{W(p;\theta-\i\pi p/2)\over W(p;\theta+\i\pi p/2)}
=\i\tanh\left({\i\pi\over4}+{\theta\over2}\right).
\label{Wproperties}
}
The functions $\bG_{ij}^{(AB)}(\theta)$ are purely trigonometric:
\subeq{\Align{
\bar G^{(\mp\pm)}_{ii}(\theta)
&=\pm2\i\sinh{\theta+\i\pi\over p_i},
\\*
\bar G^{(\pm\pm)}_{ii}(\theta)
&=2\tanh{\theta\over2}
\sinh{\theta+\i\pi\over p_i},
\\
\bar G^{(\mp\pm)}_{i,i+1}(\theta)
&=\bar G^{(\mp\pm)}_{i+1,i}(\theta)=1,
\\*
\bar G^{(\pm\pm)}_{i,i+1}(\theta)
&=\bar G^{(\mp\mp)}_{i+1,i}(\theta)
=-\i\coth\left({\i\pi p_{i+1}\over4}\pm{\theta\over2}\right).
}}
It means that the form factors of the sausage model can be expressed in
terms of multiple integrals of trigonometric functions with two
different periods.

It is important to make the following remark about the contours. To fix
the contours it is necessary to consider the product in the trace and
look at the poles of this expression. There are poles that appear from
the operator products~(\ref{opprods}), (\ref{opprods-sausage}) and the
extra factors in the screening operators. The prescription for the
integration contours with respect to those poles was already described
in the main part of the paper. The additional poles that arise due to
the trace should be treated as follows. Namely, the contours should go
below the excessive poles at the points $\theta_j+\i\delta_j$ with
$\delta_j>0$ and above the excessive poles at the points
$\theta_j+\i\delta_j$ with $\delta_j<0$. For example, the trace
$$
\llangle V^{(-)}(\theta)I^{(+)}(\xi)\rrangle
{\pi\over\sinh{\xi-\theta-\i\pi\over p}}
={\pi\bar C^2\bar W^{(-+)}(\xi-\theta)
\over\sinh{\xi-\theta-\i\pi\over p}}
=-{\i\pi\bar C^2
\over2\sinh{\xi-\theta+\i\pi\over p}\sinh{\xi-\theta-\i\pi\over p}}
$$
possesses poles at $\xi=\theta\pm\i\pi$, though the product
$$
V^{(-)}(\theta)I^{(+)}(\xi)
{\pi\over\sinh{\xi-\theta-\i\pi\over p}}
=\lcolon V^{(-)}(\theta)I^{(+)}(\xi)\rcolon\,
\i r_p^{-2}\Gamma\left(\i(\xi-\theta+\i\pi)\over\pi p\right)
\Gamma\left(-{\i(\xi-\theta-\i\pi)\over\pi p}+1\right)
$$
possesses a pole at $\xi=\theta-\i\pi$ only. It means that the
integration contour for $\xi$ must go below both poles at
$\theta\pm\i\pi$. In contrast, in the expression
$$
\llangle V^{(+)}(\theta)I^{(-)}(\xi)\rrangle
{\pi\over\sinh{\xi-\theta-\i\pi\over p}}
={\pi\bar C^2\bar W^{(+-)}(\xi-\theta)
\over\sinh{\xi-\theta-\i\pi\over p}}
={\i\pi\bar C^2\over2\sinh{\xi-\theta+\i\pi\over p}
\sinh{\xi-\theta-\i\pi\over p}}
$$
the integration contour must go above both poles at
$\xi=\theta\pm\i\pi$, because the only pole at $\xi=\theta+\i\pi$
coming from the extra factors survives after removing the trace sign.

As an example of application of the integral representation, we
calculate the one-particle form factor $\llangle Z_0(\theta)\rrangle$
in the sausage model. This form factor
$$
\llangle Z_0(\theta)\rrangle
={\langle0|\e^{\i a\varphi+b\chi}|\theta,0\rangle
\over\langle0|\e^{\i a\varphi+b\chi}|0\rangle}
$$
is just a constant. Nevertheless, it gives the first nontrivial
contribution to the infrared asymptotics of correlation functions.

Using the fact that $\bar W^{(++)}(\theta)=\bar W^{(--)}(\theta)$, we
obtain
$$
\Aligned{
\llangle Z_0(\theta)\rrangle
&=-c\sqrt{2\cos{\pi\over p}}
\int{d\xi\over2\pi\i}\,
\left\langle\!\!\left\langle
\left(V^{(+)}\e^{-\kappa}-\i V^{(-)}\e^\kappa\right)
\left(I^{(+)}\e^\kappa-\i I^{(-)}\e^{-\kappa}\right)
\right\rangle\!\!\right\rangle
{\pi\e^{k(\theta-\xi)}\over\sinh{\xi-\theta-\i\pi\over p}}
\\
&=\i c\bar C^2\sqrt{2\cos{\pi\over p}}
\left(
\int_{\cC_{+-}}{d\xi\over2\pi\i}\,\bar W^{(+-)}(\xi-\theta)
{\pi\e^{-2\kappa+k(\theta-\xi)}\over\sinh{\xi-\theta-\i\pi\over p}}
+\int_{\cC_{-+}}{d\xi\over2\pi\i}\,\bar W^{(-+)}(\xi-\theta)
{\pi\e^{-2\kappa+k(\theta-\xi)}\over\sinh{\xi-\theta-\i\pi\over p}}
\right).
}
$$
The contours $\cC_{+-}$ and $\cC_{-+}$ are chosen according to the
above mentioned rule. As we have seen just now, the contour $\cC_{+-}$
goes above the poles at $\xi=\theta\pm\i\pi$, while the contour
$\cC_{-+}$ goes below the poles at these points.

With the substitution $\xi\to\xi+\theta$ we obtain
$$
\llangle Z_0(\theta)\rrangle
=-{\pi c\bar C^2\over2}\sqrt{2\cos{\pi\over p}}
\left(
\int_{\cC_{+-}}{d\xi\over2\pi}\,
{\e^{-k\xi-2\kappa}\over\sinh{\xi+\i\pi\over p}\sinh{\xi-\i\pi\over p}}
-\int_{\cC_{-+}}{d\xi\over2\pi}\,
{\e^{-k\xi+2\kappa}\over\sinh{\xi+\i\pi\over p}\sinh{\xi-\i\pi\over p}}
\right).
$$
The integration can be done by residues, if we suppose e.~g.\ $\Im
k<0$. With these integrands, which are periodic functions times an
exponential function, it is easy to sum up the residues with the result
\eq{
\llangle Z_0(\theta)\rrangle
={p^{1/2}\sin{\pi\over p}\over\sqrt{\sin{2\pi\over p}}}
{\sin\pi{a\over\alpha}\over\sin\pi\alpha a}\sin\pi\beta b,
\qquad
p=2\alpha^2=2\beta^2+2.
\label{Z0formfactor}
}
In the limit $p\to2$ we have $\llangle
Z_0(\theta)\rrangle\to\sqrt{2\pi}\,b$ in consistency with the free boson
result. Note, that the result~(\ref{Z0formfactor}) is invariant with
respect to the reflection transformation $b\to1/\beta-b$ (see
Ref.~\cite{Baseilhac:1998eq}).

%%%%%%%%%%%%%%%%%%%%%%%%%%%%%%%%%%%%%%%%%%%%%%%%%%%%%%%%%%%%%%%%%%%%%%%%

\section{Check of the Zamolodchikov--Faddeev algebra relations}
\label{appendix-commut-gen}

The relation (\ref{HZcommut}) for the commutator
$[H,Z_{\ve\ve'}(\theta)]$ of the corner Hamiltonian and a vertex
operator can be checked straightforwardly. Here we verify the
commutation relations~(\ref{Zcommut}) of two vertex operators and prove
the expression~(\ref{cidef}) for the normalization constants $c_i$ by
checking the relation~(\ref{Znorm}).

We start with the simplest commutation relation
$$
Z_{++}(\theta_1)Z_{++}(\theta_2)
=-S_{p_1}(\theta_{12})^{++}_{++}S_{p_2}(\theta_{12})^{++}_{++}
Z_{++}(\theta_2)Z_{++}(\theta_1),
$$
where $\theta_{12}=\theta_1-\theta_2$ and we used the fact that the $S$
matrix $S_{p_1p_2}(\theta)$ is a tensor product of two sine-Gordon $S$
matrices~(\ref{SGSmatrix}). Substituting Eq.~(\ref{Z++sf}) we obtain
$$
V_3(\theta_1)V_3(\theta_2)
=-S_{p_1}(\theta_{12})^{++}_{++}S_{p_2}(\theta_{12})^{++}_{++}
V_3(\theta_2)V_3(\theta_1).
$$
Reduce the l.h.s.\ and the r.h.s.\ to normal products with the help of
Eq.~(\ref{VVprod}):
$$
g_{33}(-\theta_{12})\,\lcolon V_3(\theta_1)V_3(\theta_2)\rcolon
=-S_{p_1}(\theta_{12})^{++}_{++}S_{p_2}(\theta_{12})^{++}_{++}
g_{33}(\theta_{12})
\,\lcolon V_3(\theta_1)V_3(\theta_2)\rcolon
$$
i.e.
\eq{
-S_{p_1}(\theta)^{++}_{++}S_{p_2}(\theta)^{++}_{++}
={g_{33}(-\theta)\over g_{33}(\theta)}.
\label{Sgg}
}
The last identity can be obtained using Eq.~(\ref{SGSmatrix}).

Consider now another commutation relation
$$
Z_{++}(\theta_1)Z_{-+}(\theta_2)
=-S_{p_1}(\theta_{12})^{+-}_{+-}S_{p_2}(\theta_{12})^{++}_{++}
Z_{-+}(\theta_2)Z_{++}(\theta_1)
-S_{p_1}(\theta_{12})^{-+}_{+-}S_{p_2}(\theta_{12})^{++}_{++}
Z_{++}(\theta_2)Z_{-+}(\theta_1).
$$
Using Eqs.~(\ref{HZsf}), (\ref{Ssf}), (\ref{VVprod}), (\ref{VIprod}),
(\ref{rhoomegaalg}) we obtain that it is equivalent to the following
equality
$$
\Aligned{
&\sum_{A=\pm}\int{d\xi\over2\pi\i}\,
\lcolon\hV_3(\theta_1)\hV_3(\theta_2)\hI^{(A)}_1(\xi)\rcolon\,
{g_{33}(\theta_{12})
w^{(A)}_{31}(\xi-\theta_1)w^{(A)}_{31}(\xi-\theta_2)
\over\sinh{\xi-\theta_2-\i\pi/2\over p_1}}
\\
&\quad=S_{p_1}(\theta_{12})^{+-}_{+-}S_{p_2}(\theta_{12})^{++}_{++}
\sum_{A=\pm}\int{d\xi\over2\pi\i}\,
\lcolon\hV_3(\theta_1)\hV_3(\theta_2)\hI^{(A)}_1(\xi)\rcolon\,
{g_{33}(-\theta_{12})
w^{(A)}_{31}(\theta_1-\xi)w^{(A)}_{31}(\xi-\theta_2)
\over\sinh{\xi-\theta_2-\i\pi/2\over p_1}}
\\
&\qquad-S_{p_1}(\theta_{12})^{-+}_{+-}S_{p_2}(\theta_{12})^{++}_{++}
\sum_{A=\pm}\int{d\xi\over2\pi\i}\,
\lcolon\hV_3(\theta_1)\hV_3(\theta_2)\hI^{(A)}_1(\xi)\rcolon\,
{g_{33}(-\theta_{12})
w^{(A)}_{31}(\xi-\theta_1)w^{(A)}_{31}(\xi-\theta_2)
\over\sinh{\xi-\theta_1-\i\pi/2\over p_1}}
}
$$
with
$$
\hV_i(\theta)=V_i(\theta)\,\e^{{k_{i+1}+k_{i+2}\over2}\theta},
\qquad
\hI^{(+)}_i(\xi)=I^{(+)}_i(\xi)\,\e^{\kappa_i},
\qquad
\hI^{(-)}_i(\xi)=-\i I^{(-)}_i(\xi)\,\e^{-\kappa_i}.
$$
We note that operator part equal to
$\lcolon\hV_3(\theta_1)\hV_3(\theta_2)\hI^{(A)}_1(\xi)\rcolon$ is the
same in all terms, therefore it is sufficient to prove the identity for
the remaining functions. Here and later we need the identities
\eq{
{w^{(A)}_{i-1,i}(-\theta)\over w^{(A)}_{i-1,i}(\theta)}
={\sinh{\theta+\i\pi/2\over p_i}\over\sinh{\theta-\i\pi/2\over p_i}},
\qquad
{w^{(A)}_{i+1,i}(-\theta)\over w^{(A)}_{i+1,i}(\theta)}
={\cosh{\theta+\i\pi/2\over p_i}\over\cosh{\theta-\i\pi/2\over p_i}},
\qquad
{\bg^{(BA)}_{ii}(-\theta)\over\bg^{(AB)}_{ii}(\theta)}
={\sinh{\theta-\i\pi\over p_i}\over\sinh{\theta+\i\pi\over p_i}}.
\label{wgratios}
}
Dividing the functions in the l.h.s.\ and r.h.s.\ by the numerator of
the function in the l.h.s.\ and using Eqs.~(\ref{sfSmatrix}),
(\ref{SGSmatrix}) and (\ref{Sgg}) we obtain
$$
\Aligned{
{1\over\sinh(x-y_2-z/2)}
&=-{\sinh(y_1-y_2)\over\sinh(z-y_1+y_2)}{\sinh(x-y_1+z/2)\over\sinh(x-y_1-z/2)}
{1\over\sinh(x-y_2-z/2)}
\\
&\quad
+{\sinh z\over\sinh(x-y_1+y_2)}{1\over\sinh(x-y_1-z/2)}
}
$$
with $x=\xi/p$, $y_i=\theta_i/p$, $z=\i\pi/p$. This identity can be
checked straightforwardly.

The other commutation relations can be checked in the same way. The
commutation relation
$$
Z_{-+}(\theta_1)Z_{++}(\theta_2)
=-S_{p_1}(\theta_{12})^{+-}_{-+}S_{p_2}(\theta_{12})^{++}_{++}
Z_{-+}(\theta_2)Z_{++}(\theta_1)
-S_{p_1}(\theta_{12})^{-+}_{-+}S_{p_2}(\theta_{12})^{++}_{++}
Z_{++}(\theta_2)Z_{-+}(\theta_1)
$$
is valid due to the identity
$$
\Aligned{
{1\over\sinh(x-y_1-z/2)}{\sinh(x-y_2+z/2)\over\sinh(x-y_2-z/2)}
&=-{\sinh z\over\sinh(z-y_1+y_2)}{1\over\sinh(x-y_2-z/2)}
{\sinh(x-y_1+z/2)\over\sinh(x-y_1-z/2)}
\\
&\quad
+{\sinh(y_1-y_2)\over\sinh(x-y_1+y_2)}{1\over\sinh(x-y_1-z/2)}.
}
$$
The commutation relation
\eq{
Z_{-+}(\theta_1)Z_{-+}(\theta_2)
=-S_{p_1}(\theta_{12})^{--}_{--}S_{p_2}(\theta_{12})^{++}_{++}
Z_{-+}(\theta_2)Z_{-+}(\theta_1)
\label{ZmpZmpcommut}
}
contains a two-fold integral. There is an important fact concerning the
integration contours. Each product of vertex operators in this equation
splits into four terms containing different operator products
$V(\theta_1)\*V(\theta_2)\*\hI^{(A)}(\xi_1)\*\hI^{(B)}(\xi_2)$,
$A,B=\pm$. For each pair $(A,B)$ we can chose the contour
$\cC_1^{(A,B)}$ for $\xi_1$ and the contour $\cC_2^{(A,B)}$ for
$\xi_2$, so that it satisfy the pole avoiding rules for all three
operator products in Eq.~(\ref{ZmpZmpcommut}). The important fact is
that $\cC_1^{(A,B)}$ can be deformed into $\cC_2^{(B,A)}$ without
meeting poles. (For $A=B$ it simply means that $\cC_1^{(A,A)}$ and
$\cC_2^{(A,A)}$ are up to a deformation the same contour.) This means
that we can symmetrize the integral in the integration variables. After
this symmetrization we arrive to the identity
\eq{
f(x_1,x_2)+f(x_2,x_1){\sinh(x_2-x_1-z)\over\sinh(x_2-x_1+z)}=0
\label{fx1x2eq0}
}
with
$$
\Aligned{
f(x_1,x_2)
&={1\over\sinh(x_1-y_1-z/2)}{1\over\sinh(x_2-y_2-z/2)}
{\sinh(x_1-y_2+z/2)\over\sinh(x_1-y_2-z/2)}
\\
&\quad
-{1\over\sinh(x_1-y_2-z/2)}{1\over\sinh(x_2-y_1-z/2)}
{\sinh(x_1-y_1+z/2)\over\sinh(x_1-y_1-z/2)},
}
$$
which is checked straightforwardly.

The relations proved above contain the vertex operators
$Z_{\ve\ve'}(\theta)$ with $\ve'=+$ only. That is why the proofs only
involved the screening operator $S_1(\theta)$. The check of relations
that contain the vertex operators with $\ve=+$ go essentially in the
same line. The proofs of these relations have to do with the screening
operator $S_2(\theta)$ only, and the identities to be checked are
obtained from the above ones by the substitution $y_i\to y_i+\i\pi/2$.
The remaining relations for the vertex operators with generic $\ve$,
$\ve'$ follow from the proved ones and commutativity of the screening
operators $S_1(\theta)$ and $S_2(\theta)$.

At last, we check the normalization condition~(\ref{Znorm}). It is
enough to verify it for the product
\Multline{
Z_{++}(\theta')Z_{--}(\theta)
=c_1c_2\sum_{A,B=\pm}
\int_{\cC_1}{d\xi_1\over2\pi\i}\int_{\cC_2}{d\xi_2\over2\pi\i}\,
\lcolon\hV_3(\theta')\hV_3(\theta)\hI_1(\xi_1)\hI_2(\xi_2)\rcolon
\\*
\times
g_{33}(\theta-\theta')w^{(A)}_{31}(\xi_1-\theta')
{\pi w^{(A)}_{31}(\xi_1-\theta)\over\sinh{\xi_1-\theta-\i\pi/2\over p_1}}
w^{(B)}_{32}(\xi_2-\theta')
{\i\pi w^{(B)}_{32}(\xi_2-\theta)\over
\cosh{\xi_2-\theta-\i\pi/2\over p_1}}\bg^{(++)}_{12}(\xi_2-\xi_1).
\notag
}
As a rule, the pole at $\theta'=\theta+\i\pi$ is known to be related to
pinching contours between poles~\cite{Lukyanov93}. An elaborate
analysis of the poles shows that the only pinching situation takes
place at the term with $A=B=+$. The contour $\cC_1$ goes below the pole
at $\theta'-{\i\pi\over2}$ coming from $w^{(+)}_{31}$ and above the
pole at $\xi_2+{\i\pi p_2\over2}$ coming from $\bg^{(++)}_{12}$. The
contour $\cC_2$ for $\xi_2$ goes, in turn, above the pole at
$\theta+{\i\pi\over2}-{\i\pi p_2\over2}$ from $\pi
w^{(B)}_{32}(\xi_2-\theta)/\cosh{\xi_2-\theta-\i\pi/2\over p_1}$. We
have to push the contour $\cC_1$, e.g., upward. The residue at
$\xi_1=\theta'-{\i\pi\over2}$ contribute the singular term, while the
remaining integral along the contour that goes far from poles does not
contribute it, because $\xi_2$ will not be pinched between the points
$\xi_1-{\i\pi p_2\over2}$ and $\theta+{\i\pi\over2}-{\i\pi p_2\over2}$.
In the residue we push $\cC_2$, e.g., downward. The residue at
$\xi_2=\theta+{\i\pi\over2}-{\i\pi p_2\over2}$ contributes the
singularity.

For the operator part it is easy to get
$$
\lcolon\hV(\theta+\i\pi)\hV(\theta)
\hI^{(+)}_1(\theta+\ipihalf)\hI^{(+)}_2(\theta+\ipihalf
-{\textstyle{\i\pi p_2\over2}})\rcolon
=1.
$$

Finally, we obtain
$$
Z_{++}(\theta')Z_{--}(\theta)
=-{\i\over\theta'-\theta-\i\pi}
\pi^3c_1c_2r_{p_1}^{-2}r_{p_2}^{-2}g_{33}(-\i\pi)
{\Gamma\left(-{1\over p_1}\right)\Gamma\left(-{1\over p_2}\right)
\over
\Gamma\left(1+{1\over p_1}\right)\Gamma\left(1+{1\over p_2}\right)}
+O(1),
$$
where $r_p$ is defined in Eq.~(\ref{wpz}). Comparing with the
normalization condition~(\ref{Znorm}) we obtain the value of the
product~$c_1c_2$. This calculation does not fix $c_1$ and $c_2$
separately. But if we demand all the operators $Z^i_{\ve\ve'}(\theta)$
to be normalized by Eq.~(\ref{Znorm}), we get all the products
$c_ic_{i+1}$ (which are the result of cyclic permutations of subscripts
in the expression for $c_1c_2$) and obtain Eq.~(\ref{cidef}).

%%%%%%%%%%%%%%%%%%%%%%%%%%%%%%%%%%%%%%%%%%%%%%%%%%%%%%%%%%%%%%%%%%%%%%%%

\section{Vertex operators in the sausage model}
\label{appendix-sausage}

Here the operators $Z_+(\theta)$ and $Z_0(\theta)$ will be obtained
directly from the general model~(\ref{sfmodel}) by taking the limit
$p_2\to0$. The operator $Z_-(\theta)$ will be found from the
commutation relations of the ZF algebra for the sausage
model. The proof of the commutation relations in this case has some
additional complication related to the pole at $\xi=\theta$ in the
product $V^{(\pm)}(\theta)I^{(\pm)}(\xi)$. It will be shown how to
manage this pole. At the end, the normalization constant $c$ in
Eq.~(\ref{HZsausage}) will be derived.

Consider the general model~(\ref{sfmodel}) for $p_2<1$ and calculate the
products
\eq{
Z_{\ve+}(\theta'+\ipihalf(1-p_2))
Z_{\ve'-}(\theta-\ipihalf(1-p_2))
\label{Z+Z-prod}
}
in the vicinity of the point $\theta'=\theta$. The residue of this pole
will give the first bound state of two fundamental particles.

We begin with the product
\AlignStar{
{}&Z_{++}(\theta'+\delta)Z_{+-}(\theta-\delta)
\\
&\qquad=-\rho c_2\int{d\xi\over2\pi\i}\,V_3(\theta'+\delta)
V_3(\theta-\delta)
(I^{(+)}_2(\xi)\e^{\kappa_2}-\i I^{(-)}_2(\xi)\e^{-\kappa_2})
{\pi\e^{{k_1+k_2\over2}(\theta+\theta')-k_2\xi}
\over\sinh{\xi-\theta\over p_2}}
\\
&\qquad
=-\rho c_2\int{d\xi\over2\pi\i}\,
\lcolon V_3(\theta'+\delta)
V_3(\theta-\delta)
(I^{(+)}_2(\xi)\e^{\kappa_2}-\i I^{(-)}_2(\xi)\e^{-\kappa_2})\rcolon\,
\e^{{k_1+k_2\over2}(\theta+\theta')-k_2\xi}
\\
&\qquad\quad\times
g_{33}(\theta-\theta'-2\delta)
w^{(+)}_{32}(\xi-\theta'-\delta)w^{(+)}_{32}(\xi-\theta+\delta)
{\pi\over\sinh{\xi-\theta\over p_2}}
}
with $\delta={\i\pi\over2}(1-p_2)$. The integration contour is pinched
between the pole at the point $\xi=\theta'$ of the function
$w(p_2,1/2|\xi-\theta'-\delta)$ and that at $\xi=\theta$ of the
function $\sinh^{-1}{\xi-\theta\over p_2}$. Pushing the contour
downward and calculating the residue of the second pole we obtain
$$
\Aligned{
{}\span
Z_{++}(\theta'+\delta)Z_{+-}(\theta-\delta)
=\rho{\i C\over\theta'-\theta}\,
\lcolon V_3(\theta+\delta)V_3(\theta-\delta)
(I^{(+)}_2(\theta)\e^{\kappa_2}-\i I^{(-)}_2(\theta)\e^{-\kappa_2})
\rcolon\,
\e^{k_1\theta}
+O(1),
\\
\span
C={\sqrt\pi\,G(p_2,-\i\pi)\over G(p_1,-2\delta)G(p_2,-2\delta)}.
}
$$
It is straightforward to find that
$$
\Aligned{
\lcolon V_3(\theta+\delta)V_3(\theta-\delta)I^{(+)}_2(\theta)\rcolon
&\to V^{(+)}(\theta),
\\
\lcolon V_3(\theta+\delta)V_3(\theta-\delta)I^{(-)}_2(\theta)\rcolon
&\to V^{(-)}(\theta)
}
$$
as $\delta\to\i\pi/2$ ($p_2\to0$). In this limit the constant $C$
remains finite:
$$
C\to 2\pi^{-1/2}G^{-1}(p_1,-\i\pi),
\qquad
p_2\to0.
$$
Here we used the identity
\eq{
{G(p;-\i\pi)\over G(p;-\i\pi+\i\pi p)}
={2\over\sqrt{\pi(1-p)}}
{\Gamma\left(1-{p\over2}\right)\over\Gamma\left(1-p\over2\right)}.
\label{G1id}
}
As a result, we obtain
$$
Z_{++}(\theta'+\delta)Z_{+-}(\theta-\delta)
\to{\i C\over\theta'-\theta}\rho V(k_1,\kappa_1|\theta)+O(1).
$$
It is easy to check that
\eq{
{c_1C\over c}=\Gamma_{2,1},
\qquad
p_2=0,
\label{c1Coverc}
}
with the constant $c$ given by Eq.~(\ref{cdef}) with $p=p_1$ and
$\Gamma_{2,1}$ given by Eq.~(\ref{p20limitcoeffs}). It proves that
$B^{2,1}_+(\theta)=c_1^{-1}Z_+(k,\kappa|\theta)$ with
$Z_+(k,\kappa|\theta)$ defined by Eq.~(\ref{Z+sausage}).

Consider now the products~(\ref{Z+Z-prod}) for $\ve'+\ve=0$. For
$\ve'=+$, $\ve=-$ we have
\AlignStar{
{}&Z_{++}(\theta'+\delta)Z_{--}(\theta-\delta)
\\
&\qquad=c_1c_2
V_3(\theta'+\delta)V_3(\theta-\delta)
\int_{\cC_1}{d\xi_1\over2\pi}\int_{\cC_2}{d\xi_2\over2\pi}\,
(I^{(+)}_1(\xi_1)\e^{\kappa_2}-\i I^{(-)}_1(\xi_1)\e^{-\kappa_2})
(I^{(+)}_2(\xi_2)\e^{\kappa_2}-\i I^{(-)}_2(\xi_2)\e^{-\kappa_2})
\\*
&\qquad\quad\times
{-\pi^2\e^{{k_1+k_2\over2}(\theta+\theta')-k_1\xi_1-k_2\xi_2}
\over
\sinh{\xi_1-\theta-\i\pi p_2/2\over p_1}
\sinh{\xi_2-\theta\over p_2}}
\\
&\qquad
=c_1c_2\sum_{A,B=\pm}
\int_{\cC_1}{d\xi_1\over2\pi\i}\int_{\cC_2}{d\xi_2\over2\pi\i}\,
\lcolon\hV_3(\theta'+\delta)\hV_3(\theta-\delta)
\hI^{(A)}_1(\xi_1)\hI^{(B)}_2(\xi_2)\rcolon
\\*
&\qquad\quad\times
w^{(A)}_{13}(\xi_1-\theta'-\delta)
{\pi w^{(A)}_{13}(\xi_1-\theta+\delta)\over
\sinh{\xi_1-\theta-\i\pi p_2/2\over p_1}}
w^{(+)}_{23}(\xi_2-\theta'-\delta)
{\pi w^{(+)}_{23}(\xi_1-\theta+\delta)\over
\sinh{\xi_2-\theta\over p_2}}
\bg^{(AB)}_{12}(\xi_2-\xi_1).
}
There are two pinching points as $\theta'\to\theta$. The first one is
the same as in the case $\ve'=\ve=+$: the contour $C_1$ in this case is
also pinched between the poles at the points $\theta'$ and
$\theta$. The second one is in the term $A=B=-$ and involves both
integration variables: the contour $C_1$ goes below the point
$\theta'-{\i\pi p_2\over2}$ and above the point $\xi_2+{\i\pi
p_2\over2}$, while the contour $C_2$ goes above the point $\theta-\i\pi
p_2$. We want to move the contour $C_1$ so that it would go above the
points $\theta+\i\pi p_2/2$ and $\theta'-\i\pi p_2/2$. So we have to
push it through the pole $\theta+\i\pi p_2/2$ in the term with $A=+$,
$B=-$, and through the pole $\theta'-\i\pi p_2/2$ in the term with
$A=B=+$. After all these operations we obtain
$$
Z_{++}(\theta'+\delta)Z_{--}(\theta-\delta)
={\i c_1C\over\theta'-\theta}(J_0+J_1+J_2+J_3)+O(1)
$$
with
\AlignStar{
J_0
&=\int{d\xi\over2\pi\i}\,
\lcolon V(\theta+\delta)V(\theta-\delta)
\hI_2(\theta)\rcolon\,\hI_1(\xi)
{\pi\over\sinh{\xi-\theta-\i\pi p_2/2\over p_1}}
\\
J_1
&=-C'{p_1\over p_2}\lcolon V(\theta+\delta)V(\theta-\delta)
\hI^{(+)}_1(\theta-{\textstyle{\i\pi p_2\over2}})
\hI^{(+)}_2(\theta)\rcolon,
\\
J_2
&=-C'{p_1\over1-p_2}\lcolon V(\theta+\delta)V(\theta-\delta)
\hI^{(+)}_1(\theta-{\textstyle{\i\pi p_2\over2}})
\hI^{(-)}_2(\theta)\rcolon,
\\
J_3
&={C'\over\Gamma\left(1-{p_2\over p_1}\right)}{p_1\over p_2}
\lcolon V(\theta+\delta)V(\theta-\delta)
\hI^{(+)}_1(\theta-{\textstyle{\i\pi p_2\over2}})
\hI^{(+)}_2(\theta-\i\pi p_2)\rcolon,
\\
C'
&=\pi r_{p_1}^{-2}{\Gamma\left(1-{1-p_2\over p_1}\right)
\Gamma\left(1-{p_2\over p_1}\right)
\over\Gamma\left(1+{1\over p_1}\right)},
}
where the integration contour in $J_0$ goes below the point
$\theta-\i\pi+\i\pi p_2/2$ and above the points $\theta+\i\pi p_2/2$,
$\theta-\i\pi p_2/2$. Similarly, for $\ve'=-$, $\ve=+$ we have
$$
Z_{-+}(\theta'+\delta)Z_{+-}(\theta-\delta)
={\i c_1C\over\theta'-\theta}(J'_0+J'_1+J'_2+J'_3)+O(1).
$$
Here the operators $J'_0$ and $J'_i$ ($i=1,2,3$) have the form
\AlignStar{
J'_0
&=-\int{d\xi\over2\pi\i}\,
\lcolon V(\theta+\delta)V(\theta-\delta)
\hI_2(\theta)\rcolon\,\hI_1(\xi)
{\pi\over\sinh{\xi-\theta-\i\pi(2-p_2)/2\over p_1}}
{\sinh{\xi-\theta+\i\pi(2-p_2)/2\over p_1}
\over\sinh{\xi-\theta-\i\pi p_2/2\over p_1}},
\\
J'_i
&=J_i{\sinh{\i\pi(1-p_2)\over p_1}\over\sinh{\i\pi\over p_1}}
\qquad
(i=1,2,3),
}
where the integration contour in $J'_0$ goes above the points
$\theta'+\i\pi-\i\pi p_2/2$, $\theta+\i\pi p_2/2$, $\theta-\i\pi
p_2/2$.

Now we should take the limit $p_2\to0$. In this limit the constant $C'$
remains finite:
$$
C'=\pi r_{p_1}^{-2}
{\Gamma\left(1-{1\over p_1}\right)
\over\Gamma\left(1+{1\over p_1}\right)},
\qquad
p_2=0.
$$
It means that
$$
J_1+J_3
=\sqrt{\pi}C'p_1p_2^{-1/2}Z'_0(\theta)+O(p_2^{1/2}).
$$

We want to obtain an expression for $B^{2,1}_\rA(\theta)$ in the limit
$p_2\to0$. We should consider the difference
$$
Z_{++}(\theta'+\delta)Z_{--}(\theta-\delta)
-Z_{-+}(\theta'+\delta)Z_{+-}(\theta-\delta).
$$
Due to the relation
$$
1+{\sinh{\xi-\theta+\i\pi(2-p_2)/2\over p_1}
\over\sinh{\xi-\theta-\i\pi(2-p_2)/2\over p_1}}
={2\sinh{\xi-\theta\over p_1}\cosh{\i\pi(2-p_2)\over2p_1}
\over\sinh{\xi-\theta-\i\pi(2-p_2)/2\over p_1}}
={2\sinh{\xi-\theta\over p_1}\cosh{\i\pi\over p_1}
\over\sinh{\xi-\theta-\i\pi\over p_1}}+O(p_2)
$$
we have
\eq{
c_1C(J_0-J'_0)=-{c_1C\over c}\sqrt{2\cos{\pi\over p_1}}\>Z_0(\theta)+O(p_2)
=-\Gamma K^{2,1}_\rA Z_0(\theta)+O(p_2).
\label{J0-J'0}
}
Similarly, from the relation
$$
1-{\sinh{\i\pi(1-p_2)\over p_1}\over\sinh{\i\pi\over p_1}}
={2\sinh{\i\pi p_2\over2p_1}\cosh{\i\pi(2-p_2)\over2p_1}
\over\sinh{\i\pi\over p_1}}
=O(p_2)
$$
we obtain
$$
J_1+J_3-J'_1-J'_3=O(p_2^{1/2}),
\qquad
J_2-J'_2=O(p_2).
$$
Hence, the only finite contribution comes from the difference
$J_0-J'_0$. The equations~(\ref{J0-J'0}), (\ref{c1Coverc}) prove that
$B^{2,1}_\rA(\theta)=-Z_0(k,\kappa|\theta)$ with $Z_0(k,\kappa|\theta)$
defined in Eq.~(\ref{Z0sausage}).

To obtain an expression for $B^{2,1}_\rS(\theta)$ we should consider
the sum
$$
Z_{++}(\theta'+\delta)Z_{--}(\theta-\delta)
+Z_{-+}(\theta'+\delta)Z_{+-}(\theta-\delta).
$$
It is easy to check that
$$
1-{\sinh{\xi-\theta+\i\pi(2-p_2)/2\over p_1}
\over\sinh{\xi-\theta-\i\pi(2-p_2)/2\over p_1}}
=-{2\cosh{\xi-\theta\over p_1}\sinh{\i\pi(2-p_2)\over2p_1}
\over\sinh{\xi-\theta-\i\pi(2-p_2)/2\over p_1}}
=-{2\cosh{\xi-\theta\over p_1}\sinh{\i\pi\over p_1}
\over\sinh{\xi-\theta-\i\pi\over p_1}}+O(p_2)
$$
and hence,
$$
J_0+J'_0=O(1).
$$
Similarly, from the relation
$$
1+{\sinh{\i\pi(1-p_2)\over p_1}\over\sinh{\i\pi\over p_1}}
={2\cosh{\i\pi p_2\over2p_1}\sinh{\i\pi(2-p_2)\over2p_1}
\over\sinh{\i\pi\over p_1}}
=2+O(p_2)
$$
we have
$$
J_1+J_3+J'_1+J'_3=2\sqrt{\pi}C'p_1p_2^{-1/2}Z'_0(\theta)+O(p_2^{1/2}),
\qquad
J_2+J'_2=O(1).
$$
It is straightforward to find that
$$
\lim_{p_2\to0}2\sqrt{\pi}cC'p_1=\lim_{p_2\to0}p_2^{1/2}K^{2,1}_\rS.
$$
Together with Eq.~(\ref{c1Coverc}) it proves the
expression~(\ref{Z0prime}) for $B^{2,1}_\rS(\theta)$.

Derivation of the operator $Z_-(\theta)$ by taking the limit
of Eq.~(\ref{Z+Z-prod}) for $\ve'=\ve=-$ would be very complicated. In
fact, we do not need it at all. It is natural to suppose that the
operator $Z_-(\theta)$ has the form of a linear combination:
\eq{
C_{++}S^2_+(\theta)V(\theta)
+C_{+-}S_+(\theta)V(\theta)S_-(\theta)
+C_{-+}S_-(\theta-2\pi\i)V(\theta)S_+(\theta+2\pi\i)
+C_{--}V(\theta)S^2_-(\theta).
\label{VSScombination}
}
All four terms in this combination are different: though the integrands
are identical, the integration contours are not the same. To see it, we
consider, for example, the difference
$$
\Delta(\theta)
=S_+(\theta)V(\theta)S_-(\theta)-S^2_+(\theta)V(\theta).
$$
We can expand it as
$$
\Delta(\theta)
=\sum_{A,B,C}\Delta^{(ABC)}(\theta),
$$
where
$$
\Aligned{
\Delta^{(ABC)}(\theta)
&=c^2\hV^{(A)}(\theta)
\int_{\cC_+}{d\xi_1\over2\pi\i}\int_{\cC_-}{d\xi_2\over2\pi\i}\,
\hI^{(B)}(\xi_1)\hI^{(C)}(\xi_2)
{\pi\over\sinh{\xi_1-\theta-\i\pi\over p}}
{\pi\over\sinh{\xi_2-\theta-\i\pi\over p}}
\\
&\quad
-c^2\hV^{(A)}(\theta)
\int_{\cC_+}{d\xi_1\over2\pi\i}\int_{\cC_+}{d\xi_2\over2\pi\i}\,
\hI^{(B)}(\xi_1)\hI^{(C)}(\xi_2)
{\pi\over\sinh{\xi_1-\theta-\i\pi\over p}}
{\pi\over\sinh{\xi_2-\theta-\i\pi\over p}},
\\
\span
\hV^{(+)}(\theta)=\i V^{(+)}(\theta)\e^{-\kappa+k\theta},
\qquad
\hV^{(-)}(\theta)=V^{(-)}(\theta)\e^{\kappa+k\theta}.
}
$$
The operator $\Delta^{(ABC)}(\theta)$ is a difference of two terms,
which are distinguished by the integration contour for the
variable~$\xi_2$. For this reason the calculation amounts to finding
the residue at the point $\xi_2=\theta$:
$$
\Delta^{(ABC)}(\theta)
=\Res_{\xi_2=\theta}c\hV^{(A)}(\theta)
\int_{\cC_+}{d\xi_1\over2\pi\i}\,
\hI^{(B)}(\xi_1)\hI^{(C)}(\xi_2)
{\pi\over\sinh{\xi_1-\theta-\i\pi\over p}}
{\pi\over\sinh{\xi_2-\theta-\i\pi\over p}}
$$
The only nonzero contribution comes from the terms with $A=C$. As in
this case
\eq{
\Res_{\xi=\theta}\hV^{(+)}(\theta)\hI^{(+)}(\xi)
=-\Res_{\xi=\theta}\hV^{(-)}(\theta)\hI^{(-)}(\xi)
=D\equiv\pi pr_p^{-2}{\Gamma(1-1/p)\over\Gamma(1/p)},
\label{resVI}
}
we obtain that integrands for the operators $\Delta^{(+++)}(\theta)$
and $-\Delta^{(-+-)}(\theta)$ coincide. The same fact takes place for
the operators $\Delta^{(---)}(\theta)$ and $-\Delta^{(+-+)}(\theta)$.
If we consider the contour in the variable $\xi_1$ for the last pair,
it is the same for the operators $\Delta^{(---)}(\theta)$ and
$\Delta^{(+-+)}(\theta)$. It means that
$\Delta^{(---)}(\theta)+\Delta^{(+-+)}(\theta)=0$. We have a different
situation for the operators $\Delta^{(+++)}(\theta)$ and
$\Delta^{(-+-)}(\theta)$. In this case, both integrands possess poles
at $\xi_1=\theta-\i\pi$, but the origin of these poles is not the same.
In the term $\Delta^{(-+-)}(\theta)$ it comes from the product
$\hV^{(-)}(\theta)\hI^{(+)}(\xi_1)$ and the contour for the variable
$\xi_1$ must go below it (see Fig.~\ref{figure-VIIcontours-sausage}).
In the term $\Delta^{(+++)}(\theta)$ the pole arises from the product
$\hI^{(+)}(\xi_1)\hI^{(+)}(\theta)$ and the contour for the variable
$\xi_1$ must go above it. It means that
$$
\Aligned{
\Delta(\theta)=\Delta^{(+++)}(\theta)+\Delta^{(-+-)}(\theta)
&=c^2\Res_{\xi_1=\theta-\i\pi}\Res_{\xi_2=\theta}
\hV^{(-)}(\theta)\hI^{(+)}(\xi_1)\hI^{(-)}(\xi_2)
{\pi\over\sinh{\xi_1-\theta-\i\pi\over p}}
{\pi\over\sinh{\xi_2-\theta-\i\pi\over p}}
\\
&=-c^2D{\pi^2p\over\sinh{\i\pi\over p}}\hI^{(+)}(\theta-\i\pi).
}
$$
It proves the first line of Eq.~(\ref{VSSdiffs}). The second line is
proved similarly.

To fix the coefficients $C_{AB}$ we have to analyze the commutation
relations for $Z_I(\theta)$. It will be shown below that the only
solution consistent with the commutation
relations is the solution~(\ref{Z-sausage}), i.~e.
\eq{
C_{+-}=\const,\qquad C_{--}=C_{++}=C_{-+}=0.
\label{C+-solution}
}

Now we can prove the commutation relations of the ZF algebra
for the sausage model. This problem is more difficult than in the case
of the operators $Z_{\ve\ve'}(\theta)$. The proof of the relations for
$Z_+(\theta_1)Z_+(\theta_2)$, $Z_+(\theta_1)Z_0(\theta_2)$, and
$Z_0(\theta_2)Z_+(\theta_1)$ is, in principle, the same and reduces to
the identities
\AlignStar{
S(\theta)^{++}_{++}
&=\bg_{33}(\theta)/\bg_{33}(-\theta).
\\
{1\over\sinh(x-y_2-z)}
&={\sinh(y_1-y_2)\over\sinh(y_1-y_2-2z)}
{1\over\sinh(x-y_2-z)}{\sinh(x-y_1+z)\over\sinh(x-y_1-z)}
\\*
&\quad
-{\sinh2z\over\sinh(y_1-y_2-2z)}
{1\over\sinh(x-y_1-z)}
}
According to Eq.~(\ref{VSequal}) a single contour can cross the pole at
$\xi=\theta$. Hence, the contour in the l.~h.~s.\ of the commutation
relation can be deformed into that in the r.~h.~s. The situation with
\eq{
Z_0(\theta_1)Z_0(\theta_2)
=S(\theta_1-\theta_2)^{+-}_{\0\0}Z_-(\theta_2)Z_+(\theta_1)
+S(\theta_1-\theta_2)^{-+}_{\0\0}Z_+(\theta_2)Z_-(\theta_1)
+S(\theta_1-\theta_2)^{00}_{00}Z_0(\theta_2)Z_0(\theta_1)
\label{Z0Z0commut}
}
is more complicated. First, as it was mentioned in
Sec.~\ref{sec-sausage}, the contours in the definition of the operator
$Z_-(\theta_i)$ cannot cross the pole at $\theta_i$. Second, the
contour in the l.~h.~s.\ in the operator $Z_0(\theta_1)$ goes above the
point~$\theta_2$ and the contour in~$Z_0(\theta_2)$ goes
below~$\theta_1$. On the other hand, in the r.~h.~s.\ a contour must go
above the point~$\theta_1$ if it belongs to the operator
$Z_I(\theta_2)$ and below~$\theta_2$ if it belongs to~$Z_I(\theta_1)$.

We shall prove this identity in two steps. On the first step we shall
consider all integrations so as if they would be done along the same
contour and show that the symmetrized integrands of the l.~h.~s.\ and
the r.~h.~s.\ coincide. On the second step we shall move all contours
to a single one and show that the residues due to the encountered poles
cancel each other.

For any product $X$ of the operators $Z_I(\theta_1)$ and
$Z_J(\theta_2)$ denote by $[X]$ this product, but with the contours in
the screening operators going for example above $\theta_1$ and below
$\theta_2$. We denote as
$$
\Delta[X]=X-[X].
$$
On the first step we prove that
\Align{
[Z_0(\theta_1)Z_0(\theta_2)]
&=S(\theta_1-\theta_2)^{+-}_{\0\0}[Z_-(\theta_2)Z_+(\theta_1)]
+S(\theta_1-\theta_2)^{-+}_{\0\0}[Z_+(\theta_2)Z_-(\theta_1)]
\notag
\\*
&\quad
+S(\theta_1-\theta_2)^{00}_{00}[Z_0(\theta_2)Z_0(\theta_1)].
\label{Z0Z0commutbracket}
}
On the second step we prove the equation
\Align{
\Delta[Z_0(\theta_1)Z_0(\theta_2)]
&=S(\theta_1-\theta_2)^{+-}_{\0\0}
\Delta[Z_-(\theta_2)Z_+(\theta_1)]
+S(\theta_1-\theta_2)^{-+}_{\0\0}
\Delta[Z_+(\theta_2)Z_-(\theta_1)]
\notag
\\
&\quad+S(\theta_1-\theta_2)^{00}_{00}
\Delta[Z_0(\theta_2)Z_0(\theta_1)].
\label{Z0Z0commutDelta}
}
Combining Eqs.~(\ref{Z0Z0commutbracket}) and (\ref{Z0Z0commutDelta}) we
obtain Eq.~(\ref{Z0Z0commut}).

The first step is performed just as in
Appendix~\ref{appendix-commut-gen} and reduces the proof of
Eq.~(\ref{Z0Z0commutbracket}) to a check of the identity of the
form~(\ref{fx1x2eq0}) with
$$
\Aligned{
f(x_1,x_2)
&={-2\cosh z\over\sinh(x_1-y_2-z)\sinh(x_2-y_2-z)}
{\sinh(x_1-y_2+z)\over\sinh(x_1-y_2-z)}
\\*
&\quad
-{\sinh(y_1-y_2)\sinh2z\over\sinh(y_1-y_2-z)\sinh(y_1-y_2-2z)}
{1\over\sinh(x_1-y_2-z)\sinh(x_2-y_2-z)}
\\
&\qquad\times
{\sinh(x_1-y_2+z)\over\sinh(x_1-y_2-z)}
{\sinh(x_2-y_1+z)\over\sinh(x_2-y_1-z)}
\\
&\quad
-{\sinh(y_1-y_2)\sinh2z\over\sinh(y_1-y_2-z)\sinh(y_1-y_2-2z)}
{1\over\sinh(x_1-y_1-z)\sinh(x_2-y_1-z)}
\\
&\quad
-{\sinh z\sinh2z+\sinh(y_1-y_2)\sinh(y_1-y_2-z)
\over\sinh(x_1-y_1-z)\sinh(x_2-y_1-z)}
{-2\cosh z\over\sinh(x_1-y_2-z)\sinh(x_2-y_1-z)}
\\*
&\qquad\times
{\sinh(x_1-y_1+z)\over\sinh(x_1-y_1-z)}.
}
$$

This computation would lead to the proof for operators with the same
contours. In our case, however, the contours are different and on the
second step we should take this fact into account.

Now we calculate $\Delta[Z_0(\theta_1)Z_0(\theta_2)]$. We have to
proceed in the same way as in the calculation of the operator
$\Delta(\theta)$ above. We skip the complete analysis and write the
contributions that come from the deformation of the contours. We have
to push the contour $C_1$ attached to $Z_0(\theta_1)$ down across the
point $\theta_2$ and the contour $C_2$ attached to $Z_0(\theta_2)$ up
across $\theta_1$. The result will have the form
$$
\Aligned{
\Delta[Z_0(\theta_1)Z_0(\theta_2)]
&=-2c^2\cos{\pi\over p}
\\
&\quad\times
\biggl(
\Res_{\xi_1=\theta_2}\Res_{\xi_2=\theta_2+\i\pi}
\hV(\theta_1)\hI^{(+)}(\xi_1)\hV^{(+)}(\theta_2)\hI^{(-)}(\xi_2)
{\pi\over\sinh{\xi_1-\theta_1-\i\pi\over p}}
{\pi\over\sinh{\xi_2-\theta_2-\i\pi\over p}}
\\
&\quad
+\Res_{\xi_1=\theta_1-\i\pi}\Res_{\xi_2=\theta_1}
\hV^{(-)}(\theta_1)\hI^{(+)}(\xi_1)\hV(\theta_2)\hI^{(-)}(\xi_2)
{\pi\over\sinh{\xi_1-\theta_1-\i\pi\over p}}
{\pi\over\sinh{\xi_2-\theta_2-\i\pi\over p}}
\biggr)
}
$$
Using Eq.~(\ref{resVI}) and the commutation relations~(\ref{VIcommut}) we
obtain
$$
\Delta[Z_0(\theta_1)Z_0(\theta_2)]
=-2\pi^2c^2Dp\cos{\pi\over p}\>
(\hI^{(-)}(\theta_2+\i\pi)\hV(\theta_1)
-\hV(\theta_2)\hI^{(+)}(\theta_1-\i\pi))
{1\over\sinh{\theta_1-\theta_2+\i\pi\over p}}
{\sinh{\theta_1-\theta_2\over p}
\over\sinh{\theta_1-\theta_2-2\pi\i\over p}}.
$$
Similarly, for the terms of the r.h.s.\ of Eq.~(\ref{Z0Z0commutDelta})
we find
\AlignStar{
\Delta[Z_0(\theta_2)Z_0(\theta_1)]
&=0,
\\
\Delta[Z_-(\theta_2)Z_+(\theta_1)]
&=c^2\Res_{\xi_1=\theta_2}\Res_{\xi_2=\theta_2+\i\pi}
\hV^{(+)}(\theta_2)\hI^{(+)}(\xi_1)\hI^{(-)}(\xi_2)\hV(\theta_1)
{\pi\over\sinh{\xi_1-\theta_2-\i\pi\over p}}
{\pi\over\sinh{\xi_2-\theta_2-\i\pi\over p}}
\\
&=-\pi^2c^2Dp\hI^{(-)}(\theta_2+\i\pi)\hV(\theta_1)
{1\over\sinh{\i\pi\over p}},
\\
\Delta[Z_+(\theta_2)Z_-(\theta_1)]
&=c^2\Res_{\xi_1=\theta_2-\i\pi}\Res_{\xi=\theta_2}
\hV^{(-)}(\theta_2)\hI^{(+)}(\xi_1)\hI^{(-)}(\xi_2)\hV(\theta_1)
{\pi\over\sinh{\xi_1-\theta_2-\i\pi\over p}}
{\pi\over\sinh{\xi_2-\theta_2-\i\pi\over p}}
\\
&=\pi^2c^2Dp\hI^{(+)}(\theta-\i\pi)\hV(\theta_2)
{1\over\sinh{\i\pi\over p}}
}
for the operator $Z_-(\theta)$ defined by Eq.~(\ref{Z-sausage}),
i.~e.\ for the case~Eq.~(\ref{C+-solution}). Taking into account
Eq.~(\ref{Smatrix-sausage}) for the explicit form of the $S$ matrix of
the sausage model, we obtain Eq.~(\ref{Z0Z0commutDelta}), which
completes the proof of Eq.~(\ref{Z0Z0commut}). The proof of this
identity is a basic one, because it fixes the form of the
operator~(\ref{Z-sausage}) from that of the operators~(\ref{Z+sausage})
and~(\ref{Z0sausage}). Indeed, if we would allow all terms in
Eq.~(\ref{VSScombination}) the value
$\Delta[Z_\pm(\theta_2)Z_\mp(\theta_1)]$ would contain some extra
terms, which are absent in $\Delta[Z_0(\theta_1)Z_0(\theta_2)]$.

The commutation relations for $Z_\pm(\theta_1)Z_\mp(\theta_2)$ can be
proved in the same way. For $Z_0(\theta_1)Z_-(\theta_2)$ and
$Z_-(\theta_1)Z_0(\theta_2)$ the proof is a little more complicated: it
remains one integration in products like
$\Delta[Z_0(\theta_1)\*Z_-(\theta_2)]$. First, we need to make sure
that the integrations in both sides of the equation are taken along the
same contour. Then we have to check that the integrands in both sides
coincide identically. One more complication appears in the commutation
relation for $Z_-(\theta_1)Z_-(\theta_2)$. There is a double
integration in this case and we have to symmetrize the integrands in
the integration variables before comparing them. We have checked all
these relations, but we omit the calculations as they are very
cumbersome.

The normalization constant $c$ in Eq.~(\ref{HZsausage}) can be
calculated as follows. Consider the operator product
$Z_+(\theta')Z_-(\theta)$. It can be shown that the only contribution
to the residue at the pole $\theta'=\theta+\i\pi$ comes from the
product
$$
\Aligned{{}
&c^2\int_{\cC_+}{d\xi_1\over2\pi}{d\xi_2\over2\pi}\,
\hV^{(-)}(\theta')\hV^{(+)}(\theta)\hI^{(+)}(\xi_1)\hI^{(-)}(\xi_2)
{\pi\over\sinh{\xi_1-\theta-\i\pi\over p}}
{\pi\over\sinh{\xi_2-\theta-\i\pi\over p}}
\\
&\qquad
=c^2\int_{\cC_+}{d\xi_1\over2\pi}{d\xi_2\over2\pi}\,
\lcolon\hV^{(-)}(\theta')\hV^{(+)}(\theta)
\hI^{(+)}(\xi_1)\hI^{(-)}(\xi_2)\rcolon
\\
&\qquad\quad\times
{\bg^{(-+)}(\theta-\theta')\bg^{(+-)}(\xi_2-\xi_1)
\over\bg^{(-+)}(\xi_1-\theta')\bg^{(++)}(\xi_1-\theta)
\bg^{(--)}(\xi_1-\theta')\bg^{(+-)}(\xi_1-\theta)}
{\pi\over\sinh{\xi_1-\theta-\i\pi\over p}}
{\pi\over\sinh{\xi_2-\theta-\i\pi\over p}}
}
$$
There are two effects here. First, there is pinching of the contour for
the variable $\xi_1$ between the points $\xi_1=\theta$ and
$\xi_1=\theta'-\i\pi$ and of the contour for $\xi_2$ between the points
$\xi_2=\theta+\i\pi$ and $\xi_2=\theta'$. This gives a double pole in
$\theta'-\theta$. Second, the function $\bg^{(-+)}(\theta-\theta')$ has
a zero at $\theta-\theta'=-\i\pi$. It decreases the multiplicity of the
pole by one. Finally, we have
$$
\Aligned{{}
&Z_+(\theta')Z_-(\theta)
=-c^2g^{(-+)\prime}(-\i\pi)(\theta-\theta')
{1\over\bg^{(-+)}(\theta-\theta')\bg^{(--)}(\theta-\theta'+\i\pi)}
\\
&\qquad\quad\times
\Res_{\xi_1=\theta}\Res_{\xi_2=\theta+\i\pi}
{\pi^2\bg^{(+-)}(\xi_2-\xi_1)
\over\bg^{(++)}(\xi_1-\theta)\bg^{(+-)}(\xi_2-\theta)
\sinh{\xi_1-\theta-\i\pi\over p}\sinh{\xi_2-\theta-\i\pi\over p}}
+O(1)
\\
&\qquad
=-{\i\over\theta'-\theta-\i\pi}
c^2r_p^{-4}(\pi p)^3{\Gamma^3(1-1/p)\over\Gamma(1/p)}
+O(1).
}
$$
Demanding that the coefficient at the pole $-\i/(\theta'-\theta-\i\pi)$
was equal to~1, we obtain Eq.~(\ref{cdef}).

%%%%%%%%%%%%%%%%%%%%%%%%%%%%%%%%%%%%%%%%%%%%%%%%%%%%%%%%%%%%%%%%%%%%%%%%

\section{First bound state in the cosine-cosine model}
\label{appendix-p3zero-bs}

Here we calculate explicitly the vertex operator $B^{1,1}_+(\theta)$
corresponding to the highest component ($Q_2=2$) of the quadruplet of
mass~$M_1$. It appears in the product
\eq{
Z_{++}(\theta'+\ipihalf(1-p_1))
Z_{-+}(\theta-\ipihalf(1-p_1))
\label{Z++Z-+prod}
}
in the limit $\theta'\to\theta$. Using Eqs.~(\ref{VVprod}),
(\ref{VIprod}), (\ref{wpm01}), (\ref{wpz}), we obtain for this
quantity:
\AlignStar{
c_1r_{p_1}^{-2}g_{33}(\theta-\theta'-2\delta)
&\int{d\xi\over2\pi}\,
\biggl(
\lcolon V(\theta'+\delta)
V(\theta-\delta)I_1^{(+)}(\xi)\rcolon\,
\e^{\kappa_1+(k_1+k_2)(\theta+\theta')/2-k_1\xi}
\\*
&\times
{\Gamma\left({\i(\xi-\theta')\over\pi p_1}-{1\over2}\right)
\over\Gamma\left({\i(\xi-\theta')\over\pi p_1}-{1\over2}+{1\over p_1}
\right)}
\Gamma\left({\i(\xi-\theta)\over\pi p_1}+{1\over2}-{1\over p_1}\right)
\Gamma\left(-{\i(\xi-\theta)\over\pi p_1}+{1\over2}\right)
\\
&+\i\lcolon V(\theta'+\delta)
V(\theta-\delta)I_1^{(-)}(\xi)\rcolon\,
\e^{-\kappa_1+(k_1+k_2)(\theta+\theta')/2-k_1\xi}
\\*
&\times
{\Gamma\left({\i(\xi-\theta')\over\pi p_1}+{1\over2}\right)
\over\Gamma\left({\i(\xi-\theta')\over\pi p_1}+{1\over2}+{1\over p_1}
\right)}
\Gamma\left({\i(\xi-\theta)\over\pi p_1}+{3\over2}-{1\over p_1}\right)
\Gamma\left(-{\i(\xi-\theta)\over\pi p_1}-{1\over2}\right)
\biggr)
}
with $\delta=\ipihalf(1-p_1)$ and $r_p=\e^{(C_E+\log\pi p)/p}$. The
integration contour is pinched between the poles at $\xi=\theta'-\i\pi
p_1/2$ and $\xi=\theta-\i\pi p_1/2$ in the first term and between the
poles at $\xi=\theta'+\i\pi p_1/2$ and $\xi=\theta+\i\pi p_1/2$ in the
second one. Pulling the contour through a pole in each term we obtain
for the quantity~(\ref{Z++Z-+prod}):
\Multline{
{\i\over\theta'-\theta}
\pi^2p_1^2c_1r_{p_1}^{-2}g_{33}(-2\delta)
{\Gamma\left(1-{1\over p_1}\right)\over\Gamma\left(1\over p_1\right)}
\Bigl(
\lcolon V(\theta+\delta)V(\theta-\delta)
I_1^{(+)}(\theta-{\ts{\i\pi p_1\over2}})\rcolon
\,\e^{-\kappa_3+k_2\theta}
\\
+\i\lcolon V(\theta+\delta)V(\theta-\delta)
I_1^{(-)}(\theta+{\ts{\i\pi p_1\over2}})\rcolon
\,\e^{\kappa_3+k_2\theta}
\Bigr)+O(1).
\notag
}
The first normal ordered product here tends to $V^{(-)}(\theta)$, while
the second one tends to $V^{(+)}(\theta)$ defined in
Eq.~(\ref{Vpmsausage}). Recall that though the expressions
(\ref{absausage})--(\ref{Ssausage}) were introduces for the sausage
model with $p=p_1\ge2$, they can be analytically continued to the
cosine-cosine model case $p=p_2\le2$. Finally, in the limit $p_3\to0$
we obtain
\eq{
Z_{++}(\theta'+\delta)Z_{+-}(\theta-\delta)
\to{\i C''\over\theta'-\theta}\i\rho V(k_2,\kappa_2|\theta)+O(1)
\label{ZppZpmpfusion}
}
with $V(k,\kappa|\theta)$ defined in Eq.~(\ref{Vsausage}) and
$$
C''=\pi^{1/2}
{G(p_1,-\i\pi)\over G(p_1,-\i\pi(1-p_1))G(p_2,-\i\pi(1-p_1))}.
$$
Using the identity~(\ref{G1id}) and
\eq{
{G(p;-\i\pi)\over G(p;\i\pi-\i\pi p)}
=p\sin{\pi\over p}\cdot
{\Gamma\left(p-1\over2\right)\over\Gamma\left(p\over2\right)}
\sqrt{p-1\over\pi}
\label{G2id}
}
we obtain
\eq{
{c_2C''\over c}=\Gamma_{1,1},
\qquad
p_1+p_2=2,
\label{c2C''overc}
}
where $c$ is given by Eq.~(\ref{cdef}) with $p=p_2$. The
expressions~(\ref{ZppZpmpfusion}) and~(\ref{c2C''overc}) prove the
representation for~$B^{1,1}_+$.

The proofs of representations for the operators $B^{1,1}_\rS$ and
$B^{1,1}_-$ are in the main features the same as for the operators
$B^{2,1}_\rA$ and $B^{2,1}_-$ in the case of sausage model and we will
not reproduce them here.

%%%%%%%%%%%%%%%%%%%%%%%%%%%%%%%%%%%%%%%%%%%%%%%%%%%%%%%%%%%%%%%%%%%%%%%%

\section{Regime~II}
\label{appendix-regimeII}

In this appendix we consider the model~(\ref{sfmodel}) in the
nonunitary regime~II, where
\eq{
p_1,p_2,p_3>0,
\qquad
p_1+p_2+p_3=2.
\label{regimeIIregion}
}
In this regime the theory is completely symmetric under the
permutations of the pairs $(p_1,\varphi_1)$, $(p_2,\varphi_2)$,
$(p_3,\varphi_3)$ and possesses the $U(1)\times U(1)\times U(1)$
symmetry described by the topological charges
\eq{
Q_i=\int dx^1\,j^0_i,
\qquad
j^\mu_i={\alpha_i\over\pi}\varepsilon^{\mu\nu}\d_\nu\varphi_i.
\label{Qidef}
}
The values of the charges must satisfy the conditions
\eq{
Q_i\in\Z,\qquad Q_i+Q_j\in2\Z.
\label{Qiquantization}
}
We would like to study this regime by means of the free field
representation technique. In Sec.~\ref{sec-identification} we introduced
the vertex operators $Z^i_{\ve\ve'}(\theta)$ defined in the regions
$\rI_i$ according to Eq.~(\ref{Zisf}). We assume that these vertex
operators can be analytically continued to the
region~(\ref{regimeIIregion}). It means that the spectrum in the
regime~II consists of three quadruplets of fundamental particles
$z^i_{\ve\ve'}$, $i\in\Z_3$ ($\ve$ and $\ve'$ are eigenvalues of the
operators $Q_{i+1}$ and $Q_{i+2}$ respectively) with the masses
\eq{
m_i=M_0\sin{\pi p_i\over2},
\qquad
M_0={\mu\over\pi^2}\,
\Gamma\left(p_1\over2\right)\Gamma\left(p_2\over2\right)
\Gamma\left(p_3\over2\right),
\label{regimeIImasses}
}
and their bound states. The $S$ matrices of the fundamental particles
are derived from the commutation relations of the vertex operators
$Z^i_{\ve\ve'}(\theta)$. From the free field representation described in
the present paper we can find that
\subeq{\label{HZalgebraII}\Align{
Z^i_{\vep_1\ve'_1}(\theta_1)Z^i_{\vep_2\ve'_2}(\theta_2)
&=-\sum_{\vep_3\ve'_3\vep_4\ve'_4}
S_{p_{i+1}}(\theta_1-\theta_2)^{\ve_3\ve_4}_{\ve_1\ve_2}
S_{p_{i+2}}(\theta_1-\theta_2)^{\ve'_3\ve'_4}_{\ve'_1\ve'_2}
Z^i_{\vep_4\ve'_4}(\theta_2)Z^i_{\vep_3\ve'_3}(\theta_1),
\label{Ziicommut}
\\
Z^i_{\ve\ve'_1}(\theta_1)Z^{i+1}_{\ve'_2\ve''}(\theta_2)
&=\ve\ve''\sum_{\ve'_3\ve'_4}
\hat S_{p_{i+2}}(\theta_1-\theta_2)^{\ve'_3\ve'_4}_{\ve'_1\ve'_2}
Z^{i+1}_{\ve'_4\ve''}(\theta_2)Z^i_{\ve\ve'_3}(\theta_1)
\label{Ziip1commut}
}}
with
\eq{
\Aligned{
\hat S_p(\theta)^{++}_{++}
=\hat S_p(\theta)^{--}_{--}
&=\exp\left(
-\i\int^\infty_0{dt\over t}\,
{\tanh{\pi t\over2}\sin\theta t\over\sinh{\pi pt\over2}}
\right)
\\
&=\prod^\infty_{n=0}
{\Gamma^2\left(-{\i\theta\over\pi p}+{2n+1\over p}+{1\over2}\right)
\Gamma\left({\i\theta\over\pi p}+{2n\over p}+{1\over2}\right)
\Gamma\left({\i\theta\over\pi p}+{2n+2\over p}+{1\over2}\right)
\over
\Gamma^2\left({\i\theta\over\pi p}+{2n+1\over p}+{1\over2}\right)
\Gamma\left(-{\i\theta\over\pi p}+{2n\over p}+{1\over2}\right)
\Gamma\left(-{\i\theta\over\pi p}+{2n+2\over p}+{1\over2}\right)},
\\
\hat S_p(\theta)^{+-}_{+-}
=\hat S_p(\theta)^{-+}_{-+}
&=-{\cosh{\theta\over p}\over\cosh{\i\pi-\theta\over p}}
\hat S_p(\theta)^{++}_{++},
%\label{S*pmpm}
\qquad
\hat S_p(\theta)^{-+}_{+-}
=\hat S_p(\theta)^{+-}_{-+}
=-{\sin{\pi\over p}\over\cosh{\i\pi-\theta\over p}}
\hat S_p(\theta)^{++}_{++}.
}\label{Shat}
}
Note that the matrix $\hat S_p(\theta)$ can be expressed in terms of the
sine-Gordon $S$ matrix:
$$
\textstyle
\hat S_p(\theta)
=\i\tanh\left({\theta\over2}+{\i\pi p\over4}\right)
S_p\left(\theta+{\i\pi p\over2}\right).
\eqno({\rm\ref{Shat}}')
$$
The proof of the relations~(\ref{Ziip1commut}), (\ref{Shat}) is similar
to that of the commutation relations for $Z^3(\theta)$ given in
Appendix~\ref{appendix-commut-gen} and we omit it.

The interpretation of the result is the following. Consider the $S$
matrix of two particles $z^i$ with the rapidity $\theta_1$ and $z^j$
with the rapidity $\theta_2$. Let $\theta=\theta_1-\theta_2$. The
physical scattering process takes place for $\theta>0$. Then for $j=i$
the $S$ matrix is given by the tensor product
$S_{ii}(\theta)=-S_{p_{i+1}}(\theta)\otimes S_{p_{i+2}}(\theta)$, where
the first tensor component acts on the space related to the topological
charge $Q_{i+1}$ and the second one on the space related to the
topological charge $Q_{i+2}$. For $j=i+1$ the $S$ matrix is given by
$S_{i,i+1}(\theta)=Q_iQ_{i+1}\hat S_{p_{i+2}}(\theta)$, while for
$j=i-1$ the $S$ matrix is given by $S_{i+1,i}(\theta)=Q_iQ_{i+1}\hat
S_{p_{i+2}}^{-1}(-\theta)$, where $\hat S_{p_{i+2}}(\theta)$ acts in
the space of the topological charge~$Q_{i+2}$.

Now consider the bound states. There are two types of bound states.
Bound states of the first type are related to the poles of the $S$
matrix for two fundamental particles of the same kind and are
completely analogous to those in the regime~I, described by
Eqs.~(\ref{thetain}),~(\ref{boundstatemass}). Explicitly, for any
$p_i<1$ ($i=1,2,3$) we have series of poles of the $S$ matrices
$S_{jj}(\theta)$ with $j=i\pm1$ at the points $\theta=\i u_{i,n}$ with
$u_{i,n}$ given by Eq.~(\ref{thetain}). The masses of the corresponding
bound states $b^{j;i,n}_\nu$, $\nu=+,\rS,\rA,-$ are given by
\eq{
M_{j;i,n}=2m_j\sin{\pi p_in\over2},
\qquad
j\ne i,
\quad
n=1,2,\ldots,
\quad
np_i<1.
\label{boundstatemassIIfirst}
}
It is important to note that
\eq{
M_{j;i,1}=M_{i;j,1}=2M_0\sin{\pi p_i\over2}\sin{\pi p_j\over2},
\qquad
j\ne i.
\label{MMi1}
}
It can be checked that the quadruplets $b^{j;i,1}$ and $b^{i;j,1}$
are the same.

Bound states of the second type correspond to the poles of the $S$ matrix
$S_{ij}(\theta)$ for $i\ne j$. Let $k\ne i,j$. Then the function $\hat
S_k(\theta)$ possesses poles at the points
\eq{
\theta=\i\pi-{\i\pi\over2}p_k-\i\pi p_kn,
\qquad
n=0,1,2,\ldots,
\quad
n<{1\over p_k}-{1\over2}.
\label{ijpoles}
}
The squares of the masses of the corresponding bound states read
\eq{
m^2_{k,n}
=M^2_0\left(\sin^2{\pi p_i\over2}+\sin^2{\pi p_j\over2}
-2\sin{\pi p_i\over2}\sin{\pi p_j\over2}
\cos\left({\pi p_k\over2}+\pi p_kn\right)\right).
\label{ijboundstatemasses}
}
We denote these bound states as $z^{k,n}_{\ve\ve'}$, $\ve,\ve'=\pm$.
Their multiplet structure is the same as that of the fundamental
particle $z^k$. Moreover,
$$
m_{k,0}=M_0\sin{\pi p_k\over2}=m_k,
$$
and the lightest bound states $z^{k,0}_{\ve\ve'}$ coincide with the
fundamental particles~$z^k_{\ve\ve'}$. The other bound state
quadruplets $z^{k,n}$, $n\ge1$ appear if $p_k<2/3$. Evidently, the
masses of these states are in the range
$$
m_k<m_{k,n}<m_i+m_j,
\qquad
n=1,2,\ldots,
\quad
n<{1\over p_k}-{1\over2}.
$$

We suppose to give a detailed description of the field theory
corresponding to the regime~II in another publication.

%%%%%%%%%%%%%%%%%%%%%%%%%%%%%%%%%%%%%%%%%%%%%%%%%%%%%%%%%%%%%%%%%%%%%%%%

\end{document}